\documentclass[11pt]{article}
\usepackage{epic,eepic,amsmath,amssymb,amsfonts}
\usepackage{amscd}
\begin{document}
\newtheorem{theorem}{Theorem}[section]
\newtheorem{lemma}[theorem]{Lemma}
\newtheorem{conjecture}[theorem]{Conjecture}
\newtheorem{corollary}[theorem]{Corollary}
\newtheorem{definition}[theorem]{Definition}
\newtheorem{assumption}[theorem]{Assumption}
\newtheorem{proposition}[theorem]{Proposition}
\newtheorem{problem}[theorem]{Problem}
\newtheorem{remark}[theorem]{Remark}
\newtheorem{example}[theorem]{Example}
\newcommand\eps[2] {\varepsilon({#1},{#2})}
\newcommand\epspm[2]{\varepsilon^\pm({#1},{#2})}
\newcommand\epsmp[2]{\varepsilon^\mp({#1},{#2})}
\newcommand\epsm[2]{\varepsilon^-({#1},{#2})}
\newcommand\epsp[2]{\varepsilon^+({#1},{#2})}
\def\emptyset{\varnothing}
\def\setminus{\smallsetminus}
\def\sys{\Delta}
\def\Diff{{\mathrm{Diff}}}
\def\Vir{{\mathrm{Vir}}}
\def\Aut{{\mathrm{Aut}}}
\def\End{{\mathrm{End}}}
\def\Ker{{\mathrm{Ker}}}
\def\Mor{{\mathrm{Mor}}}
\def\opp{{\mathrm{opp}}}
\def\Rep{{\mathrm{Rep}}}
\def\Tr{{\mathrm{Tr}}}
\def\tr{{\mathrm{tr}}}
\def\supp{{\mathrm{supp}}}
\def\ind{{\mathrm{ind}}}
\def\gen{{\mathrm{gen}}}
\def\id{{\mathrm{id}}}
\def\dual{{\mathrm{dual}}}
\def\ch{{\mathrm{ch}}}
\def\A{{\mathcal{A}}}
\def\B{{\mathcal{B}}}
\def\C{{\mathbb{C}}}
\def\E{{\mathcal{E}}}
\def\G{{\mathcal{G}}}
\def\K{{\mathcal{K}}}
\def\LL{{\mathcal{L}}}
\def\M{{\mathcal{M}}}
\def\N{{\mathbb{N}}}
\def\OO{{\mathcal{O}}}
\def\R{{\mathbb{R}}}
\def\U{{\mathcal{U}}}
\def\Q{{\mathbb{Q}}}
\def\Z{{\mathbb{Z}}}
\def\a{{\alpha}}
\def\e{{\varepsilon}}
\def\g{{\mathfrak g}}
\def\la{{\lambda}}
\def\si{{\sigma}}
\def\de{{\delta}}
\def\isom{{\cong}}
\newcommand{\Ad}{\mathop{\mathrm{Ad}}\nolimits}
\newcommand{\Hom}{\mathop{\mathrm{Hom}}\nolimits}
\newcommand{\coker}{\mathop{\mathrm{coker}}\nolimits}
\def\qed{{\unskip\nobreak\hfil\penalty50
\hskip2em\hbox{}\nobreak\hfil$\square$
\parfillskip=0pt \finalhyphendemerits=0\par}\medskip}
\def\proof{\trivlist \item[\hskip \labelsep{\bf Proof.\ }]}
\def\endproof{\null\hfill\qed\endtrivlist\noindent}

\title{Conformal Field Theory,\\
Tensor Categories and\\
Operator Algebras}
\author{
{\sc Yasuyuki Kawahigashi}\footnote{Supported in part by 
Research Grants and the Grants-in-Aid
for Scientific Research, JSPS.}\\
{\small Graduate School of Mathematical Sciences}\\
{\small The University of Tokyo, Komaba, Tokyo, 153-8914, Japan}
\\[0,05cm]
{\small and}
\\[0,05cm]
{\small Kavli IPMU (WPI), the University of Tokyo}\\
{\small 5-1-5 Kashiwanoha, Kashiwa, 277-8583, Japan}\\
{\small e-mail: {\tt yasuyuki@ms.u-tokyo.ac.jp}}}
\maketitle{}
\tableofcontents

\section{Introduction}

Quantum field theory is a vast theory in physics which has
many deep connections to various fields of mathematics.
Here we are interested in analytic aspects of
quantum field theory typically represented by Wightman 
axioms.

In classical field theory, a field is some kind of function
on a spacetime.  In quantum mechanics, numbers are replaced
with operators, so we consider operator-valued
functions on a spacetime, but it turns out that we have
to deal with something like a $\delta$-function, so we
consider operator-valued distributions on a spacetime.
Such an operator-valued distribution on a spacetime
is called a quantum field.
From this viewpoint, one quantum field theory consists
of a spacetime, its symmetry group and a family of
operator-valued distributions on the spacetime.
The {\sl Wightman axioms} \cite{SW}
give a mathematical axiomatization of such objects.

We have another approach to quantum field theory
based on operator algebras, which is called
{\sl algebraic quantum field theory} \cite{Ha}.  We now
explain this idea.  The operator-valued
distributions are technically difficult to handle,
since distributions are more difficult than
functions and we have to deal with unbounded
operators.  In algebraic quantum field theory,
we deal with only bounded linear operators.
These operators appears as ``observables'', so 
observables are more emphasized than states in
this approach.

Consider a (bounded) spacetime region $O$ and
an operator-valued distribution $\Phi$.  Take
a test function $f$ supported in $O$.  Then
$\langle \Phi, f\rangle$, the application of
$\Phi$ to the test function $f$, is a
(possibly unbounded) operator and it represents
an ``observable'' in the spacetime region $O$
if it is self-adjoint.  In one quantum field 
theory, we have many operator-valued distributions,
and we have many test functions, so we have
many observable on $O$ arising in this way.
We consider an operator algebra $\A(O)$ of bounded
linear operators generated
by these operators.  (From an unbounded self-adjoint
operator, we obtain a bounded operator through
exponentiation.)  In this way, we have a family
$\{\A(O)\}$ of operator algebras parameterized
by spacetime regions $O$.  We then impose mathematical
axioms on this which are ``natural'' from a physical
viewpoint.  We explain three of basic axioms here.

First, for a larger spacetime region, we have
more test functions from a mathematical viewpoint
and more observables from a physical viewpoint,
so we have a larger operator algebra.  Second,
if we have two spacelike separated regions, we
have no interactions between the two even at the
speed of light, so an operator representing an
observable in one commutes with that in the other.
This property is called locality, or Einstein
causality.  Finally, we have a projective
unitary representation of the spacetime symmetry
group and a certain covariance property of the
family of operator algebras with respect to it.

A family of operator algebras satisfying these
axioms is our mathematical
object in algebraic quantum field theory.  We try
to construct examples of such families, classify
them and study relations among their various
properties.

A framework where many researchers have worked is
a Minkowski space with the Poincar\'e group.
However, from an axiomatic viewpoint, we have been
unable to construct an example different from free
fields, so we do not have much progress recently.
We have seen much progress in the last 30 years on
the $(1+1)$-dimensional Minkowski space with a
higher symmetry, conformal symmetry, and this is
conformal field theory we would like to explain in
this text.

Our spacetime to start with is the 
$(1+1)$-dimensional Minkowski space, where
the space coordinate is $x$ and the time coordinate
is $t$.  We have
a certain decomposition machinery and can
restrict the theory to two light rays $\{x=\pm t\}$.
Then each light rays plays the role of
a ``spacetime'', where
the space and the time are mixed into one dimension.
We further compactify each line by adding the
point at $\infty$ so that we can deal with higher
symmetries including those moving $\infty$.
We thus reach the ``spacetime'' $S^1$, the 
one-dimensional circle.  Our spacetime region is
now an arc contained in $S^1$.  We also have
to specify the spacetime symmetry group, and
we choose the orientation preserving diffeomorphism
group $\Diff(S^1)$.  This is an infinite dimensional
Fr\'echet Lie group and represents a large symmetry.
With these choices, our quantum field theory is 
called {\sl chiral conformal field theory} \cite{BPZ}.
We say ``chiral'' because we have only one of the
two light rays.

We also have another mathematical theory to deal
with chiral conformal field theory, that is, theory of
vertex operator algebras.  This approach is based on
algebraic axiomatization of Fourier expansions of
a family of operator-distributions
on the circle $S^1$.
It is not a purpose here
to develop its full theory, but we would like to
compare this approach to that based on operator algebras.
It is expected that the operator algebraic approach
to chiral conformal field theory is mathematically
equivalent to the approach using vertex operator
algebras, at least in nice situations, so we
focus on the relations between the two approaches.

We also consider conformal field theory on the entire
$(1+1)$-dimensional Minkowski space.  Such a theory
is called full conformal field theory.  It is also
interesting to consider the theory on the half
Minkowski space $\{x>0\}$.  This is called a
boundary conformal field theory.  We present
outlines of these theories at the end.

Many important results are scattered
in literature, and sometimes different names are used
for the same notion, and sometimes the same name
mean different notions, so it is not easy for a
beginner to grasp the structure of the theory.
Our aim here is to present a clear outline of the
entire theory.  

For those who are interested in the operator algebraic
approach to conformal field theory from other topics
such as vertex operator algebras or quantum groups,
it is often hard to understand the argument
due to technical difficulty.  Here
we present rather ideas than rigorous proofs or technical
details so that the entire structure is comprehensible
without knowing technical details of operator algebras.

This text is based on the lectures given by the
author in the fall of 2014 at the University of
Tokyo.  This text is also partly based on \cite{K2}.

We refer a reader to \cite{DMS} for general conformal
field theory, \cite{G2} for background materials,
\cite{FFRS1}, \cite{FFRS2}, \cite{FuRS1}, \cite{FuRS2},
\cite{FuRS3}, for another approach to full conformal
field theory and tensor categories,
and \cite{R4} for a recent review on the operator
algebraic approach to conformal field theory.

The author thanks Y.-Z. Huang, R. Longo and Y. Ogata
for their helpful comments.  The author also thanks a referee
for many helpful comments.

\section{Basics of Operator Algebras}

We prepare basic facts about operator algebras which are
necessary for studying conformal field theory.  
We do not include proofs.
As standard references, we list the textbooks \cite{T1,T2,T3}.

\subsection{$C^*$-algebras and von Neumann algebras}

We provide some minimal basics on theory of operator algebras.
For simplicity, we assume all Hilbert spaces appearing in 
the text are separable.  Our convention for notations is as follows.
We use $A, B, \dots, M, N, \dots$ for operator algebras,
$a,b,\dots,u,v,\dots,x,y,z$ for operators,
$H,K,\dots$ for Hilbert spaces and
$\xi,\eta,\dots$ for vectors in Hilbert spaces.

Let $H$ be a complex Hilbert space and $B(H)$ be the set of all
bounded linear operators on $H$.  We have a natural
$*$-operation $x\mapsto x^*$ on $B(H)$.
It is common to use seven topologies on $B(H)$ as in 
\cite[Section II.2]{T1}, but here we need only two of them as follows.

\begin{definition}{\rm 
(1) The {\sl norm topology} on $B(H)$ is induced by the operator norm
$\|x\|=\sup_{\|\xi\|\le1} \|x\xi\|$.

(2) We define convergence $x_i\to x$ in the 
{\sl strong operator topology}
when we have $x_i\xi \to x\xi$ for all $\xi\in H$.
}\end{definition}

Note that the {\sl strong} operator topology is {\sl weaker} than the
norm topology.  This is because we have another topology called
the weak operator topology, which is weaker than the strong
operator topology.  The norm convergence is uniform convergence
on the unit ball of the Hilbert space and the strong operator
convergence is pointwise convergence on the Hilbert space.

\begin{definition}{\rm 
(1) Let $M$ be a subalgebra of $B(H)$ which is closed in the
$*$-operation and contains the identity operator $I$.  We
say $M$ is a {\sl von Neumann algebra} if $M$ is closed in the
strong operator topology.

(2) Let $A$ be a subalgebra of $B(H)$ which is closed in the
$*$-operation.  We
say $A$ is a {\sl $C^*$-algebra} if $A$ is closed in the
norm topology.
}\end{definition}

By this definition, a von Neumann algebra is automatically
a $C^*$-algebra, but a von Neumann algebra is quite different
from ``ordinary'' $C^*$-algebras, so we often think that
operator algebras have two classes, von Neumann algebras
and $C^*$-algebras.

We have a natural notion of isomorphisms for $C^*$-algebras
and von Neumann algebras.  An isomorphism of a $C^*$-algebra
onto another has norm continuity automatically.
(See \cite[Corollary I.5.4]{T1}.)
An isomorphism of a von Neumann algebra onto another
has appropriate continuity automatically.
(See \cite[Corollary III.3.10]{T1}.)

A commutative $C^*$-algebra containing the multiplicative
unit is isomorphic to $C(X)$, where $X$ is a compact
Hausdorff space and $C(X)$ means the algebra of all complex-valued
continuous functions.  A commutative $C^*$-algebra without
a multiplicative
unit is isomorphic to $C_0(X)$, the algebra of all
complex-valued continuous
functions on a locally compact Hausdorff space $X$
vanishing at infinity.
(See \cite[Theorem I.4.4]{T1}.)
A commutative von Neumann algebra is isomorphic to
$L^\infty(X,\mu)$, where $(X,\mu)$ is a measure space.
(See \cite[Proposition XIII.1.2]{T3}.)

Easy examples are as follows.

\begin{example}{\rm 
Let $H$ be $L^2([0,1])$.  The polynomial algebra
$\C[x]$ acts on $H$ by left multiplication.
The image of this representation is a $*$-subalgebra
of $B(H)$.  Its norm closure is isomorphic to
$C([0,1])$ and its closure in the strong operator
topology is isomorphic to $L^\infty([0,1])$.
}\end{example}

If a $C^*$-algebra is finite dimensional, then it is also
a von Neumann algebra, and it is isomorphic to
$\bigoplus_{j=1}^k M_{n_j}(\C)$, where $M_n(\C)$ is the
$n\times n$-matrix algebra.
(See \cite[Theorem I.11.9]{T1}.)

\begin{definition}{\rm
For $X\subset B(H)$, we set
$$X'=\{y\in B(H)\mid xy=yx \textrm{ for all }x\in X\}.$$
We call $X'$ the commutant of $X$.
}\end{definition}

We have the following proposition for von Neumann algebras.
(See \cite[Proposition II.3.9]{T1}.)

\begin{proposition}
Let $M$ be a subalgebra of $B(H)$ closed under the $*$-operation
and containing $I$.  Then the double commutant $M''$ is equal
to the closure of $M$ in the strong operator topology.
\end{proposition}

Note that taking the commutant is a purely algebraic operation, but
the above Proposition says it contains information on the topology.

For von Neumann algebras $M\subset B(H)$ and $N\subset \B(K)$, we 
have natural operations of the direct sum $M\oplus N\subset
B(H\oplus K)$ and the tensor product $M\otimes N\subset B(H\otimes K)$.

We have the following proposition.
(See \cite[Proposition II.3.12]{T1}.)

\begin{proposition}
The following conditions are equivalent for a von Neumann
algebra $M$.

(1) The von Neumann algebra $M$ is not isomorphic to the direct sum
of two von Neumann algebras.

(2) The center $M\cap M'$ of $M$ is $\C I$.

(3) Any two-sided ideal of $M$ closed in the strong operator topology
is equal to $0$ or $M$.
\end{proposition}

A natural name for such a von Neumann algebra would be a simple
von Neumann algebra, but for a historic reason, this name is not
used and such a von Neumann algebra is called a {\sl factor}
instead.

\subsection{Factors of types I, II and III}

The matrix algebra $M_n(\C)$ is a factor and the algebra $B(H)$ is also
a factor.  The former is called a factor of type I$_n$, and the latter is
called a factor of type I$_\infty$ if $H$ is infinite dimensional.
We introduce another example of a factor.

\begin{example}{\rm
\label{21fac}
For $x\in M_2(\C)\otimes\cdots\otimes M_2(\C)$, we consider the
embedding $x\mapsto x\otimes I_2\in M_2(\C)\otimes\cdots\otimes M_2(\C)
\otimes M_2(\C)$, where $I_2$ is the identity matrix in $M_2(\C)$.
We identify $M_2(\C)\otimes\cdots\otimes M_2(\C)$ with $M_{2^k}(\C)$, 
where $k$ is the number of the factorial components $M_2(\C)$ so
that the above embedding is compatible with this identification.
Let $\tr$ be the usual trace $\Tr$ on $M_{2^k}(\C)$ divided 
by $2^k$.  Then this $\tr$ is compatible with the embedding
$M_{2^k}(\C)$ into $M_{2^{k+1}}(\C)$.  Let $A$ be the increasing
union of $M_{2^k}(\C)$ with respect to this embedding.  This is
a $*$-algebra and the 
linear functional $\tr$ is well-defined on $A$.

Setting $(x,y)=\tr(y^*x)$ for $x,y\in A$, we make $A$ a 
pre-Hilbert space.  (We use a convention that an inner product is
linear in the first variable.)  Let $H$ be its completion.  For
$x\in A$, let $\pi(x)$ be the multiplication operator
$y\mapsto xy$ on $A$.  This is extended to a bounded
linear operator on $H$ and we still denote the extension
by $\pi(x)$.  Then $\pi$ is a $*$-homomorphism from $A$ 
into $B(H)$.  The norm closure of $\pi(A)$ is a $C^*$-algebra
called the type $2^\infty$-UHF algebra or the CAR algebra.
(The abbreviations UHF and CAR stand for
``Uniformly Hyperfinite'' and ``Canonical Anticommutation
Relations'', respectively.)
The closure $M$ of $\pi(A)$ in the strong operator
topology is a factor and
it is called the {\sl hyperfinite type II$_1$ factor}.
(Here the name ``hyperfinite'' means that we have an
increasing union of finite dimensional von Neumann algebras
which is dense in the strong operator topology.  A
hyperfinite type II$_1$ factor is unique up to
isomorphism \cite[Theorem XIV.2.4]{T3}.  Sometimes, the terminology 
{\sl AFD}, standing for ``approximately
finite dimensional'', is used instead of ``hyperfinite''.)

The linear functional $\tr$ is extended to 
$M$ and satisfies the following properties.
\begin{enumerate}
\item We have $\tr(xy)=\tr(yx)$ for $x,y\in M$.
\item We have $\tr(x^*x)\ge0$ for $x\in M$ and if $\tr(x^*x)=0$,
then we have $x=0$.
\item We have $\tr(I)=1$.
\end{enumerate}
(See \cite[Section XIV.2]{T3}.)
}\end{example}

If an infinite dimensional von Neumann algebra has a
linear functional $\tr$ satisfying the above three
conditions, then it is called a type II$_1$ factor.
Such a linear functional is unique on each
type II$_1$ factor and called a {\sl trace}.  
(See \cite[Section V.2]{T1}.)
There are many type II$_1$ factors which are not
hyperfinite.

A type II$_\infty$ factor is a tensor product of
a type II$_1$ factor and $B(H)$ for an infinite
dimensional Hilbert space $H$.

\begin{definition}{\rm
Two projections $p,q$ in a von Neumann algebra
are said to be {\sl equivalent}
if we have $u$ in the von Neumann algebra
satisfying $p=uu^*$ and $q=u^*u$.
}\end{definition}

If $uu^*$ is a projection, then $u^*u$ is also
automatically a projection, and such $u$ is called
a {\sl partial isometry}.

\begin{definition}{\rm
A factor is said to be of type III if
any two non-zero projections in it are 
equivalent and it is not isomorphic to $\C$.
}\end{definition}

This definition is different from the usual
definition of a type III factor, but means
the same condition since we consider only
separable Hilbert spaces.  (See
\cite[Definition V.1.17]{T1} and
\cite[Proposition V.1.39]{T1}.)

Two equivalent projections are analogous to two
sets having the same cardinality in set theory.
Then the property analogous to the above in
set theory would be that any two non-empty
subsets have the same cardinality for a set
which is not a singleton.  Such a condition
is clearly impossible in set theory. Still,
based on this analogy, we interpret that the
above property for a type III factor manifests
a very high level of infiniteness.  Because of
this analogy, a type III factor is also
called purely infinite.

The following is an example of a type III factor.

\begin{example}{\rm
Fix $\lambda$ with $0<\lambda<1$ and set
$\phi_\lambda: M_2(\C)\to \C$ by
$$\phi_\lambda\left(\left(
\begin{array}{cc}
a & b \\
c & d
\end{array}\right)\right)=\frac{a}{1+\lambda}
+\frac{d\lambda}{1+\lambda}.$$

Let $A$ be the same as in Example \ref{21fac}.
The linear functionals
$\phi_\lambda\otimes\cdots\otimes\phi_\lambda$
on $M_2(\C)\otimes\cdots\otimes M_2(\C)$
are compatible with the embedding, so 
$\phi^\lambda=\bigotimes \phi_\lambda$ is well-defined on
$A$.  We set the inner product on $A$ by
$(x,y)=\phi^\lambda(y^*x)$ and set $H$ be its completion.
Let $\pi(x)$ be the left multiplication of $x$ on $A$,
then it is extended to a bounded linear operator on $H$
again.  The extension is still denoted by $\pi(x)$.
The norm closure of $\pi(A)$ is isomorphic to the
$2^\infty$ UHF algebra in Example \ref{21fac}.
The closure $M$ of $\pi(A)$ in the strong operator
topology is a type III factor, and we have 
non-isomorphic von Neumann algebras for different
values of $\lambda$.  They are called the {\sl Powers
factors}.  (See \cite[Section XVIII.1]{T3}.)
}\end{example}

It is non-trivial that Powers factors are of
type III.  Here we give a rough idea why this 
should be the case.  On the one hand,
two equivalent projections
are regarded as ``having the same size''.
On the other hand, now the functional
$\phi^\lambda$ is also involved in measuring
the size of projections.  The two projections
$$\left(
\begin{array}{cc}
1 & 0 \\
0 & 0
\end{array}\right),
\left(
\begin{array}{cc}
0 & 0 \\
0 & 1
\end{array}\right)$$
are equivalent, but have different ``sizes'' according
to $\phi^\lambda$.  Because of this incompatibility,
we do not have a consistent way of measuring sizes of
projections, and it ends up that all nonzero projections
are of ``the same size'' in the sense of equivalence.

Connes has refined the class of type III factors into
those of type III$_\lambda$ factors with $0\le\lambda\le1$.
The Powers factors as above are of type III$_\lambda$ 
with $0 < \lambda <1$.  If $M$ and $N$ are the Powers
factors of type III$_\lambda$ and III$_\mu$, respectively,
and $\log\lambda\log\mu$ is irrational, then 
$M\otimes N$ is a factor of type III$_1$.  The isomorphism
class of $M\otimes N$ does not depend on $\lambda$ and $\mu$
as long as $\log\lambda/\log\mu$ is irrational, and this
factor is called the {\sl Araki-Woods factor} of type III$_1$.
The factor which appears in conformal field theory is this
one.  (See \cite[Chapter XII]{T2}.)  The Powers and Araki-Woods
factors are hyperfinite.  There are many type III factors
which are not hyperfinite, but they do not appear in
conformal field theory.

For a von Neumann algebra $M\subset B(H)$ and a unit
vector $\xi\in H$ with $\overline{M\xi}=\overline{M'\xi}=H$,
we have the {\sl modular operator} $\Delta_\xi$, a positive and
possibly unbounded operator, and the {\sl modular conjugation}
$J_\xi$, an antiunitary involution, on $H$.
For $x\in M$ and $t\in\R$, we have 
$\sigma_\xi(x)=\Ad(\Delta_\xi^{it})(x)\in M$.
The one parameter automorphism group $\sigma_t$
is called the {\sl modular automorphism group}
of $M$ with respect to $\xi$.  This is the Tomita-Takesaki
theory and classification of type III factors into
type III$_\lambda$, $0\le\lambda\le1$, is based on this.
(See \cite[Chapter VI]{T2}
for more details in a more general setting.)

\subsection{Dimensions and modules}

First consider a trivial example of $M_2(\C)$.  We would like
to find the ``most natural'' Hilbert space on which $M_2(\C)$
acts.  One might think it is clearly $\C^2$, but from our
viewpoint of infinite dimensional operator algebras, it is
not the right answer.  Instead, let $M_2(\C)$ act on itself
by the left multiplication and put a Hilbert space
structure on $M_2(\C)$ so that a natural system $\{e_{ij}\}$
of matrix units gives an orthonormal basis.  Then the
commutant of the left multiplication of $M_2(\C)$ is
exactly the right multiplication of $M_2(\C)$ and thus
the left and right multiplications are now symmetric.  This
is the ``natural'' representation from our viewpoint, and
we would like to consider its infinite dimensional analogue.

Let $M$ be a type II$_1$ factor with $\tr$.  Put an inner 
product on $M$ by $(x,y)=\tr(y^*x)$ and denote its completion
by $L^2(M)$.  The left and right multiplications by an element 
of $M$ on $M$ extend to bounded linear operators on $L^2(M)$.
We say that $L^2(M)$ is a left $M$-module and also
a right $M$-module.
(As long as we consider separable Hilbert spaces, any module
of a factor of type II or III gives a representation which
has appropriate continuity
automatically.  See \cite[Theorem 5.1]{T1}.)

Let $p$ be a projection in $M$.  Then $L^2(M)p$ is naturally
a left $M$-module.  For projections $p_n\in M$,
we define $\dim_M \bigoplus_n L^2(M)p_n=\sum_n \tr(p_n)$.
Then it turns out that any left $M$-module $H$ is unitarily
equivalent to this form and this number $\dim_M H\in [0,\infty]$
is well-defined.  It is called the {\sl dimension}
of a left $M$-module $H$,
and is a complete invariant up to unitary equivalence.
Note that we have $\dim_M L^2(M)=1$.
(See \cite[Section V.3]{T1}, where the dimension is
called the coupling constant under a more general setting.)

For a type III factor $M$, any two nonzero left $M$-modules
are unitarily equivalent.
(See \cite[Corollary V.3.2]{T1}.)

In this sense, representation theory of a type II$_1$ factor
is dictated by a single number, the dimension, and that of
a type III factor is trivial.  Note that a left module of
a type II or III factor is never irreducible.

\subsection{Subfactors}

Let $M$ be a type II$_1$ factor with $\tr$.  Suppose
$N$ is a von Neumann subalgebra of $M$ and $N$ is also
a factor of type II$_1$.  We say $N\subset M$ is a
{\sl subfactor}.  (The unit of $N$ is assumed to be the
same as that of $M$.)  The Hilbert space $L^2(M)$ is a left
$M$-module, but it is also a left $N$-module and we
have $\dim_N L^2(M)$.  This number is called the 
{\sl index} of the subfactor and denoted by $[M:N]$.
The index value is in $[1,\infty]$.
The celebrated theorem of Jones \cite{J1} is as follows.
(Also see \cite[Theorem 9.16]{EK}.)

\begin{theorem}
The set of the index values of subfactors is equal to
$$\{ 4\cos^2\frac{\pi}{n}\mid n=3,4,5,\dots\}
\cup [4,\infty].$$
\end{theorem}

There have been many results on the case $M$ is hyperfinite,
when $N$ is automatically hyperfinite.  It is often 
assumed that the index value is finite.
A subfactor $N\subset M$ is said to be 
{\sl irreducible} if we have $N'\cap M=\C$.  Irreducibility
of a subfactor is also often assumed.

A subfactor is an analogue of an inclusion
$L^\infty(X,{\mathcal{B}_1},\mu)\subset 
L^\infty(X,{\mathcal{B}_2},\mu)$ of commutative von Neumann
algebras where ${\mathcal{B}_1}$ is a $\sigma$-subalgebra
of $\mathcal{B}_2$ on the space $X$ and $\mu$ is a probability
measure.  That is, a smaller commutative
von Neumann algebra means that
we have less measurable sets.  For
$f\in L^\infty(X,{\mathcal{B}_2},\mu)$, we regard it
as an element in $L^2(X,{\mathcal{B}_2},\mu)$ and
apply the orthogonal projection $P$ onto 
$L^2(X,{\mathcal{B}_1},\mu)$.  Then $Pf$ is in
$L^\infty(X,{\mathcal{B}_1},\mu)$, and this map
from $L^\infty(X,{\mathcal{B}_2},\mu)$ onto
$L^\infty(X,{\mathcal{B}_1},\mu)$ is called a
{\sl conditional expectation}.  For a subfactor
$N\subset M$ of type II$_1$, we have a similar map
$E:M\to N$ satisfying the following properties.
\begin{enumerate}
\item $E(x^*x) \ge 0$ for all $x\in M$.
\item $E(x) = x$ for all $x \in N$.
\item $\tr(xy) = \tr(E(x)y)$ for all $x \in M$, $y \in N$.
\item $E(axb) = aE(x)b$ for all $x \in M$, $a, b \in N$.
\item $E(x^*) = E(x)^*$ for all $x \in M$.
\item $\|E(x)\| \le \|x\|$ for all $x \in M$.
\end{enumerate}
This map $E$ is also called the {\sl conditional
expectation} from $M$ onto $N$.  Actually, properties
1, 4 and 5 follow from 2 and 6.  
(See \cite[Theorem III.3.4]{T1}.)

Pimsner and Popa \cite{PP} proved that
we have $E(x)\ge\frac{1}{[M:N]}x$ for all
positive $x\in M$ and the 
coefficient $\frac{1}{[M:N]}$ is the
best possible for this inequality
under the convention $1/\infty=0$.
(See \cite[Theorem 9.48]{EK}.)
A general linear map $E$ from a von Neumann algebra 
$M$ onto a von Neumann subalgebra $N$ satisfying
the above properties 2 and 5 is also called
a {\sl conditional expectation}.  A conditional
expectation $E$ is said to be {\sl faithful}
if $E(x)=0$ for a positive $x\in M$ implies $x=0$.
A conditional expectation is said to be {\sl normal}
if it satisfies appropriate continuity.  In this
text, we simply say a conditional expectation 
for a normal faithful conditional expectation.

Kosaki \cite{K} extended the definition of 
the index of a subfactor to the index of a conditional
expectation $E:M\to N$ for a subfactor of type III.
If such a conditional expectation does not exist,
we interpret that the index of $N$ in $M$ is $\infty$.
If the subfactor $N'\cap M$ is irreducible and
we have a conditional expectation from $M$ to $N$,
then such a conditional expectation is unique, so
we call its index the index of the subfactor,
$[M:N]$.  Many results on indices of type III factors
are parallel to those of type II factors.
If one conditional expectation $E:M\to N$ has a finite
index, the all other conditional expectations
from $M$ onto $N$ have finite indices, and we have
the unique conditional expectations achieving the
minimum value of the indices.  We define $[M:N]$ to
be the index of this conditional expectation.
We have the following results.
(See \cite{Hi} for details.) 

\begin{proposition}
\label{tensor}
(1) For two subfactors
$N\subset M$ and $P\subset Q$, we have
$[M\otimes Q:N\otimes P]=[M:N][Q:P]$.

(2) For a subfactor $N\subset M$, we have
$[M:N]=[N':M']$.
\end{proposition}

\subsection{Bimodules and relative tensor products}

Let $M$ be a type II factor with $\tr$.  The Hilbert space
$L^2(M)$ is a left $M$-module and a right $M$-module.
Furthermore, the left action of $M$ and the right action
of $M$ commute, so this is an $M$-$M$ bimodule.
We consider a general $M$-$N$ bimodule ${}_M H_N$ for
type II$_1$ factors $M$ and $N$.
For a bimodule ${}_M H_N$,
we have $\dim H_N$ defined in a similar way to the
definition of $\dim_M H$.  If we have
$\dim_M H \dim H_N < \infty$, we say that the bimodule
is of finite type.  We consider only bimodules of finite
type. 

Let $M,N,P$ be type II$_1$ factors and 
consider a general $M$-$N$ bimodule ${}_M H_N$
and an $N$-$P$ bimodule ${}_N K_P$.  Then we can
define a {\sl relative tensor product}
${}_M H\otimes_N K_P$, which is an $M$-$P$ bimodule.
This is again of finite type.
We have 
$${}_M L^2(M)\otimes_M H_N\isom 
{}_M H\otimes_N L^2(N)_N\isom {}_M H_N.$$
(See \cite[Section 9.7]{EK}.)

For an $M$-$N$ bimodule ${}_M H_N$, we have the
{\sl contragredient} (or {\sl conjugate}) bimodule
${}_N \bar H_M$.  As a Hilbert space, it consists of
the vectors of the form $\bar\xi$ with $\xi\in H$
and has operations $\overline{\xi+\eta}=\bar\xi+\bar\eta$
and $\overline{\a\xi}=\bar\a \bar\xi$.  The bimodule
operation is given by $x\cdot\bar\xi\cdot y=
\overline{y^*\cdot\xi\cdot x^*}$, where $x\in N$ and $y\in M$.
This is again of finite type.

For $M$-$N$ bimodules ${}_M H_N$ and ${}_M K_N$,
we say that a bounded linear map $T:H\to K$ is an
{\sl intertwiner} when we have $T(x\xi y)=xT(\xi)y$
for all $x\in M$, $y\in N$, $\xi\in H$.  We denote
the set of all the intertwiners from $H$ to $K$ by
$\Hom({}_M H_N, {}_M K_N)$.
We say that ${}_M H_N$ is {\sl irreducible} if
$\Hom({}_M H_N, {}_M H_N)=\C I$.  We have a natural
notion of a direct sum
${}_M H_N\oplus {}_M K_N$.  

A bimodule ${}_M H_N$  
decomposes into a finite direct sum of irreducible
bimodules, because we assume ${}_M H_N$ is of
finite type.  (See \cite[Proposition 9.68]{EK}.)

Start with a subfactor $N\subset M$ of type II$_1$ 
with $[M:N]<\infty$.  Then the $N$-$M$ bimodule 
${}_N L^2(M)_M$ is of finite type.
The finite relative tensor products of ${}_N L^2(M)_M$
and ${}_M L^2(M)_M$ and their irreducible decompositions
produce four kinds of bimodule, $N$-$N$, $N$-$M$, $M$-$N$
and $M$-$M$.  They are all of finite type.
We have only finitely many irreducible
bimodules for one of them up to isomorphisms only
if we have only finitely many irreducible bimodules
for all four kinds.  When this finiteness condition
holds, we say the subfactor $N\subset M$ is of
{\sl finite depth}.  If the index is less than 4,
the subfactor is automatically of finite depth.
(See \cite[Section 9]{EK} for more details.)

Consider a type II$_1$ subfactor $N\subset M$ of
finite depth and pick a representative from each
of finitely many isomorphism classes of the $N$-$N$
bimodules arising in the above way.  For each
such ${}_N X_N$, we have $\dim_N X=\dim X_N$.
For such ${}_N X_N$ and ${}_N Y_N$, the relative
tensor product ${}_N X\otimes_N Y_N$ is isomorphic
to $\bigoplus_{j}^k n_j {}_N {Z_j}_N$, where
$\{{}_N {Z_j}_N\}$ is the set of the representatives.
This gives {\sl fusion rules} and the bimodule
${}_N L^2(M)_N$ plays the role of the identity
for the relative tensor product.  
(Note that the name ``fusion'' sometimes 
means the relative tensor product operation
is commutative, but we do not assume this here.)
For each ${}_N {Z_j}_N$, we have $k$ with
$\overline{{}_N {Z_j}_N}\isom{}_N {Z_k}_N$.
We also have the {\sl Frobenius reciprocity},
$\dim\Hom({}_N X\otimes_N Y_N, {}_N Z_N)=
\dim\Hom({}_N X, {}_N Z\otimes_N \bar Y_N)$.
(See \cite[Section 9.8]{EK}.)
The $N$-$N$ bimodules isomorphic to finite direct
sums of these representative $N$-$N$ bimodule make
a {\sl unitary fusion category}, which is an
abstract axiomatization of this system of
bimodules and is some kind of a {\sl tensor category}.  
A basic model of unitary fusion category is that of
finite dimensional unitary representations of a
finite group.  We recall the definitions for unitary
fusion categories as follows.  (See \cite{BK},\cite{CE}.)

\begin{definition}{\rm
A category $\mathcal C$ is called an {\sl abelian}
 category over $\C$ is we have the following.
\begin{enumerate}
\item All $\Hom(U,V)$ are $\C$-vector spaces and the
compositions 
$$\Hom(V,W)\times\Hom(U,V)\to\Hom(U,W), \quad (\phi,\psi)
\mapsto\phi\circ\psi$$
are $\C$-bilinear, where $U,V,W$ are objects in $\mathcal C$.
\item We have a zero objects $0$ in $\mathcal C$ with
$\Hom(0,V)=\Hom(V,0)=0$ for all objects $V$ in $\mathcal C$.
\item We have finite direct sums in $\mathcal C$.
\item Every morphism $\phi\in\Hom(U,V)$ has a kernel
$\ker\phi\in\Mor\mathcal C$ and a cokernel $\coker\phi
\in\Mor\mathcal C$.
\item Every morphism is the composition of an 
epimorphism followed by a monomorphism.
\item If $\ker\phi=0$, then we have $\phi=\ker(\coker\phi)$
and if $\coker\phi=0$, then we have $\phi=\coker(\ker\phi)$.
\end{enumerate}
}\end{definition}

\begin{definition}{\rm
An object $U$ in an abelian category $\mathcal C$ is called
{\sl simple} if any injection $V\hookrightarrow U$ is either 0 or 
an isomorphism.

An abelian category $\mathcal C$ is called {\sl semisimple}
if any object $V$ is isomorphic to a direct sum of simple
ones, $V\isom \bigoplus_i n_i V_i$, where
$V_i$ are simple objects, $n_i$ are multiplicities and
only finitely many $n_i$ are nonzero.
}\end{definition}

\begin{definition}{\rm
An abelian category $\mathcal C$ is called a
{\sl monoidal category} if we have the following.
\begin{enumerate}
\item A bifunctor $\otimes:\mathcal C\times
\mathcal C\to \mathcal C$.
\item A functorial isomorphism $\a_{UVW}$ from
$(U\otimes V)\otimes W$ to $U\otimes(V\otimes W)$.
\item A unit object $\mathbf 1$ in $\mathcal C$ and functorial
isomorphisms
$\la_V:{\mathbf 1}\otimes V\isom V$ and
$\rho_V:V\otimes {\mathbf 1}\isom V$.
\item If $X_1$ and $X_2$ are two expressions
obtained from $V_1\otimes V_2\otimes\cdots\otimes V_n$
by inserting $\mathbf 1$'s and brackets.  Then
all isomorphisms composed of $\a$'s, $\la$'s, $\rho$'s
and their inverses are equal.
\item The functor $\otimes$ is bilinear on the space
of morphisms.
\item The object $\mathbf 1$ is simple and $\End({\mathbf 1})=\C$.
\end{enumerate}}
\end{definition}

\begin{definition}{\rm
Let $\mathcal C$ be a monoidal category and $V$ be an object
in $\mathcal C$.  A {\sl right dual} to $V$ is an object $V^*$
with two morphisms $e_V:V^*\otimes V\to{\mathbf 1}$ 
and $i_V:{\mathbf 1}\to V\otimes V^*$
such that we have
$(\id_V\otimes e_V)(i_V\otimes\id_V)=\id_V$ and
$(e_V\otimes\id_{V^*})(\id_{V^*}\otimes i_V)=\id_{V^*}$.

Similarly, we define a left dual of $V$ to be ${}^*V$ with
morphisms $e'_V:V\otimes{}^*V\to{\mathbf 1}$ and
$i'_V:{\mathbf 1}\to {}^*V\otimes V$ satisfying
similar axioms.}
\end{definition}

\begin{definition}{\rm
A monoidal category is called {\sl rigid} if every object
has right and left duals.

A {\sl tensor category} is a rigid abelian monoidal category.

A {\sl fusion category} is a semisimple tensor category
with finitely many simple objects and finite dimensional
spaces of morphisms.}
\end{definition}

\begin{definition}{\rm 
A fusion category $\mathcal C$ over $\C$ is said to
be unitary if we have the following conditions.
\begin{enumerate}
\item We have a Hilbert space
structure on each $\Hom$ space.
\item We have a contravariant endofunctor $*$ 
on $\mathcal C$ which is the identity on objects.
\item We have $\|\phi\psi\|\le \|\phi\|\;\|\psi\|$ and
$\|\phi^*\phi\|=\|\phi\|^2$ for each morphism $\phi,\psi$ where
$\phi$ and $\psi$ are composable.
\item We have $(\phi\otimes \psi)^*=\phi^*\otimes \psi^*$
for each morphism $\phi,\psi$.
\item All structure isomorphisms for
simple objects are unitary.
\end{enumerate}
}\end{definition}

For any such $N$-$N$ bimodule
${}_N H_N$, we automatically have
$\dim_N H=\dim H_N$.
The index value of a subfactor of finite depth is
automatically a cyclotomic integer 
by \cite[Theorem 8.51]{ENO}.

Conversely, we have the following theorem.

\begin{theorem}
\label{fusion}
Any abstract unitary fusion category is realized
as that of $N$-$N$ bimodules arising from
some (not necessarily irreducible)
subfactor $N\subset M$ with finite index
and finite depth, where $N$ and $M$ are hyperfinite
type II$_1$ factors.
\end{theorem}

This is a slight generalization of \cite[Theorem 12.19]{EK}
since such bimodules produce quantum $6j$-symbols in
the sense of \cite[Section 12.2]{EK}.

\subsection{Classification of subfactors with small indices}

Popa's celebrated
classification theorem \cite{P1} say that if 
a hyperfinite type II$_1$ subfactor $N\subset M$ has
a finite depth, then the finitely many irreducible
bimodules of the four kinds and the intertwiners between
their tensor products contain complete information
on the subfactor and recover $N\subset M$.

Classification of subfactors up to index 4 was announced
in \cite{O1} as follows.  (See also \cite[Theorem 11.24]{EK}.)

\begin{theorem}
\label{ocn}
The hyperfinite II$_1$ subfactors with index less than
4 are labelled with the Dynkin diagrams
$A_n$, $D_{2n}$, $E_6$ and $E_8$.  Subfactors corresponding
to $A_n$ and $D_{2n}$ are unique and those corresponding
to $E_6$ and $E_8$ have two isomorphism classes each.
\end{theorem}

Classification of subfactors with index equal to $4$
has been achieved in \cite{P1}.

Recently, we have classification of
subfactors with finite depth up to index $5$.  See
\cite{JMS} for details.  Many of them are related to
conformal field theory and quantum groups, but we 
see some exotic examples which have been so far unrelated
to them.  Up to index $5$, we have three such exotic subfactors,
the Haagerup subfactor \cite{AH}, the Asaeda-Haagerup
subfactor \cite{AH} and the extended Haagerup subfactor
\cite{BMPS}.

This is a very important active topic of the current
research, but we refrain from going into details here.

\subsection{Bimodules and endomorphisms}

In this section, $M$ is a type III factor.  We present another
formulation of the bimodule theory which is more useful
in conformal field theory.

We can also define $L^2(M)$
as a completion of $M$ with respect to some inner
product arising from some positive linear functional on $M$.  
Then the left action of $M$ is defined usually, and we can
also define the right action of $M$ on $L^2(M)$ using
the modular conjugation in the
Tomita-Takesaki theory.  We then have an $M$-$M$
bimodule ${}_M L^2(M)_M$, and the commutant of the
left action of $M$ is exactly the right action of $M$.
(See \cite[Section IX.1]{T2}.)

Consider an $M$-$M$ bimodule $H$.  The left actions of
$M$ on $H$ and $L^2(M)$ are unitarily equivalent
since $M$ is a type III factor.  So by changing $H$
within the equivalence class of left $M$-modules, 
we may and do assume that $H=L^2(M)$  and the left
actions of $M$ on $H$ and $L^2(M)$ are the same.
Now consider the right action of $M$ on $H=L^2(M)$.  It
must commute with the left action of $M$, but this 
commutant is exactly the right action of $M$ on $L^2(M)$,
so this means that a general right action of $M$ on $H$
is given by a homomorphism of $M$ into $M$, that is,
an endomorphism of $M$.  (We consider only unital homomorphisms
and endomorphisms in this text.)  Conversely, if we have an
endomorphism $\lambda$ of $M$, then we can define an
$M$-$M$ bimodule $L^2(M)$ with the standard left
action and the right action given by 
$x\cdot\xi\cdot y=x\xi\lambda(y)$.  In this way, considering
bimodules and considering endomorphisms are the same.
We now see the corresponding notions of various ones
in the setting of bimodules.  We write $\End(M)$ for
the set of all endomorphisms of $M$.

Two endomorphisms $\lambda_1$ and $\lambda_2$ of $M$
are said to be unitarily equivalent if we have
a  unitary $u$ with $\Ad(u)\cdot \lambda_1=\lambda_2$.
The unitary equivalence of endomorphisms corresponds
to the isomorphism of bimodules.  A unitary equivalence
class of endomorphisms is called a {\sl sector}.  This
name comes from {\sl superselection sectors} which 
appear later in this text.  We write $[\la]$ for
the sector of $\la$.

For two endomorphisms $\lambda_1$ and $\lambda_2$ of $M$,
we define the direct sum $\lambda_1\oplus \lambda_2$ as follows.
Since $M$ is a factor of type III, we have isometries
$V_1,V_2\in M$ with $V_1V_1^*+V_2V_2^*=I$.  Then we
set $(\lambda_1\oplus \lambda_2)(x)=V_1\lambda_1(x)V_1^*+
V_2\lambda_2(x)V_2^*$.  The unitary equivalence class
of $\lambda_1\oplus\lambda_2$ is well-defined, and this
direct sum of endomorphisms corresponds to
the direct sum of bimodules.

An intertwiner in the setting of endomorphisms is given by 
$$\Hom(\lambda_1,\lambda_2)=
\{T\in M\mid T\lambda_1(x)=\lambda_2(x)T 
\textrm{ for all }x\in M\}.$$
For two endomorphisms $\lambda_1,\lambda_2$, we
set $\langle \lambda_1,\lambda_2\rangle=\dim
\Hom(\lambda_1,\lambda_2)$.

The relative tensor product of bimodules 
corresponds to composition of endomorphisms.
The contragredient bimodule corresponds to
the {\sl conjugate endomorphism}.  The conjugate endomorphism
of $\lambda$ is denoted by $\bar\lambda$ and it is well-defined
{\sl only} up to unitary equivalence.  The conjugate endomorphism
is also given using
the {\sl canonical endomorphism} in \cite[Section 2]{L2} arising from
the modular conjugation in the Tomita-Takesaki theory.
The canonical endomorphism for a subfactor $N\subset M$
corresponds to the bimodule ${}_M L^2(M)\otimes_N L^2(M)_M$.
The dual canonical endomorphism for a subfactor $N\subset M$
is an endomorphism of $N$ corresponding to the 
bimodule ${}_N M_N$.

An endomorphism $\lambda$ of $M$ is said to be
{\sl irreducible} if $\lambda(M)'\cap M=\C$.  This corresponds
to irreducibility if bimodules.  The {\sl index}
of $\lambda$ is the index $[M:\lambda(M)]$.  We set
the {\sl dimension} of $\lambda$ to be $[M:\lambda(M)]^{1^2}$
and write $d(\lambda)$ or $d_\la$.  Note that an endomorphism with
dimension 1 is an automorphism.  We have
$d(\lambda_1\oplus\lambda_2)=d(\lambda_1)+d(\lambda_2)$ and
$d(\lambda_1\lambda_2)=d(\lambda_1)d(\lambda_2)$.
(See \cite[Theorem 5.5]{L1} and 
\cite[Theorem 2.1]{L3} for details.)

As a counterpart of $M$-$N$ bimodule, we consider $M$-$N$
morphisms.  Let $M$, $N$ be type III factors.  (We do not
assume $N\subset M$.)  As a general bimodule ${}_M H_N$
is isomorphic to ${}_M L^2(M)$ as a left $M$-module, so
the right action of $N$ on $H$ gives a homomorphism from
$N$ to $M$.  Based on this observation, we say a homomorphism
from $N$ to $M$ is an $M$-$N$ {\sl morphism} and denote
the set of $M$-$N$ morphisms by $\Mor(N,M)$.  (Be careful
about the order of $M$ and $N$.)  Two $M$-$N$ morphisms
$\lambda_1$ and $\lambda_2$ are said to be unitarily equivalent
when we have a unitary $u\in N$ with $\Ad(u)\cdot\lambda_1=\lambda_2$.
A unitary equivalence class of $M$-$N$ morphisms is called
an $M$-$N$ sector.  We $M$-$N$ morphism $\lambda_1$, $N$-$P$
morphism $\lambda_2$ and $M$-$P$ morphism $\lambda_3$, we 
also have the Frobenius reciprocity 
$\Hom(\lambda_1\lambda_2,\lambda_3)\isom
\Hom(\lambda_1,\lambda_3\bar\lambda_2)$.
For a subfactor $N\subset M$ of type III, let $\iota$ be the
inclusion map $N\to M$, which is an $M$-$N$ morphism.  Then
we have that $\bar\iota\iota$ is the dual canonical endomorphism
as $N$-$N$ morphisms.
(See \cite{I1} for details.)

Suppose we have a finite set $\{\lambda_i\mid i=0,1,\dots,n\}$
of endomorphisms of finite dimensions of $M$ with $\lambda_0$
being the identity automorphism.  Suppose we have the
following conditions.
\begin{enumerate}
\item Different $\lambda_i$ and $\lambda_j$ are not unitarily
equivalent.
\item The composition $\lambda_i\lambda_j$ is unitarily 
equivalent to $\bigoplus_{k=1}^n m_k \lambda_k$, where
$m_k$ is the multiplicity of $\lambda_k$.
\item For each $\lambda_i$, its conjugate $\bar\lambda_i$ is
unitarily equivalent to some $\lambda_j$.
\end{enumerate}
Then the set of endomorphisms of $M$ unitarily equivalent
to finite direct sums of $\{\lambda_i\}$ gives a unitary
fusion category.  This is a counterpart of the unitary
fusion category of bimodules.
Conversely, any abstract
unitary fusion category is realized
as that of endomorphisms 
of the type III$_1$ Araki-Woods factor.
This is a direct consequence of Theorem \ref{fusion}.

\subsection{A $Q$-system and an extension of a factor}

We deal with abstract characterization of a
subfactor $N\subset M$ in terms of tensor categories.

Suppose we have a type II$_1$ subfactor $N\subset M$
with finite index.  Then the multiplication map
$S: {}_N L^2(M)\otimes_N L^2(M)_N\to {}_N L^2(M)_N$
extending $x\otimes y\mapsto xy$ for $x,y\in M$ 
exists.  It is a bimodule intertwiner and satisfies the 
associativity $S(\id\otimes S)=S(S\otimes\id)$.
Conversely, if we have an intertwiner
$S: {}_N L^2(M)\otimes_N L^2(M)_N\to {}_N L^2(M)_N$
with associativity, it essentially recovers $M$
with the product structure. Longo's $Q$-system
in \cite{L4} gives a precise formulation of this
and this bimodule version was later given in \cite{Ma}.
Since what we use in conformal field theory is
the original version based on endomorphisms, we
introduce the definition in
\cite[Theorem 6.1]{L4} as follows.

\begin{definition}{\rm 
A {\sl $Q$-system} $(\lambda, v, w)$ is a triple of an
endomorphism of $M$ and isometries $v\in\Hom(\id,\lambda)$,
$w\in\Hom(\lambda,\lambda^2)$ satisfying the following identities:
\begin{align*}
v^* w &=\lambda(v^*)w \in\R_+,\\
\lambda(w)w&=w^2.
\end{align*}
}\end{definition}

If $N\subset M$ is a subfactor with finite index, the 
associated canonical endomorphism $\lambda$ gives a $Q$-system
for appropriate $v\in \Hom(\id,\lambda)$, $w\in\Hom(\lambda,\lambda^2)$.
Conversely, any $Q$-system determines a subfactor $N$ of $M$
such that $\lambda$ is the canonical endomorphism for
$N\subset M$ and $N$ is given by
$N=\{x\in M\mid wx=\lambda(x)w\}$.  We then have
$M=Nv$.  The $Q$-system also determines a larger
factor $M_1\supset M$ such that the dual canonical
endomorphism for the subfactor $M\subset M_1$ is $\lambda$.
Note that the intertwiner $w$
corresponds to the intertwiner $S$ in the bimodule
setting, and the second condition on $w$ represents
associativity.  If $\lambda$ is an endomorphism in some
unitary fusion category of endomorphisms of $M$, then
the intertwiners are also in the category, and
the conditions make sense within the fusion category.

A $Q$-system in the abstract language of tensor categories
is the same as a $C^*$-Frobenius algebra, a special version of
a special symmetric Frobenius algebra \cite{De}.
(Also see \cite{FFRS1}, \cite{FFRS2}, \cite{FuRS1}, \cite{FuRS2}
for a special symmetric Frobenius algebra.)

Theory of bimodules over II$_1$ factors and that of
endomorphisms of type III factors are parallel, but
for some purpose, one is conceptually easier than
the other, so it is convenient to have basic
understanding of the both.  It is the latter which
we use in conformal field theory.

\section{Local conformal nets}

We now present a precise formulation of chiral conformal field 
theory in the operator algebraic framework.  

After introducing basic definitions, we present elementary
properties, representation theory, the machinery of
$\a$-induction, examples and classification theory.

\subsection{Definition}

We now introduce the axioms for a {\sl local conformal net}.
We say $I\subset S^1$ is an {\sl interval} when it is a non-empty,
connected, non-dense and open subset of $S^1$.  

\begin{definition}{\rm
We say that
a family of von Neumann algebras $\{\A(I)\}$ parameterized 
by intervals $I\subset S^1$ acting on the same Hilbert space
$H$ is a local conformal net when it satisfies the following
conditions.
\begin{enumerate}
\item (Isotony)
For two intervals $I_1\subset I_2$, we have $\A(I_1)\subset \A(I_2)$.
\item (Locality) When two intervals $I_1, I_2$ satisfy
$I_1\cap I_2=\varnothing$, we have
$[\A(I_1), \A(I_2)]=0$.
\item (M\"obius covariance)
We have a unitary representation $U$ of
$PSL(2, {\mathbb{R}})$ on $H$ such that
we have $U(g) \A(I) U(g)^*=  \A(gI)$ for all
$g\in PSL(2,{\mathbb{R}})$, where $g$ acts on $S^1$
as a fractional linear transformation on $\R\cup\{\infty\}$ and 
$S^1\setminus\{-1\}$ is identified with $\R$ through
the Cayley transform $C(z)=-i(z-1)/(z+1)$.
\item (Conformal covariance) 
We have a projective unitary representation, still
denoted by $U$, of ${\mathrm{Diff}}(S^1)$
extending the unitary representation
$U$ of $PSL(2, \R)$ such that
\begin{align*}
U(g)\A(I) U(g)^* &= \A(gI),\quad  g\in{\mathrm{Diff}}(S^1), \\
U(g)x U(g)^* &=x,\quad x\in \A(I),\ g\in{\mathrm{Diff}}(I'),
\end{align*}
where $I'$ is the interior of the complement of $I$ and
${\mathrm{Diff}}(I')$ is the set of diffeomorphisms of $S^1$
which are the identity map on $I$.
\item (Positive energy condition)
The generator of the restriction of $U$ to the rotation
subgroup of $S^1$, the {\sl conformal Hamiltonian}, is
positive.
\item (Existence of the vacuum vector)
We have a unit vector $\Omega\in H$, called the
{\sl vacuum vector}, such that $\Omega$ is fixed
by the representation $U$ of 
$PSL(2, {\mathbb{R}})$ and
$(\bigvee_{I\subset S^1} \A(I))\Omega$ is dense in  $H$,
where $\bigvee_{I\subset S^1} \A(I)$ is the von Neumann
algebra generated by $\A(I)$'s.
\item (Irreducibility) The von Neumann algebra
$\bigvee_{I\subset S^1}\A(I)$  is $B(H)$.
\end{enumerate}
}\end{definition}

The convergence in $\Diff(S^1)$ is defined by 
uniform convergence of all the derivatives.

We say $\{\A(I)\}$ is a local M\"obius covariant net when
we drop the conformal covariance axiom.

The name ``net'' originally meant that the set of spacetime
regions are directed with respect to inclusions, but now
the set of intervals in $S^1$ is not directed, so this
name is not appropriate, but has been widely used.  Another
name ``pre-cosheaf'' has been used in some literatures.

If the Hilbert space is $1$-dimensional and all
$\A(I)$ are just $\C$, all the axioms are clearly satisfied,
but this example is of no interest, so we exclude this
from a class of local conformal nets.

Locality comes from the fact that we have no interactions
between two spacelike separated regions in the $(1+1)$-dimensional
Minkowski space.  Now because of the restriction procedure to
two light rays, the notion of spacelike separation takes this
simple form of disjointness.

The projective unitary representation of $\Diff(S^1)$ in
conformal covariance extending the unitary representation
of $PSL(2,\R)$ is unique if it exists, by \cite{CW}.

The positive energy condition is our counterpart to
what is called the spectrum condition in quantum field
theory on the higher dimensional Minkowski space.

Irreducibility condition is equivalent to the uniqueness
of the $PSL(2,\R)$-invariant vector up to scalar, and 
is also equivalent to factoriality of each algebra $\A(I)$.
(See \cite[Proposition 1.2]{GL} for a proof.)

Note that a subfactor $N\subset M$ produces the Jones
tower/tunnel
$$\cdot \subset N_2\subset N_1\subset N\subset M\subset
M_1\subset M_2\subset\cdots$$
as in \cite[Definition 9.24, Definition 9.43]{EK}.
If we set the interval $I_t$, $t\in (0,\pi)$, to be the
arc between $(1,0)$ and $(\cos t,\sin t)$ on the unit
circle on the $xy$-plane, then the family $\{\A(I_t)\}_t$ of
the von Neumann algebras is a continuous analogue of the
Jones tower.

It would be better to have some easy examples here, but
unfortunately, there are no easy examples one can present
immediately without preparations, so we postpone examples
to a later section.

We have the following consequences from the axioms.

\begin{theorem}
(the Reeh-Schlieder theorem)
For each interval $I\subset S^1$,
both $\A(I)\Omega$ and $\A(I)'\Omega$ are dense in $H$,
where $\A(I)'$ is the commutant of $\A(I)$.
\end{theorem}

The positive energy condition is used essentially for
a proof of this theorem through analytic continuation.
See \cite[Corollary 2.8]{FG} or \cite[Theorem 6.2.3]{Ba} 
for a proof.

Let $S^1$ be the unit circle in the complex plane 
and $I_1$ the upper open half circle.  Let
$C:S^1\to{\mathbb{R}}\cup\{\infty\}$ be the 
Cayley transform $C(z)=-i(z-1)(z+1)^{-1}$.
We define a one-parameter 
diffeomorphism group $\Lambda_{I_1}(s)$ by
$C\Lambda_{I_1}(s)C^{-1}x=e^s x$.  Define
$r_{I_1}$ by $r_{I_1}(z)=\bar z$ for $z \in S^1$.
For a general interval $I\subset S^1$,
choose $g\in PSL(2,{\mathbb{R}})$ so that
$I=gI_1$ and we define $\Lambda_I=g\Lambda_{I_1}g^{-1}$,
$r_I=gr_{I_1}g^{-1}$.  (They are independent of 
the choice of $g$.)
The action of $r_{I_1}$ on
$PSL(2,{\mathbb{R}})$ defined a semi-direct product
$PSL(2,{\mathbb{R}})\rtimes{\mathbb{Z}}_2$.
Let $\Delta_I$, $J_I$ be the modular operator and
the modular conjugation of $(\A(I),\Omega)$.
Then we have an extension of $U$ appearing in the
definition of M\"obius covariance such that
$U(g)$ is a unitary or an anti-unitary depending
on whether $g$ preserves or reverses the orientation.
We still write $U$ for this extension.  
Then we have the following {\sl Bisognano-Wichmann property}.

\begin{theorem}
We have $U(\Lambda_I(2\pi t))=\Delta_I^{it}$, 
$U(r_I)=J_I$  for this $U$.
\end{theorem}

See \cite[Theorems 2.3, 2.5]{BGL} or 
\cite[Theorem 2.19]{FG} for details of this property.
This immediately
implies the following important result.

\begin{theorem}
(the Haag duality) We have $\A(I)'=\A(I')$.
\end{theorem}

The next result has been proved in \cite[page 545]{FJ}.

\begin{theorem}
(Additivity) If a family $\{I_i\}$ of intervals and
an interval $I$ satisfy
$I\subset\bigcup_i I_i$, then 
$\A(I)$ is contained in the von Neumann algebra
generated by $\{\A(I_i)\}_i$.
\end{theorem}

One has that the fixed point algebra of $\A(I)$ of
the modular automorphism group with respect to
$\Omega$ is $\C$, then this implies the following.

\begin{theorem}
Each $\A(I)$ is a factor of type III$_1$.
\end{theorem}

See \cite[Corollary 2.6]{DLR} or
\cite[Theorem 6.2.5]{Ba} for details.

We now introduce extra properties of local conformal nets.

\begin{definition}{\rm
Removing one point from an interval $I$, we obtain a disjoint
union of two intervals $I_1$ and $I_2$.
We say that the local conformal net has
{\sl strong additivity} if the von Neumann algebra
$\A(I)$ is always generated by $\A(I_1)$ and $\A(I_2$).
}\end{definition}

Many important examples satisfy strong additivity, but there
are also examples without it.  In this text, we consider
only local conformal nets with strong additivity.

\begin{definition}{\rm
Consider two intervals $I_1$, $I_2$ with
$\bar I_1 \cap \bar I_2=\varnothing$ and the map
$x\otimes y \mapsto xy$ from the algebraic tensor product
of $\A(I_1)$ and $\A(I_2)$ to the von Neumann algebra 
generated by $\A(I_1)$ and $\A(I_2)$, where $x\in \A(I_1)$
and $y\in \A(I_2)$.  We say that the local conformal net
has the {\sl split property} if this map always extends
to an isomorphism from $\A(I_1)\otimes \A(I_2)$.
}\end{definition}

It has been proved in \cite[Theorem 3.2]{DLR} that if
we have $\Tr(e^{-tL_0})<\infty$ for all $t>0$, where
$L_0$ is the conformal Hamiltonian, then the local
conformal net has the split property.
Practically all known examples of local conformal nets
satisfy the split property, and we consider only those
with this property in this text.  
If we have the split
property for $\{\A(I)\}$, then each $\A(I)$ is 
a hyperfinite type III$_1$ factor, which is isomorphic
to the Araki-Woods factor of type III$_1$.
(See \cite[Proposition 3.1]{DLR}
and \cite[Theorem XVIII.4.16]{T3}.)

This means that in our setting the isomorphism class of
each von Neumann algebra $\A(I)$ is unique for any interval
and any local conformal net.  So each $\A(I)$ has no
information on conformal field theory, and it is the
relative positions of the algebras $\A(I)$ that contain
information of conformal field theory.

We remark that conformal field theory is supposed to be
a certain scaling limit of statistical mechanical models.
There have been many works on the operator algebraic
approach to quantum statistical mechanics, so we expect
that it is possible to realize this passage from 
quantum statistical mechanics to conformal field theory
in the rigorous framework using operator algebras, but
there has been no such work so far.

\subsection{Superselection sectors and braiding}

An important tool to study local conformal nets is
their representation theory.

Each $\A(I)$ of a local conformal net acts on the Hilbert
space $H$ from the beginning by definition, but consider 
representations of a family $\{\A(I)\}$ of factors on the
common Hilbert space $H_\pi$.  That is, we consider a 
family $\pi$ of representations
$\pi_I:\A(I)\to B(H_\pi)$ such that the restriction of
 $\pi_{I_2}$ to $\A(I_1)$  is equal to $\pi_{I_1}$ for
$I_1\subset I_2$.  Note that $H_\pi$ does not have
a vacuum vector in general.  The original identity
representation on $H$ is called the {\sl vacuum
representation}.

For this notion of a representation, it is easy to
define an irreducible representation, the direct
sum of two representations and unitary equivalence
of two representations.  A unitary equivalence class
of representations is called a {\sl superselection
sector} or a {\sl DHR (Doplicher-Haag-Roberts) sector}.

We also introduce a notion of a covariant representation
as follows.

\begin{definition}{\rm
A representation $\pi$ is said to be
M\"obius [resp.~conformal] covariant if 
we have projective unitary positive energy 
representation $U_\pi$ of the universal cover
$\widetilde{PSL(2,\R)}$ [resp.~$\widetilde{\Diff(S^1)}$]
such that $\pi_{gI}(U(g)xU(g)^*)=
U_\pi(g)\pi_I(x)U_\pi(g)^*$ for all $x\in\A(I)$.
Here the positive energy condition means that the
generator of the one-parameter unitary group arising
from the rotation subgroup of $\Diff(S^1)$ is positive.
}\end{definition}

Any irreducible representation of a local conformal net
is automatically conformal covariant
by \cite[Proposition 2.2]{Ca}. (Also see \cite{DFK}.)

We would like to define a notion of a tensor
product of two representations.  This is a non-trivial
task, and an answer has been given in the
Doplicher-Haag-Roberts theory \cite{DHR1,DHR2},
which was originally developed for quantum field
theory on the 4-dimensional Minkowski space.
The Doplicher-Haag-Roberts theory adapted to conformal
field theory is given as follows.  (See \cite{FRS1,FRS2}.
Also see \cite[Appendix B]{KLM} for relations to the
representation theory of nets on $\R$.)

Take a representation
$\pi=\{\pi_I\}$ of a local conformal net $\{\A(I)\}$.
Fix an arbitrary interval $I_0\subset S^1$ and
consider the representation $\pi_{I'_0}$ of $\A(I'_0)$.
Since $\A(I'_0)$ is a type III factor, 
the identity representation $\A(I'_0)\hookrightarrow B(H)$
and $\pi_{I'_0}$ are unitarily equivalent.  By changing
the representation $\pi$ within the unitary equivalence
class if necessary, we may and do assume that
$H=H_\pi$ and $\pi_{I'_0}$ is the identity representation.

Take an interval $I_1$ with $\bar I_1\subset I_0$, and
then take an interval $I_2$ containing both $I_1$ and
$I'_0$.  For $x\in \A(I_1)$, we have
$\pi_{I_2}(xy)=\pi_{I_2}(yx)$  for any $y\in
\A(I'_0)$.  This implies
$\pi_{I_1}(x)y=y\pi_{I_1}(x)$, and thus 
the image of $\A(I_1)$ by $\pi_{I_0}$ is contained in
$\A(I'_0)'=\A(I_0)$.  Additivity now implies that
the image of $\A(I_0)$ by $\pi_{I_0}$ is contained in
$\A(I_0)$, that is, $\pi_{I_0}$ gives an endomorphism
of $\A(I_0)$.  We say that this representation $\pi$
is {\sl localized} in $I_0$.

We can construct a universal $C^*$-algebra $C^*(\A)$ from
the local conformal net $\{\A(I)\}$ with the
property that $C^*(\A)$ contains each $\A(I)$ and
for any representation $\{\pi_I\}$
of  $\{\A(I)\}$, we have a representation $\pi$
of $C^*(\A)$ with $\pi|_{\A(I)}=\pi_I$.  We identify
a representation of $\{\A(I)\}$ with the
corresponding representation of $C^*(\A)$.
(See \cite[page 382]{F} for details.)
Let $\pi_0$ be
the representation of $C^*(\A)$ corresponding to the
vacuum representation of $\{\A(I)\}$.
For a representation $\pi=\{\pi_I\}$ of $\{\A(I)\}$,
we construct an endomorphism $\lambda$ of $C^*(\A)$ so 
that $\pi$ is unitarily equivalent to $\pi_0\cdot\lambda$.
The endomorphism $\lambda$ is given by a family $\{\lambda_I\}$
of homomorphisms $\lambda_I:\A(I)\to C^*(\A)$ with
$\lambda_{I_2}|_{\A(I_1)}=\lambda_{I_1}$ for $I_1\subset I_2$.

Start with a representation $\pi$ of $\{\A(I)\}$ 
localized in an interval $I_0$.  We construct the
corresponding family $\{\lambda_I\}$ of homomorphisms
$\lambda_I:\A(I)\to C^*(\A)$.  If $I\cup I_0$ is not dense in $S^1$,
we choose an interval $I_1\supset I\cup I_0$ and set
$\lambda_I=\pi_{I_1}|_{\A(I)}$.  This is independent
of $I_1$.  If $I\cup I_0$ is dense in $S^1$,
choose an interval $I_1\subset I_0\cap I'$ and
let $\pi'$ be a representation of $\{\A(I)\}$ which
is localized in $I_1$ and unitarily equivalent to $\pi$.
The unitary equivalence gives a unitary $V\in B(H)$ with
$\Ad(V)\cdot\pi'(x)=\pi(x)$ for all $x\in C^*(\A)$, and
the Haag duality implies $V\in\pi_0(\A(I_0))$.
Then we have a unitary $U\in\A(I_0)$ with $\pi_0(U)=V$.
We then set $\lambda_I=\Ad(U)|_{\A(I)}$.  This is
independent of $I_1,\pi',V$.  In this way, we obtain
a family $\lambda=\{\lambda_I\}$ of homomorphisms
$\lambda_I:\A(I)\to C^*(\A)$.  This also gives an endomorphism,
still denoted by $\lambda$, of $C^*(\A)$.  (See \cite[page 383]{F}
for details.)

For the family $\lambda=\{\lambda_I\}$ of homomorphisms,
we say $\lambda$ is {\sl localized} in $I_0$ if we have
$\lambda_{I'_0}=\id$ on $\A(I'_0)$.  We say that $\lambda$
is {\sl transportable} if for all intervals $I_1,I$
with $I\supset I_0\cup I_1$, we have a unitary
$U\in \A(I)$ so that
$\Ad(U)\cdot\lambda$ is localized in $I_1$.
Such a unitary $u$ is called a {\sl transporter}.
We say that an endomorphism of $C^*(\A)$ is  a
{\sl DHR endomorphism} if it is of the form
$\Ad(U)\cdot\lambda'$ where $U$ is a unitary in $C^*(\A)$
and $\lambda'$ is localized transportable endomorphism
of $C^*(\A)$.  Considering  a representation of $\{\A(I)\}$
and considering a DHR endomorphism of $C^*(\A)$ are
equivalent.  Two representations of $\{\A(I)\}$
are unitarily equivalent if and only if 
one of the
corresponding DHR endomorphisms of $C^*(\A)$ is
an inner perturbation of the other.

We see that a composition of two DHR
endomorphisms is again a DHR endomorphism.  This
defines a {\sl tensor product} operation of 
two representations.  This also gives a tensor product
of superselection sectors.

We define the {\sl dimension} of 
a DHR endomorphism $\lambda$ to be the square root of
the index $[\A(I_0):\lambda_{I_0}(\A(I_0))]$ when $\lambda$ is
localized in $I_0$.  This is independent of $I_0$ by
\cite[Proposition 2.1]{GL}.  So the dimension is additive
and multiplicative with respect
to the direct sum and the tensor product of representations.

For any DHR endomorphism $\lambda$ of $C^*(\A)$, we localize
it in the fixed interval $I_0$ and have an endomorphism
$\lambda_{I_0}$ of $\A(I_0)$.  By
\cite[Theorem 2.3]{GL}, considering these endomorphisms
with intertwiners in $\A(I_0)$ is equivalent to
considering the representations of $\{\A(I)\}$ with
intertwiners.  This gives a notion of the contragredient
representation of a local conformal net.
So we have a unitary tensor category
consisting of endomorphisms $\A(I_0)$ and this gives
{\sl the} representation category of the local
conformal net $\{\A(I)\}$.  We write $\Rep(\A)$ for this.
When we have an 
endomorphism of $\A(I_0)$ arising from a DHR endomorphism
of $C^*(\A)$ localized in $I_1\subset I_0$, we say it is
localized in $I_1$.
We next introduce the braiding structure on this unitary
tensor category.  We first have the following
lemma as in \cite[Lemma 2.1]{BE1}.  This is clear
with strong additivity, but actually holds without it.

\begin{lemma}
Let $\lambda_1,\lambda_2$ be the endomorphisms
of $\A(I_0)$ localized in $I_1,I_2\subset I_0$,
respectively, with
$I_1\cap I_2=\emptyset$.  Then we have
$\la_1\cdot\la_2=\la_2\cdot\la_1$.
\end{lemma}

Suppose we have $I_1,I_2\subset I_0$ with $I_1\cap I_2=\emptyset$
and also suppose $\la,\mu$ are localized in $I\subset I_0$.
Choose unitaries $U_1$, $U_2$ in $\A(I_0)$ so that
$\Ad(U_1)\cdot\la$ and $\Ad(U_2)\cdot\mu$ are
localized in $I_1, I_2$, respectively.
Because of the above Lemma, it is now clear that $\la$
and $\mu$ commute up to unitary equivalence.
Set $\e(\la,\mu)=\mu(U_1^*)U_2^*U_1\la(U_2)$.
As in \cite[Lemma 2.2]{BE1}, $\e(\la,\mu)$ depends
only on the order of $I_1$ and $I_2$.  When $I_2$ is
forward of $I_1$ in the counterclockwise order, we 
write $\e^+(\la,\mu)$, and for the other case, we
write $\e^-(\la,\mu)$.  These are called
{\sl statistic operators}.
We then have the following results as in \cite[Lemma 2.3,
Proposition 2.5]{BE1}.

\begin{theorem}
Let $\la,\mu,\nu$ be the localized
endomorphisms of $\A(I_0)$.  We have the following relations.
\begin{align*}
\Ad(\e^\pm(\la,\mu))\cdot\la\cdot\mu&=\mu\cdot\la,\\
\e^\pm(\la.\mu)&\in\A(I_0),\\
\e^+(\la,\mu)&=\e^-(\mu,\la)^+,\\
\e^\pm(\la\cdot\mu,\nu)&=\e^\pm(\la,\nu)\la(\e^\pm(\mu,\nu)),\\
\e^\pm(\la,\mu\cdot\nu)&=\mu(\e^\pm(\la,\nu))\e^\pm(\la,\mu)),\\
\nu(t)\e^\pm(\la,\nu)&=\e^\pm(\mu,\nu)t,\quad t\in \Hom(\la,\mu),\\
t\e^\pm(\nu,\la)&=\e^\pm(\nu,\mu)\nu(t),\quad t\in \Hom(\la,\mu).
\end{align*}
\end{theorem}

The last two identities imply the following
{\sl braiding fusion equations} (BFE's).

\begin{corollary}
Let $\la,\mu,\nu,\rho$ be the localized
endomorphisms of $\A(I_0)$.  We have the following identities
for $s\in\Hom(\la\cdot\mu,\nu)$.
\begin{align*}
\rho(s)\e^\pm(\la,\rho)\la(\e^\pm(\mu,\rho))&=\e^\pm(\nu,\rho)s,\\
s\la(\e^\pm(\rho,\mu))\e^\pm(\rho,\la)&=\e^\pm(\rho,\nu)\rho(s).
\end{align*}
\end{corollary}

In this way, DHR endomorphisms localized in $I_0$ gives
a unitary braided tensor category of endomorphisms of $\A(I_0)$
in the following sense.  (See \cite[Chapter 1]{BK},
\cite[Section I.1.2]{Tu} for a general theory on
braided tensor categories.)

\begin{definition}{\rm
Let $\mathcal C$ be a monoidal category with 
functorial isomorphisms $\si_{VW}:V\otimes W\to W\otimes V$
for all objects $V,W$ in $\mathcal C$.

For given objects $V_1,V_2,\dots,V_n$ in $\mathcal C$, we consider
all expressions of the form $$((V_{i_1}\otimes V_{i_2})
\otimes({\mathbf 1}\otimes V_{i_3}))\otimes\cdots\otimes V_{i_n}$$
obtained from $V_{i_1}\otimes V_{i_2}\otimes\cdots\otimes V_{i_n}$
by inserting some $mathbf 1$'s and brackets. where $(i_1,i_2,\dots,i_n)$
is a permutation of $\{1,2,\dots,n\}$.  To any composition
of $\a$'s, $\la$'s, $\rho$'s, $\si$'s and their inverses acting
on the element in the above tensor product, we assign an element of 
the braid group $B_n$ with the standard generators $b_i$ 
$(i=1,2,\dots,n-1)$ satisfying $b_ib_j=b_j b_i$ for $|i-j|>1$ and
$b_i b_{i+1}b_i=b_{i+1}b_ib_{i+1}$ as follows.
To $\a$, $\la$ and $\rho$, we assign $1$ and to 
$\si_{V_{i_k}V_{i_{k+1}}}$ the generator $b_k$.

The category $\mathcal C$ is called a 
{\sl braided tensor category} if for any two expressions
$X_1, X_2$ of the above form and any isomorphism $\phi:X_1\to X_2$
obtained by composing $\a$'s, $\la$'s, $\rho$'s, $\si$'s and
their inverses, $\phi$ depends only on its image in the braid
group $B_n$.
}\end{definition}

It is sometimes written as DHR$(\A)$.

\subsection{Graphical intertwiner calculus}

We now explain a method to describe these relations involving
braiding graphically as in \cite[Section 3]{BEK1}.
We start a general setting of endomorphisms of
a type III factor $M$ which is
an abstract version of the set of irreducible DHR endomorphisms.

\begin{definition}
\label{system}
{\rm We call a finite subset $\sys\subset\End(M)$ a
{\sl system of endomorphisms} if it
satisfies the following properties.
\begin{enumerate}
\item Each $\la\in \sys$ is irreducible and has finite
dimension.
\item Different elements in $\sys$ are unitarily inequivalent.
\item We have $\id_M\in \sys$.
\item For any $\la\in\sys$, we have an endomorphism
$\bar\la\in\sys$ giving the conjugate sector of $\la$.
\item The system $\sys$ is closed under composition and subsequent
irreducible decomposition.
\end{enumerate}
}\end{definition}

The rules for irreducible decompositions of the compositions
are called {\sl fusion rules}.

\begin{definition}
\label{braid}{\rm
We say that a system $\sys$ of endomorphisms is {\sl braided}
if for any pair $\la,\mu\in\sys$ there is a unitary operator
$\e(\lambda,\mu)\in\Hom(\la\mu,\mu\la)$ satisfying
the identities 
\begin{equation}
\eps {\id_M}\mu=\eps\lambda{\id_M}=1,
\label{ini}
\end{equation}
and the following BFE's
\begin{equation}
\begin{array}{rl}
\rho(t)  \eps \lambda \rho 
&= \eps \mu \rho  \mu (\eps \nu \rho)  t , \\
t  \eps \rho \lambda
&= \mu (\eps \rho \nu)  \eps \rho \mu  \rho (t) , \\
\rho(t)^*  \eps \mu\rho  \mu( \eps \nu\rho )
&= \eps \la\rho  t^* , \\
t^*  \mu( \eps \rho\nu )  \eps \rho\mu
&=  \eps \rho\la  \rho(t)^* ,
\end{array}
\label{BFE}
\end{equation}
for any $\la,\mu,\nu\in\sys$ and $t\in \Hom (\lambda,\mu\nu)$.
}\end{definition}

\begin{figure}[htb]
\begin{center}
\unitlength 0.6mm
\begin{picture}(60,30)
\thinlines
\put(5,10){\dashbox{2}(50,10){$x$}}
\put(10,30){\vector(0,-1){10}}
\put(25,30){\vector(0,-1){10}}
\put(50,30){\vector(0,-1){10}}
\put(10,10){\vector(0,-1){10}}
\put(25,10){\vector(0,-1){10}}
\put(50,10){\vector(0,-1){10}}
\put(3,25){\makebox(0,0){$\la_1$}}
\put(18,25){\makebox(0,0){$\la_2$}}
\put(57,25){\makebox(0,0){$\la_n$}}
\put(3,5){\makebox(0,0){$\mu_1$}}
\put(18,5){\makebox(0,0){$\mu_2$}}
\put(57,5){\makebox(0,0){$\mu_m$}}
\put(37.5,25){\makebox(0,0){$\cdots$}}
\put(37.5,5){\makebox(0,0){$\cdots$}}
\end{picture}
\end{center}
\caption{An intertwiner $x$}
\label{boxx}
\end{figure}
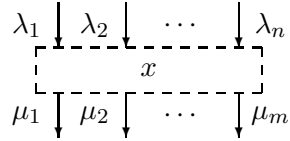

\begin{figure}[htb]
\begin{center}
\unitlength 0.6mm
\begin{picture}(80,30)
\thinlines
\put(25,10){\dashbox{2}(50,10){$x$}}
\put(10,30){\vector(0,-1){30}}
\put(30,30){\vector(0,-1){10}}
\put(45,30){\vector(0,-1){10}}
\put(70,30){\vector(0,-1){10}}
\put(30,10){\vector(0,-1){10}}
\put(45,10){\vector(0,-1){10}}
\put(70,10){\vector(0,-1){10}}
\put(3,15){\makebox(0,0){$\rho$}}
\put(23,25){\makebox(0,0){$\la_1$}}
\put(38,25){\makebox(0,0){$\la_2$}}
\put(77,25){\makebox(0,0){$\la_n$}}
\put(23,5){\makebox(0,0){$\mu_1$}}
\put(38,5){\makebox(0,0){$\mu_2$}}
\put(77,5){\makebox(0,0){$\mu_m$}}
\put(57.5,25){\makebox(0,0){$\cdots$}}
\put(57.5,5){\makebox(0,0){$\cdots$}}
\end{picture}
\end{center}
\caption{The intertwiner $\rho(x)$}
\label{rhoboxx}
\end{figure}
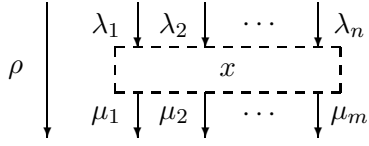

\begin{figure}[htb]
\begin{center}
\unitlength 0.6mm
\begin{picture}(80,30)
\thinlines
\put(5,10){\dashbox{2}(50,10){$x$}}
\put(10,30){\vector(0,-1){10}}
\put(25,30){\vector(0,-1){10}}
\put(50,30){\vector(0,-1){10}}
\put(10,10){\vector(0,-1){10}}
\put(25,10){\vector(0,-1){10}}
\put(50,10){\vector(0,-1){10}}
\put(70,30){\vector(0,-1){30}}
\put(3,25){\makebox(0,0){$\la_1$}}
\put(18,25){\makebox(0,0){$\la_2$}}
\put(57,25){\makebox(0,0){$\la_n$}}
\put(3,5){\makebox(0,0){$\mu_1$}}
\put(18,5){\makebox(0,0){$\mu_2$}}
\put(57,5){\makebox(0,0){$\mu_m$}}
\put(77,15){\makebox(0,0){$\rho$}}
\put(37.5,25){\makebox(0,0){$\cdots$}}
\put(37.5,5){\makebox(0,0){$\cdots$}}
\end{picture}
\end{center}
\caption{The intertwiner $x$}
\label{boxxrho}
\end{figure}
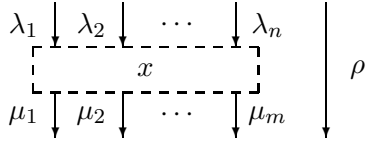

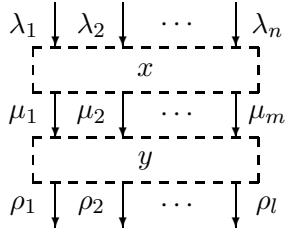
\begin{figure}[htb]
\begin{center}
\unitlength 0.6mm
\begin{picture}(80,50)
\thinlines
\put(5,30){\dashbox{2}(50,10){$x$}}
\put(5,10){\dashbox{2}(50,10){$y$}}
\put(10,50){\vector(0,-1){10}}
\put(25,50){\vector(0,-1){10}}
\put(50,50){\vector(0,-1){10}}
\put(10,30){\vector(0,-1){10}}
\put(25,30){\vector(0,-1){10}}
\put(50,30){\vector(0,-1){10}}
\put(10,10){\vector(0,-1){10}}
\put(25,10){\vector(0,-1){10}}
\put(50,10){\vector(0,-1){10}}
\put(3,45){\makebox(0,0){$\la_1$}}
\put(18,45){\makebox(0,0){$\la_2$}}
\put(57,45){\makebox(0,0){$\la_n$}}
\put(3,25){\makebox(0,0){$\mu_1$}}
\put(18,25){\makebox(0,0){$\mu_2$}}
\put(57,25){\makebox(0,0){$\mu_m$}}
\put(3,5){\makebox(0,0){$\rho_1$}}
\put(18,5){\makebox(0,0){$\rho_2$}}
\put(57,5){\makebox(0,0){$\rho_l$}}
\put(37.5,45){\makebox(0,0){$\cdots$}}
\put(37.5,25){\makebox(0,0){$\cdots$}}
\put(37.5,5){\makebox(0,0){$\cdots$}}
\end{picture}
\end{center}
\caption{Product $yx$ of diagrammatically
composable intertwiners $x$ and $y$}
\label{x-top-y}
\end{figure}

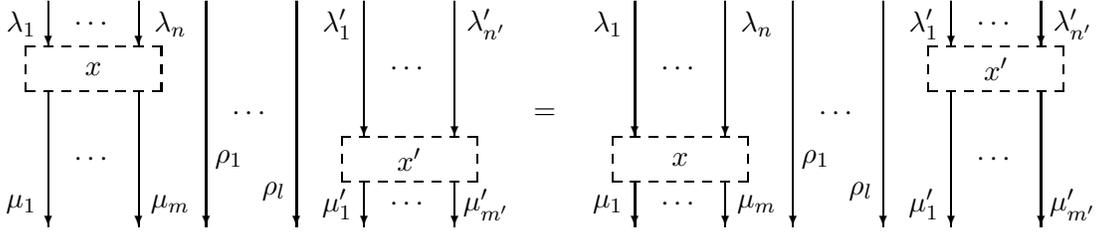
\begin{figure}[htb]
\begin{center}
\unitlength 0.6mm
\begin{picture}(240,50)
\thinlines
\put(5,30){\dashbox{2}(30,10){$x$}}
\put(75,10){\dashbox{2}(30,10){$x'$}}
\put(10,50){\vector(0,-1){10}}
\put(30,50){\vector(0,-1){10}}
\put(10,30){\vector(0,-1){30}}
\put(30,30){\vector(0,-1){30}}
\put(45,50){\vector(0,-1){50}}
\put(65,50){\vector(0,-1){50}}
\put(80,50){\vector(0,-1){30}}
\put(100,50){\vector(0,-1){30}}
\put(80,10){\vector(0,-1){10}}
\put(100,10){\vector(0,-1){10}}
\put(20,45){\makebox(0,0){$\cdots$}}
\put(20,15){\makebox(0,0){$\cdots$}}
\put(55,25){\makebox(0,0){$\cdots$}}
\put(90,35){\makebox(0,0){$\cdots$}}
\put(90,5){\makebox(0,0){$\cdots$}}
\put(4,45){\makebox(0,0){$\la_1$}}
\put(37,45){\makebox(0,0){$\la_n$}}
\put(4,5){\makebox(0,0){$\mu_1$}}
\put(37,5){\makebox(0,0){$\mu_m$}}
\put(50,15){\makebox(0,0){$\rho_1$}}
\put(60,8){\makebox(0,0){$\rho_l$}}
\put(74,45){\makebox(0,0){$\la_1'$}}
\put(107,45){\makebox(0,0){$\la_{n'}'$}}
\put(74,5){\makebox(0,0){$\mu_1'$}}
\put(107,5){\makebox(0,0){$\mu_{m'}'$}}
\put(120,25){\makebox(0,0){$=$}}
\put(135,10){\dashbox{2}(30,10){$x$}}
\put(205,30){\dashbox{2}(30,10){$x'$}}
\put(140,50){\vector(0,-1){30}}
\put(160,50){\vector(0,-1){30}}
\put(140,10){\vector(0,-1){10}}
\put(160,10){\vector(0,-1){10}}
\put(175,50){\vector(0,-1){50}}
\put(195,50){\vector(0,-1){50}}
\put(210,50){\vector(0,-1){10}}
\put(230,50){\vector(0,-1){10}}
\put(210,30){\vector(0,-1){30}}
\put(230,30){\vector(0,-1){30}}
\put(150,35){\makebox(0,0){$\cdots$}}
\put(150,5){\makebox(0,0){$\cdots$}}
\put(185,25){\makebox(0,0){$\cdots$}}
\put(220,45){\makebox(0,0){$\cdots$}}
\put(220,15){\makebox(0,0){$\cdots$}}
\put(134,45){\makebox(0,0){$\la_1$}}
\put(167,45){\makebox(0,0){$\la_n$}}
\put(134,5){\makebox(0,0){$\mu_1$}}
\put(167,5){\makebox(0,0){$\mu_m$}}
\put(180,15){\makebox(0,0){$\rho_1$}}
\put(190,8){\makebox(0,0){$\rho_l$}}
\put(204,45){\makebox(0,0){$\la_1'$}}
\put(237,45){\makebox(0,0){$\la_{n'}'$}}
\put(204,5){\makebox(0,0){$\mu_1'$}}
\put(237,5){\makebox(0,0){$\mu_{m'}'$}}
\end{picture}
\end{center}
\caption{Vertical translation intertwiners $x$ and $x'$}
\label{xx'}
\end{figure}

We fix and consider a braided
system of endomorphisms $\sys\subset\End(M)$. We 
represent intertwiners
by ``wire diagrams'' where the (oriented) wires represent
endomorphisms $\la\in\sys$. This works as follows.
For an intertwiner
$x\in\Hom(\la_1\la_2\cdots\la_n,\mu_1\mu_2\cdots\mu_m)$,
we draw a (dashed) box with $n$ (downward) incoming wires
labelled by $\la_1,\dots,\la_n$ and $m$ (downward) outgoing
wires $\mu_1,\dots,\mu_m$ as in Figure \ref{boxx},
$\la_i,\mu_j\in\sys$.
%
This means the diagrammatic representation of $x$ does not
only specify it as an operator, it also specifies the intertwiner
space to which it belongs. 
If a morphism $\rho\in\sys$ is applied to $x$, then
$\rho(x)\in\Hom(\rho\la_1\la_2\cdots\la_n,\rho\mu_1\mu_2\cdots\mu_m)$
is represented graphically by adding a straight wire on the left
as in Figure \ref{rhoboxx}.
%
Since we can consider $x$ also as an intertwiner in
$\Hom(\la_1\la_2\cdots\la_n\rho,\mu_1\mu_2\cdots\mu_m\rho)$,
we can always add (or remove) a straight wire on the right
as in Figure \ref{boxxrho}
%
without changing the intertwiner as an operator.
We say that intertwiners
$x\in\Hom(\la_1\la_2\cdots\la_n,\mu_1\mu_2\cdots\mu_m)$
and
$y\in\Hom(\nu_1\nu_2\cdots\nu_k,\rho_1\rho_2\cdots\rho_l)$,
$\rho_j\in\sys$,
are {\sl diagrammatically composable} if $m=k$ and
$\mu_i=\nu_i$ for all $i=1,2,\dots,m$. Then the composed
intertwiner
$yx\in\Hom(\la_1\la_2\cdots\la_n,\rho_1\rho_2\cdots\rho_l)$
is represented graphically by putting the wire diagram
for $x$ on top of that for $y$ as in Figure \ref{x-top-y}.
(Note that some authors use an opposite convention where
we compose intertwiners from the bottom to the top.)
%
Now let also
$x'\in\Hom(\la_1'\la_2'\cdots\la_{n'}',\mu_1'\mu_2'\cdots\mu_{m'}')$
with $\la_i',\mu_j'\in\sys$. The intertwining property of $x$
give the identity
$\mu_1\mu_2\cdots\mu_m \rho_1\rho_2\cdots\rho_l(x')x
= x\la_1\la_2\cdots\la_n \rho_1\rho_2\cdots\rho_l(x')$,
and this is diagrammatically given in Figure \ref{xx'}.
%
We thus have some freedom in translating intertwiner boxes vertically
without changing the represented intertwiner.

\begin{figure}[htb]
\begin{center}
\unitlength 0.6mm
\begin{picture}(170,20)
\thinlines
\put(0,0){\dashbox{2}(170,20){}}
\put(10,20){\vector(0,-1){20}}
\put(25,20){\vector(0,-1){20}}
\put(60,20){\vector(0,-1){20}}
\put(75,20){\vector(1,-1){20}}
\put(95,20){\line(-1,-1){8}}
\put(83,8){\vector(-1,-1){8}}
\put(110,20){\vector(0,-1){20}}
\put(125,20){\vector(0,-1){20}}
\put(160,20){\vector(0,-1){20}}
\put(5,10){\makebox(0,0){$\rho_1$}}
\put(20,10){\makebox(0,0){$\rho_2$}}
\put(55,10){\makebox(0,0){$\rho_n$}}
\put(100,5){\makebox(0,0){$\la$}}
\put(70,5){\makebox(0,0){$\mu$}}
\put(115,10){\makebox(0,0){$\nu_1$}}
\put(130,10){\makebox(0,0){$\nu_2$}}
\put(165,10){\makebox(0,0){$\nu_m$}}
\put(37.5,10){\makebox(0,0){$\cdots$}}
\put(147.5,10){\makebox(0,0){$\cdots$}}
\end{picture}
\end{center}
\caption{$\rho_1\rho_2\cdots\rho_n(\eps\la\mu)$}
\label{rrrepslm}
\end{figure}
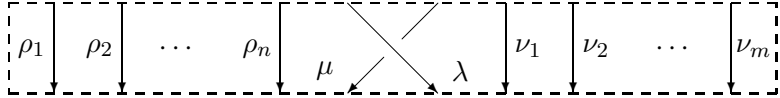

\begin{figure}[htb]
\begin{center}
\unitlength 0.6mm
\begin{picture}(170,20)
\thinlines
\put(0,0){\dashbox{2}(170,20){}}
\put(10,20){\vector(0,-1){20}}
\put(25,20){\vector(0,-1){20}}
\put(60,20){\vector(0,-1){20}}
\put(85,20){\vector(0,-1){10}}
\put(85,10){\vector(-1,-1){10}}
\put(85,10){\vector(1,-1){10}}
\put(110,20){\vector(0,-1){20}}
\put(125,20){\vector(0,-1){20}}
\put(160,20){\vector(0,-1){20}}
\put(5,10){\makebox(0,0){$\rho_1$}}
\put(20,10){\makebox(0,0){$\rho_2$}}
\put(55,10){\makebox(0,0){$\rho_n$}}
\put(90,15){\makebox(0,0){$\la$}}
\put(70,5){\makebox(0,0){$\mu$}}
\put(100,5){\makebox(0,0){$\nu$}}
\put(85,5){\makebox(0,0){$t$}}
\put(115,10){\makebox(0,0){$\nu_1$}}
\put(130,10){\makebox(0,0){$\nu_2$}}
\put(165,10){\makebox(0,0){$\nu_m$}}
\put(37.5,10){\makebox(0,0){$\cdots$}}
\put(147.5,10){\makebox(0,0){$\cdots$}}
\end{picture}
\end{center}
\caption{$\protect\sqrt[4]{\frac{d_\mu d_\nu}{d_\la}}
\rho_1\rho_2\cdots\rho_n(t)$
where $t\in\Hom(\la,\mu\nu)$ is an isometry}
\label{rrrt}
\end{figure}
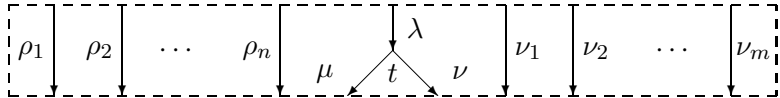

The intertwiners we consider are (sums over)
compositions of {\sl elementary intertwiners} arising
from the unitary braiding operators
$\eps\la\mu\in\Hom(\la\mu,\mu\la)$
and isometries $t\in\Hom(\la,\mu\nu)$. The wire diagrams
and boxes we deal with are  compositions
of ``elementary boxes'' representing the elementary intertwiners.
We now have to introduce some normalization convention. First,
the identity intertwiner $1\equiv1_M$ is naturally
given by the ``trivial box'' with only straight
wires of any labels. The next elementary
intertwiner is $\rho_1\rho_2\cdots\rho_n(\eps\la\mu)$
for which we have a diagram as in Figure \ref{rrrepslm}
%
where the arbitrary labels $\nu_1,\dots,\nu_m$ are irrelevant.
Similarly, the box of Figure \ref{rrrt}
%
represents the elementary intertwiner 
$d_\mu^{1/4} d_\nu^{1/4} d_\la^{-1/4}\rho_1\rho_2\cdots\rho_n(t)$,
where $t\in\Hom(\la,\mu\nu)$ is an isometry. We label the
trivalent vertex in the box with an isometry $t$ to
label elements in $\Hom(\la,\mu\nu)$.
Finally, the elementary intertwiners
$\eps\la\mu ^*=\epsm\mu\la$ and 
$d_\mu^{1/4} d_\nu^{1/4} d_\la^{-1/4}\rho_1\rho_2\cdots\rho_n(t)^*$
are similarly represented as the vertical reflections. 
Note that $\e\equiv\e^+$ represents overcrossing
and $\e^-$ undercrossing of wires.
We consider intertwiners which are products
of diagrammatically composable elementary intertwiners.
Note that if a wire
diagram represents some intertwiner $x$, then $x^*$ is
represented by the diagram obtained by vertical reflection
and reversing all the arrows. Note that our resulting wire
diagrams are then composed only from straight lines, over- and
undercrossings and trivalent vertices.

\begin{figure}[htb]
\begin{center}
\unitlength 0.6mm
\begin{picture}(145,20)
\thinlines
\put(25,20){\vector(0,-1){10}}
\put(25,10){\vector(-1,-1){10}}
\put(25,10){\vector(1,-1){10}}
\put(30,15){\makebox(0,0){$\id$}}
\put(10,5){\makebox(0,0){$\bar\la$}}
\put(40,5){\makebox(0,0){$\la$}}
\put(50,10){\makebox(0,0){$=$}}
\put(75,10){\vector(-1,-1){10}}
\put(75,10){\vector(1,-1){10}}
\put(60,5){\makebox(0,0){$\bar\la$}}
\put(90,5){\makebox(0,0){$\la$}}
\put(100,10){\makebox(0,0){$=$}}
\put(120,2){\arc{20}{3.142}{0}}
\put(110,0){\line(0,1){2}}
\put(130,0){\line(0,1){2}}
\put(135,5){\makebox(0,0){$\la$}}
\put(130,0){\vector(0,-1){0}}
\end{picture}
\end{center}
\caption{Wire diagrams for $\protect\sqrt{d_\la} r_\la$}
\label{risom}
\end{figure}
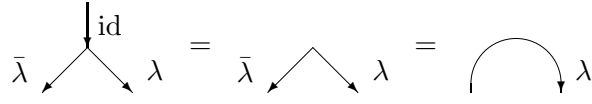

\begin{figure}[htb]
\begin{center}
\unitlength 0.6mm
\begin{picture}(170,40)
\thinlines
\put(8,5){\makebox(0,0){$\la$}}
\put(15,20){\vector(0,-1){20}}
\put(25,20){\arc{20}{3.142}{0}}
\put(45,20){\arc{20}{0}{3.142}}
\put(55,40){\line(0,-1){20}}
\put(70,20){\makebox(0,0){$=$}}
\put(85,40){\vector(0,-1){40}}
\put(92,5){\makebox(0,0){$\la$}}
\put(100,20){\makebox(0,0){$=$}}
\put(115,40){\line(0,-1){20}}
\put(125,20){\arc{20}{0}{3.142}}
\put(145,20){\arc{20}{3.142}{0}}
\put(155,20){\vector(0,-1){20}}
\put(162,5){\makebox(0,0){$\la$}}
\end{picture}
\end{center}
\caption{A topological invariance for intertwiners represented
by wire diagrams}
\label{isoinv1}
\end{figure}
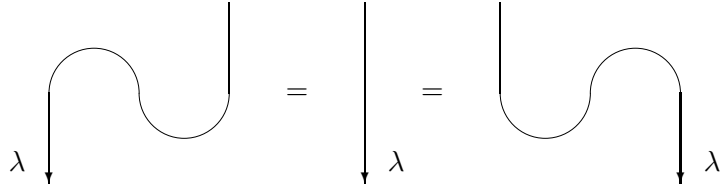

\begin{figure}[htb]
\begin{center}
\unitlength 0.6mm
\begin{picture}(80,20)
\thinlines
\put(22,19.84){\vector(1,0){0}}
\put(20,10){\circle{20}}
\put(5,10){\makebox(0,0){$\la$}}
\put(40,10){\makebox(0,0){$=$}}
\put(58,19.84){\vector(-1,0){0}}
\put(60,10){\circle{20}}
\put(75,10){\makebox(0,0){$\la$}}
\end{picture}
\end{center}
\caption{Wire diagrams for the statistical dimension $d_\la$}
\label{dla}
\end{figure}
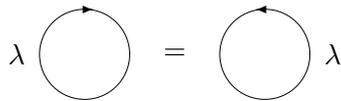

So far, we have considered only wires with downward orientation.
We now introduce also the reversed orientation in terms of conjugation
as follows. Reversing the orientation of an arrow on a wire
changes its label $\la$ to $\bar\la$. Also we usually omit
drawing a wire labelled by $\id\equiv\id_M$. For each $\lambda\in\sys$,
we fix (the common phase of) isometries
$r_\la\in\Hom(\id,\bar\la\la)$ and
${\bar r}_\la\in\Hom(\id,\la\bar\la)$ arising from the
Frobenius reciprocity $\Hom(\la,\la)\isom\Hom(\id,\bar\la\la)\isom
\Hom(\id,\la\bar\la)$ so that we have
$\la(r_\la)^*{\bar r}_\la=\bar\la({\bar r}_\la)^*r_\la=d_\la^{-1} 1$
and in turn for $\sqrt{d_\la} r_\la$ we draw one of the
equivalent diagrams in Figure \ref{risom}.
%
So the normalized isometries and their adjoints appear in wire
diagrams as ``caps'' and ``cups'', respectively.
The point is that with our normalization convention, the
relation $\la(r_\la)^*{\bar r}_\la=d_\la^{-1}1$ (and its
adjoint) gives a  topological invariance
for intertwiners represented by wire diagrams,
displayed as in Figure \ref{isoinv1}.
%
Note that then the wire diagrams in Figure \ref{dla}
%
represent the scalar $d_\la$ (i.e., the intertwiner
$d_\la 1\in\Hom(\id,\id)$).

\begin{figure}[htb]
\begin{center}
\unitlength 0.6mm
\begin{picture}(120,60)
\thinlines
\put(26.180,5){\arc{32.361}{3.142}{4.249}}
\put(10,5){\vector(0,-1){5}}
\put(28,24){\line(-2,-1){9.1}}   
\put(33.820,45){\arc{32.361}{0}{1.107}}
\put(50,60){\line(0,-1){15}}
\put(32,26){\line(2,1){9.1}}
\put(30,60){\line(0,-1){35}}
\put(30,25){\vector(0,-1){15}}
\put(30,10){\vector(1,-1){10}}
\put(30,10){\vector(-1,-1){10}}
\put(30,5){\makebox(0,0){$t$}}
\put(35,55){\makebox(0,0){$\la$}}
\put(18,5){\makebox(0,0){$\mu$}}
\put(42,5){\makebox(0,0){$\nu$}}
\put(5,5){\makebox(0,0){$\rho$}}
\put(60,25){\makebox(0,0){$=$}}
\put(86.180,5){\arc{32.361}{3.142}{4.199}}    
\put(70,5){\vector(0,-1){5}}
\put(90,25){\line(-2,-1){8.1}}   
\put(93.820,45){\arc{32.361}{0}{1.057}} 
\put(110,60){\line(0,-1){15}}
\put(90,25){\line(2,1){8.1}}
\put(90,60){\vector(0,-1){10}}
\put(90,50){\line(1,-1){10}}
\put(90,50){\line(-1,-1){10}}
\put(90,45){\makebox(0,0){$t$}}
\put(80,40){\line(0,-1){20}}
\put(100,40){\line(0,-1){10}}
\put(80,10){\line(0,1){10}}
\put(100,10){\line(0,1){20}}
\put(80,10){\vector(0,-1){10}}
\put(100,10){\vector(0,-1){10}}
\put(95,55){\makebox(0,0){$\la$}}
\put(85,5){\makebox(0,0){$\mu$}}
\put(105,5){\makebox(0,0){$\nu$}}
\put(65,5){\makebox(0,0){$\rho$}}
\end{picture}
\end{center}
\caption{The first braiding fusion equation}
\label{wireBFE1}
\end{figure}
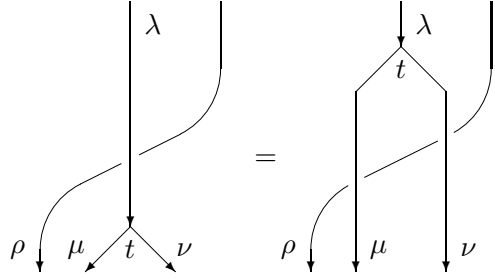

The BFE's give another topological invariance,
as in Figure \ref{wireBFE1} for the first equation.
%
The second, third and fourth equations are obtained similarly.
Up to conjugation they
can also be obtained by changing over- to undercrossings in
Figure \ref{wireBFE1}. 

The topological invariance gives us the freedom to write
down the intertwiner algebraically from a given wire diagram.
We can deform the wire diagram by finite sequences of
the above moves and then split it into
elementary wire diagrams --- in whatever way we may
decompose the wire diagrams into horizontal slices
of elementary intertwiners,
we always obtain the same intertwiner due to our topological
invariance identities.

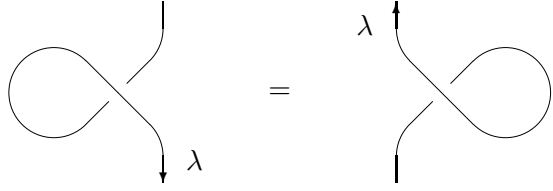
\begin{figure}[htb]
\begin{center}
\unitlength 0.6mm
\begin{picture}(120,40)
\thinlines
\put(10,20){\arc{20}{0.785}{5.498}}
\put(24.142,20){\line(-1,1){7.071}}
\put(22.142,18){\line(-1,-1){5.071}}
\put(24.142,20){\line(1,-1){7.071}}
\put(26.142,22){\line(1,1){5.071}}
\put(24.142,34.142){\arc{20}{0}{0.785}}
\put(24.142,5.858){\arc{20}{5.498}{6.283}}
\put(34.142,40){\line(0,-1){5.858}}
\put(34.142,5.858){\vector(0,-1){5.858}}
\put(41.142,5){\makebox(0,0){$\la$}}
\put(60,20){\makebox(0,0){$=$}}
\put(110,20){\arc{20}{3.927}{2.357}}
\put(95.858,20){\line(1,-1){7.071}}
\put(97.858,22){\line(1,1){5.071}}
\put(95.858,20){\line(-1,1){7.071}}
\put(93.858,18){\line(-1,-1){5.071}}
\put(95.858,34.142){\arc{20}{2.357}{3.142}}
\put(95.858,5.858){\arc{20}{3.142}{3.927}}
\put(85.858,34.142){\vector(0,1){5.858}}
\put(85.858,0){\line(0,1){5.858}}
\put(78.858,35){\makebox(0,0){$\la$}}
\end{picture}
\end{center}
\caption{Statistics phase $\omega_\la$ as a ``twist''}
\label{statph}
\end{figure}

Next, the statistics phase $\omega_\la$
is defined by the wire diagram on the
left-hand side of Figure \ref{statph}, which is equal to
the intertwiner $d_\la r_\la^* \bar\la(\eps\la\la)r_\la$.
%
The diagram on the right-hand side expresses that $\omega_\la$
can also be obtained as
$d_\la \bar\la(r_\la)^*\eps{\bar\la}{\bar\la} \bar\la (r_\la)$.
We also define a matrix element
$Y_{\la,\mu}=Y_{\mu,\la}$ of Rehren's Y-matrix \cite[Section 5]{R1}
as $d_\la d_\mu \phi_\mu(\eps \la\mu \eps \mu\la)^*
= d_\la d_\mu r_\mu^* \bar\mu (\epsm \la\mu \epsm \mu\la) r_\mu$.

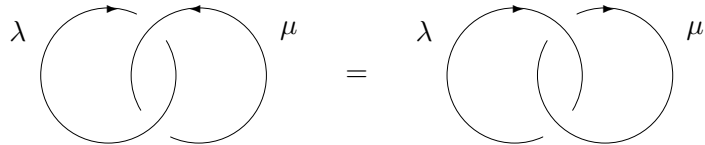
\begin{figure}[htb]
\begin{center}
\unitlength 0.6mm
\begin{picture}(160,32)
\thinlines
\put(25,15){\arc{30}{5.742}{5.142}}
\put(45,15){\arc{30}{2.601}{2.001}}
\put(27,29.8){\vector(1,0){0}}
\put(43,29.8){\vector(-1,0){0}}
\put(5,25){\makebox(0,0){$\la$}}
\put(65,25){\makebox(0,0){$\mu$}}
\put(80,15){\makebox(0,0){$=$}}
\put(115,15){\arc{30}{1.141}{0.541}}
\put(135,15){\arc{30}{4.283}{3.683}}
\put(117,29.8){\vector(1,0){0}}
\put(137,29.8){\vector(1,0){0}}
\put(95,25){\makebox(0,0){$\la$}}
\put(155,25){\makebox(0,0){$\mu$}}
\end{picture}
\end{center}
\caption{Matrix element $Y_{\la,\mu}$ of Rehren's Y-matrix
as a Hopf link}
\label{Ymatrix}
\end{figure}

We have drawn the circle $\mu$ symmetrically relative
to the straight wire $\la$ because it does not make a difference
whether we put the ``caps'' and ``cups'' for the isometry
$r_\mu$ and its conjugate $r_\mu^*$ on the left or on the
right due to the braiding fusion relations.
Since it is a scalar, we can write
$Y_{\la,\mu}={\bar r}_\mu^*Y_{\la,\mu}{\bar r}_\mu$
and therefore its expression
$d_\la d_\mu {\bar r}_\mu^*r_\la^* \bar\la
(\epsm \mu\la \epsm \la\mu) r_\la {\bar r}_\mu$ 
exactly gives the Hopf link as the wire diagram
for the matrix element $Y_{\la,\mu}$, given by the
left-hand side of Figure \ref{Ymatrix}.
%
The equality to the right-hand side is just the
relation $Y_{\la,\mu}=Y_{\la,\bar\mu}^*$.
We put $S_{\la,\mu}=Y_{\la,\mu}\sqrt{w}$
and $T_{\la,\mu}=(\sigma/|\sigma|)^{1/3}\delta_{\la,\mu}\omega_\la$,
where $w=\sum_{\la\in\sys} d^2_\la$ and 
$\sigma=\sum_{\la\in\sys} d^2_\la \omega_\la^{-1}$.
We say that $\la$ is {\sl degenerate} if we have
$Y_{\la,\mu}=d_\la d_\mu$ for all $\mu\in\sys$.
This is equivalent to the condition that we have
$\e^+(\la,\mu)=\e^-(\la,\mu)$ for all $\mu\in\sys$
by \cite[Lemma on page 352]{R1}.
We have the following proposition as in \cite[page 351]{R1}.

\begin{proposition}
The following are equivalent.

(1) The identity morphism is the only degenerate morphism
in $\sys$.

(2) The matrix $(Y_{\la,\mu})$ is invertible.
\end{proposition}

If the above two conditions hold, we can define a unitary
representation $\pi$ of $SL(2,\Z)$ by setting
$\pi\left(\left(\begin{array}{cc}
0 & -1 \\
1 & 0 \end{array}\right)\right)=S$,
$\pi\left(\left(\begin{array}{cc}
1 & 1 \\
0 & 1 \end{array}\right)\right)=T$.  
(See \cite[page 351]{R1}.)  In this case,
we say that the tensor category consisting of
endomorphisms of $M$ unitarily equivalent to
a finite direct sum of endomorphisms in $\sys$
is a unitary {\sl modular tensor category} in the
following abstract sense.
(See \cite[Chapter 3]{BK}, \cite[Section II.1.4]{Tu}
for more details on modular
tensor categories.  See \cite{RT}, \cite[Chapter IV]{Tu},
\cite{KSW} for a
$(2+1)$-dimensional {\sl topological quantum field theory}
arising from a modular tensor category.)
We also say that the braiding is {\sl non-degenerate}
in this case.  

\begin{definition}{\rm
A {\sl ribbon category} is a rigid braided tensor category
$\mathcal C$
with functorial isomorphisms $\de_V:V\to V^{**}$ satisfying
$\de_{V\otimes W}=\de_V\otimes\de_W$, $\de_{\mathbf 1}=\id$,
and $\de_{V^*}=(\de^*_V)^{-1}$ for all objects $V,W$ in
$\mathcal C$.
}\end{definition}

\begin{definition}{\rm
A {\sl modular tensor category} is a semisimple ribbon
category with the following properties.
\begin{enumerate}
\item We have only finitely many isomorphism classes of
simple objects.
\item The matrix $(Y_{\la\mu})$, defined as above for the
representatives $\la,\mu$ for the isomorphism classes
of simple objects, is invertible.
\end{enumerate}
}\end{definition}

In this case, we have the celebrated
{\sl Verlinde formula},
\begin{equation}\label{verl}
N_{\la\mu}^\nu=\sum_{\sigma}\frac
{S_{\la,\sigma}S_{\mu,\sigma}S^*_{\nu,\sigma}}{S_{0,\sigma}},
\end{equation}
where the nonnegative integer $N_{\la\mu}^\nu$ is determined
by the fusion rules
$\la\cdot\mu=\sum_\nu N_{\la\mu}^\nu \nu$.

\begin{figure}[htb]
\begin{center}
\unitlength 0.6mm
\begin{picture}(60,40)
\thinlines
\dottedline{1}(15,25)(30,10)(45,25)(15,25)
\put(30,30){\vector(0,-1){10}}
\put(30,20){\vector(-1,-1){10}}
\put(30,20){\vector(1,-1){10}}
\put(10,30){\makebox(0,0){$N$}}
\put(30,5){\makebox(0,0){$P$}}
\put(50,30){\makebox(0,0){$M$}}
\put(15,5){\makebox(0,0){$\sigma$}}
\put(30,35){\makebox(0,0){$\rho$}}
\put(45,5){\makebox(0,0){$\tau$}}
\put(30,15){\makebox(0,0){$t$}}
\end{picture}
\end{center}
\caption{The intertwiner
$\protect\sqrt[4]{\frac{d_\sigma d_\tau}{d_\rho}}t$
as a triangle}
\label{tABC}
\end{figure}
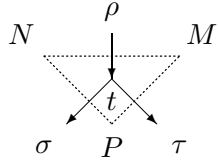

Wire diagrams can also be used for intertwiners of morphisms
between different factors. Let $M,N,P$ infinite factors,
$\rho\in\Mor(M,N)$, $\sigma\in\Mor(P,N)$,
$\tau\in\Mor(M,P)$ irreducible morphisms and
$t\in\Hom(\rho,\sigma\tau)$ an isometry. Then Figure \ref{tABC}
%
represents the intertwiner
$d_\sigma^{1/4} d_\tau^{1/4} d_\rho^{-1/4} t$.
Similarly we can draw a picture using a co-isometry.
Along the lines of the previous paragraphs, we can
similarly build up larger wire diagrams out of trivalent
vertices involving different factors. 
So far we do not have a meaningful way to cross wires with
differently labelled regions left and right, but all the
arguments listed above which do not involve braidings
can be used for intertwiners of morphisms between different
factors exactly as proceeded above. Moreover, the diagrams
may also involve wires where left and right regions are
labelled by the same factor, i.e., these wires correspond to
{\em endo}morphisms of some factor which may well
form a braided system, and then one may have crossings for
those wires.

\begin{figure}[htb]
\begin{center}
\unitlength 0.6mm
\begin{picture}(170,40)
\thinlines
\put(20,20){\vector(0,-1){20}}
\put(12.929,27.071){\line(1,-1){7.071}}
\put(27.071,27.071){\vector(-1,-1){7.071}}
\put(20,34.142){\arc{20}{2.356}{3.142}}
\put(20,34.142){\arc{20}{0}{0.785}}
\put(10,34.142){\vector(0,1){5.858}}
\put(30,40){\line(0,-1){5.858}}
\put(5,35){\makebox(0,0){$\rho$}}
\put(35,35){\makebox(0,0){$\tau$}}
\put(25,5){\makebox(0,0){$\sigma$}}
\put(16,17){\makebox(0,0){$t$}}
\put(50,20){\makebox(0,0){$:=$}}
\put(70.858,20){\vector(0,1){20}}
\put(80.858,20){\arc{20}{0.785}{3.142}}
\put(95,40){\vector(0,-1){20}}
\put(95,20){\line(-1,-1){7.071}}
\put(95,20){\line(1,-1){7.071}}
\put(95,5.858){\arc{20}{5.498}{0}}
\put(105,5.858){\vector(0,-1){5.858}}
\put(65,35){\makebox(0,0){$\rho$}}
\put(100,35){\makebox(0,0){$\tau$}}
\put(110,5){\makebox(0,0){$\sigma$}}
\put(95,15){\makebox(0,0){$t$}}
\put(125,20){\makebox(0,0){$=$}}
\put(155,20){\vector(0,-1){20}}
\put(147.929,27.071){\line(1,-1){7.071}}
\put(162.071,27.071){\vector(-1,-1){7.071}}
\put(155,34.142){\arc{20}{2.356}{3.142}}
\put(155,34.142){\arc{20}{0}{0.785}}
\put(145,34.142){\vector(0,1){5.858}}
\put(165,40){\line(0,-1){5.858}}
\put(140,35){\makebox(0,0){$\rho$}}
\put(170,35){\makebox(0,0){$\tau$}}
\put(160,5){\makebox(0,0){$\sigma$}}
\end{picture}
\end{center}
\caption{Left Frobenius reciprocity for a trivalent vertex
labelled by an isometry}
\label{Ltight}
\end{figure}
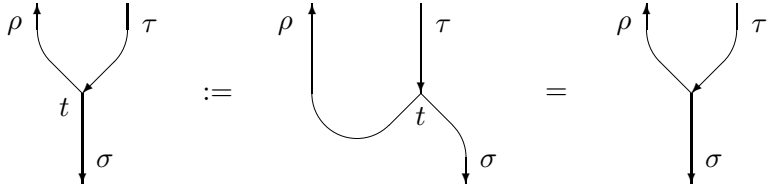

Let $M,N,P$ be type III factors, $\rho\in\Mor(P,N)$,
$\tau\in\Mor(M,N)$, $\sigma\in\Mor(M,P)$
morphisms with finite  dimensions.
For $t\in\Hom(\tau,\rho\sigma)$, we have the identity
as in Figure \ref{Ltight} from the Frobenius reciprocity.
Here the trivalent vertex in the right diagram represents
the co-isometry given by the Frobenius reciprocity.
%
%

If we have a link diagram whose components are labelled with
elements of $\sys$, the entire diagram denotes an intertwiner
from $\id$ to $\id$, that is, a scalar.  This is a $\C$-valued
{\sl regular isotopy invariant} of a {\sl colored} link,
since this number is invariant under the Reidemeister moves
of types II and III \cite[Definition 1.13]{BZ}.
(Labeling a wire is called coloring in this context.)
In general, this is not a topological invariant since
the twist $\omega_\la$ is not necessarily $1$  and
the number is not invariant under the Reidemeister move of
type I.  This is essentially
a generalization of the Jones polynomial at specific values.
(See \cite{J2} for the original Jones polynomial.)

\subsection{Complete rationality and modular tensor categories}

A unitary braided tensor category of representations of a local
conformal net is similar to that of those of a
quantum group at a root of unity.   In such a representation
theory, it is important to consider a case where we have
only finitely many irreducible representations up to
unitary equivalence.  Such finiteness is often called
{\sl rationality}.  This name comes from the fact that such
finiteness of representation theory gives rationality of
various parameters in conformal field theory.
Based on this, we introduce the following notion.

\begin{definition}{\rm 
Let $\{\A(I)\}$ be a local conformal net with the split
property.  Split the circle $S^1$ into four intervals and
label them  $I_1$, $I_2$, $I_3$, $I_4$ in the clockwise order.
If the subfactor
$\A(I_1)\vee \A(I_3) \subset (\A(I_2)\vee \A(I_4))'$ 
has a finite index, we say that the local conformal net
$\{\A(I)\}$ is {\sl completely rational}.
}
\end{definition}

The reason we call this complete rationality comes
from the following theorem \cite[Theorem 33, Corollary 37]{KLM}.

\begin{theorem}
\label{klm-main}
When a local conformal net 
is completely rational, then it
has only finitely many irreducible representations
up to unitary equivalence, and all of them have 
finite dimensions.
When this holds, the unitary braided tensor category of
finite dimensional representations of $\{\A(I)\}$ 
is a unitary modular tensor category and the index of the
above subfactor
$\A(I_1)\vee \A(I_3) \subset (\A(I_2)\vee \A(I_4))'$ 
is equal to the square sum of the
dimensions of the irreducible representations.
\end{theorem}

We have finite dimensionality of the 
irreducible representations
and this is why we have added the word ``completely''.
Note that it is difficult in general to know all the irreducible
representations, but the above theorem gives information
on representations from a subfactor defined in the vacuum
representation.

We call the index of the above subfactor
$\A(I_1)\vee \A(I_3) \subset (\A(I_2)\vee \A(I_4))'$ 
the {\sl $\mu$-index} of $\{\A(I)\}$.  This index is 
independent of the choice of $I_1, I_2, I_3, I_4$.
(See \cite[Proposition 5]{KLM}.)  It has been proved
in \cite[Theorem 5.3]{LX} that a completely rational local
conformal net is strongly additive.
(In this text, we consider only local conformal nets, but
one sometimes deal with local M\"obius covariant nets.
Then one has to include strong additivity as one of
the requirements.)

The above theorem implies that 
for a local conformal net with the split property,
all of its irreducible representations
are unitarily equivalent to the
vacuum representation if and only if the local conformal
net has $\mu$-index $1$, since the vacuum representation
has dimension $1$.
We call such a local conformal net {\sl holomorphic}.
This name comes from holomorphicity of the partition function
of a full conformal field theory.

An outline of the proof of the first half
of Theorem \ref{klm-main} in \cite{KLM} is as follows.

Let $\rho$ and $\sigma$ be finite dimensional irreducible
localized endomorphisms of
$\{\A(I)\}$ localized in $I_1$ and $I_3$, respectively.
Then $\rho\sigma|_{\A(I_1)\vee \A(I_3)}$ gives an
endomorphism of $\A(I_1)\vee \A(I_3)$.
Let $\la$ be the dual canonical endomorphism for the
subfactor
$\A(I_1)\vee \A(I_3) \subset (\A(I_2)\vee \A(I_4))'$.
The Frobenius reciprocity applied to
$\iota: \A(I_1)\vee \A(I_3)\hookrightarrow
(\A(I_2)\vee \A(I_4))'$ gives
$$\Hom(\lambda, \rho\sigma|_{\A(I_1)\vee \A(I_3)})\isom
\Hom(\iota,\iota\rho\sigma|_{\A(I_1)\vee \A(I_3)})$$ since
$\lambda=\bar\iota\iota$, and then we have an isometry 
$v \in (\A(I_2)\vee \A(I_4))'$ satisfying $vx=\rho\sigma(x)v$
for all $x\in \A(I_1)\vee \A(I_3)$ if and only if 
$\rho\sigma|_{\A(I_1)\vee \A(I_3)}$ appears in the
irreducible decomposition of
$\la$.  If $v$ satisfies the above identity,
the strong additivity implies the same identity for all
$x\in \A(I)$ for all $I$.  This gives $\sigma=\bar\rho$
as representations of $\{\A(I)\}$ again by the Frobenius
reciprocity.  Conversely, if
we have $\sigma=\bar\rho$, then we have an isometry
$v\in \A(\tilde I)$ satisfying $vx=\rho\sigma(x)v$ for
all $x\in\A(\tilde I)$ where $\tilde I$
is given by $\tilde I=I_1\cup\bar I_2 \cup I_3$.  
Since $\rho$ and $\sigma$ act trivially on
$\A(I_2)$, we have $v\in\A(I_2)'\cap \A(\tilde I)$,
and we also have 
$\A(I_2)'\cap \A(\tilde I)=(\A(I_2)\vee\A(I_4))'$.
This means $\la$ contains $\rho\sigma|_{\A(I_1)\vee \A(I_3)}$ 
as a direct summand  with multiplicity $1$.

This now implies that for
all the representations $\{\rho_i\}$ of $\{\A(I)\}$,
the direct sum
$\bigoplus_i \rho_i\bar\rho_i|_{\A(I_1)\vee \A(I_3)}$
is a direct summand in the irreducible decomposition of $\la$,
where $\rho_i$ is localized in $I_1$ and $\bar\rho_i$ is
localized in $I_3$.
Considering the representation theory of a new type of
``net'' which assigns $\A(I)\otimes\A(jI)$ to each
interval $I$ where $j$ is the complex conjugation map
on the unit circle $S^1$ in the complex plane, we
find that an irreducible direct summand of
$\la$ must be of the form
$\rho\sigma|_{\A(I_1)\vee \A(I_3)}$, so this gives
that $\bigoplus_i \rho_i\bar\rho_i|_{\A(I_1)\vee \A(I_3)}$
is actually equal to $\la$.  
We also show that we do not have any infinite dimensional
irreducible representation of $\{\A(I)\}$ by 
\cite[Corollary 39]{KLM}.  This completes the proof
of the first half
of Theorem \ref{klm-main}, because $d_\la=\mu$.
An outline of the proof 
of the second half is given after introduction of
$\a$-induction.

A conformal {\sl subnet} $\{\B(I)\}$ of $\{\A(I)\}$ is an
isotonic map assigning
$\B(I)\subset\A(I)$ to each interval $I$ satisfying
$U(g)\B(I)U(g)^* = \B(gI)$ for all $g\in\Diff(S^1)$.
Each $\B(I)$ is automatically a factor since the modular
automorphism group $\sigma_\Omega$ is ergodic on $\B(I)$.
The Reeh-Schlieder theorem
adapted to this setting implies that
the Hilbert space $H_0 = \overline{\B(I)\Omega}$
is independent of $I$.  The restriction of $\{\B(I)\}$ to
$H_0$ is then a local conformal net and we write
$\{\B_0(I)\}$ for this.  We say that
$\{\B_0(I)\}$ is a {\sl subnet} of $\{\A(I)\}$ and we often
identify $\{\B(I)\}$ with $\{\B_0(I)\}$.  We also say that
the local conformal net $\{\A(I)\}$ is an {\sl extension}
of $\{\B(I)\}$.  When we have $\B(I)'\cap \A(I)=\C$, we
say that the extension is {\sl irreducible}.  The index
$[\A(I):\B(I)]$ is independent of $I$.  We have the
following theorem.

\begin{theorem}
\label{finite-index}
Let $\{\B(I)\subset\A(I)\}$ be an inclusion of local
conformal nets with the split property having 
the index $[\A(I):\B(I)]$.  We write
$\mu_\A$ and $\mu_\B$ for the $\mu$-indices
of $\{\A(I)\}$ and $\{\B(I)\}$, respectively.  Then
we have $\mu_\B=\mu_\A[\A(I):\B(I)]^2$,
In particular, if one of the two local conformal nets
$\{\A(I)\}$, $\{\B(I)\}$ is completely rational, 
then so is the other.
\end{theorem}

An outline of the proof of Theorem \ref{finite-index} in
\cite{KLM} is as follows.  On the Hilbert space $H$
of $\{\A(I)\}$, the index of the subfactor
$\A(I_1)\vee \A(I_3) \subset (\A(I_2)\vee \A(I_4))'$
is $\mu_\A$.  The indices of the subfactor
$\B(I_1)\vee\B(I_3)\subset\A(I_1)\vee\A(I_3)$ and
$\B(I_2)\vee\B(I_4)\subset\A(I_2)\vee\A(I_4)$ are
both equal to $[\A(I):\B(I)]^2$ by the split
property and Proposition \ref{tensor} (1).
The index of the subfactor
$\B(I_1)\vee \B(I_3) \subset (\B(I_2)\vee \B(I_4))'$.
{\sl on the Hilbert $H$ space of $\{\A(I)\}$} turns out
to equal to the index of the corresponding subfactor on 
the Hilbert space $H_0$ of $\{\B(I)\}$, denoted by 
$\mu_\B$, times $[\A(I):\B(I)]^2$.
(The Hilbert space $H$ for $\{\A(I)\}$ is {\sl larger}
than the one $H_0$ for $\{\B(I)\}$, so the index of the 
subfactor is multiplied by the square of
the relative size of
$\{\A(I)\}$ with respect to $\{\B(I)\}$.  
See the proof of \cite[Proposition 24]{KLM}.)
These relations give $\mu_\A=\mu_\B[\A(I):\B(I)]^2$
with Proposition \ref{tensor} (2).
The conclusion on the split property
follows from \cite[Lemma 22]{L5}.

Suppose we have a local conformal net
$\{\A(I)\}$ and its irreducible extension $\{\B(I)\}$.
Regarding the vacuum representation of
$\{\B(I)\}$ as a representation $\{\A(I)\}$, 
we have an irreducible decomposition
$\bigoplus_\lambda n_\lambda \lambda$, where
$n_\lambda$ is an integer representing the multiplicity.
The irreducibility implies that the multiplicity
of the vacuum representation of $\{\A(I)\}$ is $1$.
This gives a $Q$-system where the endomorphism
is given as a direct sum of
irreducible representations of
$\{\A(I)\}$.  Conversely, if an endomorphism
$\bigoplus_\lambda n_\lambda \lambda$,
where $n_0=1$ and $\lambda$'s are irreducible
representations of a local conformal net
$\{\A(I)\}$, gives a $Q$-system with locality,
then it gives an extension of $\{\A(I)\}$.
We first construct an extension $\B(I)$ for
one interval $I$, then we construct the family
$\{\B(I)\}$ using transporters.
(See \cite[Theorem 4.9]{LR1} for details.
Conformal covariance of the extension does not
follow from \cite[Theorem 4.9]{LR1}, but
the arguments in the proof of
\cite[Proposition 3.7 (b)]{Ca} gives conformal 
covariance of the extension.)
The index of the extension is equal to
$\sum_\lambda n_\lambda d_\lambda$.
(Here locality of the $Q$-system corresponds to
locality of the extension.)

As we have seen, the finite dimensional
representations of a 
local conformal net give a unitary modular tensor category.
Conversely, consider the following  realization problem.

\begin{problem} 
\label{prob-tensor}
For a given unitary modular tensor category $\mathcal C$, 
find a completely  rational local conformal net whose
finite dimensional representations  gives $\mathcal C$.
\end{problem}

We believe this problem always has a positive answer,
but this is still an open problem.  The reason to expect
a positive answer is that as long as we have appropriate
amenability on operator algebras,  similar realization
problems for subfactors, group actions and so on, have
always had positive answers.  Typical examples where
amenability controls algebraic structures uniformly
are \cite{C} and \cite{P1}.  A completely rational local
conformal net is ``amenable'' in any aspect, since each
factor for each interval is injective and the 
representation category has finitely many simple objects.
(The latter should be regarded as an amenable case,
just like finite groups are amenable.)

As a particular case of the above problem, consider
unitary modular tensor categories arising from the ``quantum
doubles'' of unitary fusion categories.
The quantum double construction of Drinfel$'$d \cite{Dr}
has a subfactor counterpart and it was first
considered by Ocneanu \cite{O2} under the name of the
``asymptotic inclusion''.  (See \cite[Chapter 12]{EK}
for more details.)  As we have seen in
Theorem \ref{fusion}, any unitary fusion category
is realized as that of $M$-$M$ bimodules of a
hyperfinite subfactor $N\subset M$ of type II$_1$
with finite index and finite depth.  Then the
$M_\infty$-$M_\infty$ bimodules of the asymptotic
inclusion $M\vee (M'\cap M_\infty)\subset M_\infty$
give a unitary modular tensor category, which is the
{\sl quantum double} of the original unitary
fusion category, where the II$_1$ factor $M_\infty$ is
given as the limit of the Jones basic construction
as in  \cite[Chapter 12]{EK}.  This construction has been 
formulated in the setting of type III subfactors
using the $Q$-system in \cite[Proposition 4.10]{LR1}
and studied in detail in \cite{I2,I3}.  The
$Q$-system is described as follows.

Let $M$ be a factor of type III and suppose we have
a system $\sys=\{\la_i\}_{i=0}^n$ of irreducible
endomorphisms of $M$.  Let $M^{\mathrm{opp}}$ the
opposite algebra of $M$, where the order of the
multiplication is reversed, and $j$ be the natural
map sending $x\in M$ to $x^*\in M^{\mathrm{opp}}$.
We set $\la=\bigoplus_{i=0}^n \la_i\otimes
(j\cdot\la_i\cdot j^{-1})$.  We can then find
$v,w$ satisfying the axioms for the $Q$-system.
The index of the subfactor arising from this
$Q$-system is equal to $\sum_i d_{\la_i}^2$.

In this context, the construction of a subfactor
is often called the {\sl Longo-Rehren subfactor} and
the $Q$-system is called the {\sl Longo-Rehren
$Q$-system}.  A generalization of the asymptotic
inclusion to another direction
has been given in \cite{P2}.

A particular example of the quantum double 
construction is given as follows.  Let $G$ be
a finite group.  The group semiring consisting of
the elements of the form $\sum_{g\in G} n_g g$
where $n_g\in\{0,1,2,\dots\}$ gives the objects
of a unitary fusion category where we have
$\Hom(g,h)=\delta_{g,h}\C$.  The quantum double
of this unitary fusion category is the quantum
double of $G$ in the usual sense.  This finite
group $G$ gives another unitary braided tensor category
of finite dimensional unitary representations of
$G$.  The braiding $\e^+(\la,\mu)=\e^-(\la,\mu)$
is given by  the canonical switching of the
two tensor components.  The quantum double
of this (degenerate) unitary braided tensor category
is also the same.

We then consider the following problem.

\begin{problem}
For a given unitary fusion category, find
a completely rational local conformal net
whose finite dimensional representations 
give the quantum double of the given
unitary fusion category.
\end{problem}

If we find such a local conformal net, it is easy
to see that it has an extension, 
using the dual of the Longo-Rehren $Q$-system 
given in \cite[Section 6]{I3}, and then
the $\mu$-index of the extension is $1$
by Theorem \ref{finite-index} because the original
$\mu$-index and the square of the index of the
Longo-Rehren $Q$-system are equal.  So we expect to be able
to construct an answer to the above problem as a 
subnet of a holomorphic local conformal net.
This should be an ``orbifold construction for a
paragroup action'', but we do not know yet such
a construction.  (See Example \ref{ex-orbif1} for
the orbifold construction.)
The work \cite{EG} gives a first step to this
direction for a particular example of the
Haagerup subfactor.

\subsection{$\alpha$-induction and modular invariants}

Consider an inclusion 
$\{\A(I)\subset \B(I)\}$ of a local conformal net
and its subnet.  Suppose the index $[\B(I):\A(I)]$ 
is finite.  We are interested in the case the smaller local
conformal net $\{\A(I)\}$ is completely rational.

In classical representation theory, if we have a
representation of a group $H$ with a larger group
$G\supset H$, then we have an induced representation
of $G$.  We would like to have its analogue for
local conformal nets.
Fix an interval $I\subset S^1$.  A representation
of  a local conformal net $\{\A(I)\}$ is realized as a
localized endomorphism of $\A(I)$, so denote this 
endomorphism by
$\lambda$.  We would like to extend $\lambda$  to an
endomorphism of $\B(I)$, using the braiding
between $\lambda$ and the dual canonical endomorphism
of $\A(I)\subset \B(I)$.
This construction is called
the {\sl $\alpha$-induction}.  This was first
defined in \cite{LR1}, and examples and applications
were first given in \cite{X1} and further developed
in \cite{BE1,BE2,BE3}.  This has been unified with
Ocneanu's graphical construction \cite{O3} on the Dynkin
diagrams and generalized in \cite{BEK1,BEK2}.

The construction of the extension of the endomorphism
is as follows.  This works on a more abstract setting 
where we have only a unitary
braided tensor category of endomorphisms,
so we present the basics in this abstract setting.

Let $N\subset M$ be a type III subfactor with finite index.
Suppose we have a braided system $\sys$ of endomorphisms
of $N$ and the dual canonical endomorphism for $N\subset M$
has an irreducible decomposition within $\sys$.  We write
$\iota:N\hookrightarrow M$ for the inclusion map and
$\theta=\bar\iota\iota$ for the dual canonical endomorphism
of $N\subset M$.  We recall that we have an isometry $v\in M$
satisfying $M=Nv$ and $vx=\iota\bar\iota(x)v$ for all $x\in M$.

\begin{definition}{\rm
For $\la$ unitarily equivalent to a finite
direct sum of endomorphisms in $\sys$, we set
$$\a^\pm_\la=\bar\iota^{-1}\cdot\Ad(\e^\pm(\la,\theta))
\cdot\la\cdot\bar\iota,$$
and call it the $\a^\pm$-induction of $\la$.
}\end{definition}

It is easy to see that $\a^\pm_\la(x)=\la(x)$ for any $x\in N$.
For $v\in M$, we have $\a^\pm_\la(v)=\e^\pm(\la,\theta)^*v$
as in \cite[Lemma 3.1]{BE1}, so the map $\a^\pm_\la$ is
an endomorphism of $M$ and regarded as an $M$-$M$ sector.
It is easier to see the meaning of this construction in
the context of bimodules as follows.  Suppose we have
a braided system of $N$-$N$ bimodules for a type II$_1$
subfactor $N\subset M$ so that the bimodule ${}_N M_N$
has an irreducible decomposition within the system.  Then
for an $N$-$N$ bimodule ${}_N X_N$ in the system, we have
an isomorphism ${}_N X\otimes_N M_N\isom {}_N X\otimes_N M_N$
through the braiding.  The left hand side expression has a
right action of $M$ and the right hand side expression
has a left action of $M$.  We see that these two actions
commute and we have an $M$-$M$ bimodule.  This construction
of an $M$-$M$ bimodule from an $N$-$N$ bimodule is the
bimodule version of the $\a$-induction.

We also introduce the following property.

\begin{definition}{\rm
When we have $\e^+(\theta,\theta)v=v$, we say that
the $Q$-system for $N\subset M$ is local.
}\end{definition}

The Longo-Rehren $Q$-system arising from a braided
system $\sys$ is local \cite[Proposition 4.10]{LR1}.
If $N\subset M$ arises from an inclusion of local
conformal nets, this locality holds as in
\cite[Theorem 4.9]{LR1} from the locality of the larger
conformal net.

We have the following result as in \cite[Theorem 3.9]{BE1}.

\begin{theorem}
\label{main-be1}
If we have locality, then we have
$$\langle \a^\pm_\la,\a^\pm_\mu\rangle=
\langle \la\cdot\theta,\mu\rangle.$$
\end{theorem}

An outline of a proof is as follows.  We have
\begin{align*}
\langle {}_M {\a^\pm_\la}_M,{}_M {\a^\pm_\mu}_M \rangle&\le
\langle {}_N \bar\iota\cdot {\a^\pm_\la}_M,{}_N
\bar\iota\cdot {\a^\pm_\mu}_M \rangle\\
&=&
\langle {}_N \Ad(\e^\pm(\la,\theta))\cdot\la\cdot\bar\iota_M,
{}_N \Ad(\e^\pm(\mu,\theta))\cdot\mu\cdot\bar\iota_M, \rangle\\
&=\langle {}_N \la \bar\iota \iota_N,{}_N\mu_N\rangle.
\end{align*}
We can actually show that the locality implies an equality in the
first inequality.

We now assume that the braiding on $\sys$ is non-degenerate.
Consider the two generators 
$S=\left(\begin{array}{cc} 0 & -1 \\ 1 & 0 \end{array}\right)$,
$T=\left(\begin{array}{cc} 1 & 1 \\ 0 & 1 \end{array}\right)$,
of $SL(2,{\mathbb{Z}})$.  Since we have a 
unitary modular tensor category arising from $\sys$,
we have a finite dimensional unitary representation
of $SL(2,{\mathbb{Z}})$ as before,
whose dimension is equal to the number
of the endomorphisms in $\sys$.
We drop the symbol for a representation and simply write
$S, T$ for the images of the matrices $S, T$.
Take $\lambda,\mu\in\sys$ and set
$Z_{\lambda,\mu}=\dim\Hom(\alpha^+_\lambda,\alpha^-_\mu)$.
Then we have the following theorem (\cite[Corollary 5.8]{BEK1}).

\begin{theorem}
\label{mod-inv}
The above matrix $(Z_{\lambda,\mu})_{\lambda,\mu}$ has
the following properties.
\begin{enumerate}
\item $Z_{\lambda,\mu}\in{\mathbb{N}}=\{0,1,2,\dots\}$.
\item We have $Z_{00}=1$, where the label
$0$ denotes the vacuum representation.
\item We have $ZS=SZ$.
\item We have $ZT=TZ$.
\end{enumerate}
\end{theorem}

A proof of Theorem \ref{mod-inv} is based on graphical
expression of $Z_{\la,\mu}$ as in Figure \ref{Zgraph},
where $w=\sum_{\la\in\sys}d_\la^2$ and $b,c$ are
$N$-$M$ sectors.  We also follow the graphical
convention of \cite[Figures 33 and 41]{BEK1} here.
(We actually do not need the nondegeneracy of the
braiding here.)

\begin{figure}[htb]
\begin{center}
\unitlength 0.7mm
\begin{picture}(115,50)
\thinlines
\put(34,23){\makebox(0,0){$Z_{\la,\mu} \;= \;\displaystyle
\sum_{b,c} \;\;\frac{d_b d_c}{w d_\la d_\mu}$}}
\put(80,5){\line(0,1){40}}
\put(100,5){\line(0,1){18}}
\put(100,45){\line(0,-1){18}}
\put(80,45){\arc{5}{0}{3.142}}
\put(100,45){\arc{5}{0}{3.142}}
\put(80,5){\arc{5}{3.142}{0}}
\put(100,5){\arc{5}{3.142}{0}}
\put(72,25){\arc{5}{4.712}{1.571}}
\put(108,25){\arc{5}{1.571}{4.712}}
\put(72,25){\line(1,0){6}}
\put(108,25){\line(-1,0){26}}
\put(80,20){\vector(0,-1){0}}
\put(100,22){\vector(0,1){0}}
\thicklines
\put(77,5){\line(1,0){26}}
\put(77,45){\line(1,0){26}}
\put(72,10){\line(0,1){30}}
\put(108,10){\line(0,1){30}}
\put(77,40){\arc{10}{3.142}{4.712}}
\put(103,40){\arc{10}{4.712}{0}}
\put(77,10){\arc{10}{1.571}{3.142}}
\put(103,10){\arc{10}{0}{1.571}}
\put(90,2){\makebox(0,0){$c$}}
\put(90,48){\makebox(0,0){$c$}}
\put(69,15){\makebox(0,0){$b$}}
\put(69,35){\makebox(0,0){$b$}}
\put(111,15){\makebox(0,0){$b$}}
\put(111,35){\makebox(0,0){$b$}}
\put(76,35){\makebox(0,0){$\la$}}
\put(105,35){\makebox(0,0){$\mu$}}
\end{picture}
\end{center}
\caption{Graphical representation of $Z_{\la,\mu}$}
\label{Zgraph}
\end{figure}

A matrix $Z$ satisfying the above four conditions
is called a {\sl modular invariant}.
It is easy to see that we have only finitely many
modular invariants for each unitary modular tensor
category.  Modular invariants arising from
$SU(2)_k$ and $SU(3)_k$ have been classified
in \cite[Table 1]{CIZ}, \cite{G1}, respectively.

The following terminology comes from \cite{O3}.

\begin{definition}{\rm
When an $M$-$M$ sector appears in both of the
irreducible decompositions of the images of
$\alpha^\pm$-inductions, we say the $M$-$M$
sector is ambichiral.
}\end{definition}

The following is a combination of 
Proposition 3.1 and Theorem 4.2 in \cite{BEK2},
because the dual canonical endomorphism $\theta$
for the local case is given by
$\theta=\bigoplus_\la Z_{0,\la}\la$
by Theorem \ref{main-be1}.

\begin{theorem}
\label{ambichiral}
Suppose the $Q$-system for $N\subset M$ is local.
Let $d_1=\sum_{\la\in\sys} d_\la^2$ and
$d_2$ is the square sum of the dimensions of
the irreducible $M$-$M$ sectors.  Then
we have $d_2[M:N]^2=d_1$.
\end{theorem}

The proof in \cite{BEK2} is based on graphical
manipulations of diagrams.  We do need the nondegeneracy
of the braiding here.

In the application of the above construction we are 
interested in, we have an inclusion $\{\A(I)\subset\B(I)\}$
of a completely rational
local conformal net and its extension with finite index.
We fix an interval
$I$ and set $N=\A(I)$, $M=\B(I)$ and let $\sys$ be the system
of endomorphisms of $N$ arising from the 
irreducible DHR endomorphisms
of $\{\A(I)\}$ localized in $I$.  The embedding
$\A(I)\hookrightarrow \B(I)$ on the Hilbert space of
$\{\B(I)\}$ gives a representation of $\{\A(I)\}$, so it
decomposes within $\sys$, and it also gives the dual canonical
endomorphisms for $N\subset M$.  Then the $\a$-induction
$\a^\pm_\la$ of $\la\in\sys$ gives an $M$-$M$ sector.

We now present an outline of the proof of the second half of
Theorem \ref{klm-main} in \cite{KLM}.

Consider the unitary braided tensor category of finite
dimensional representations of a completely rational
local conformal
net $\{\A(I)\}$ and regard it as the one arising
from a braided system $\sys$ of endomorphisms of a
type III factor $N$.  Let $\{\A^{\mathrm{opp}}(I)\}$
be the local conformal net which assigns 
$\A(I)^{\mathrm{opp}}$ to each interval $I$.

Consider the Longo-Rehren $Q$-system $(\lambda,v,w)$ on 
$N\otimes N^{\mathrm{opp}}$ corresponding to $\sys$ of
the irreducible endomorphisms of $N=\A(I)$ arising
from irreducible representations of $\{\A(I)\}$. 
Since it is local,
it gives an extension $\A(I)\otimes\A^{\mathrm{opp}}(I)
\subset\B(I)$ of a local conformal net.
We apply the $\alpha$-induction for
the subfactor $\A(I)\otimes\A^{\mathrm{opp}}(I)
\subset\B(I)$.  If we have a
degenerate braiding on $\sys$, then we have a
non-trivial degenerate endomorphism $\rho$ in $\sys$,
but we then have 
$\a^+_{\rho\otimes\id}=\a^-_{\rho\otimes\id}$ by
degeneracy of the braiding and this gives a
representation of $\{\B(I)\}$ by \cite[Proposition 3.9]{LR1},
and this is different from
the identity endomorphism by Theorem \ref{main-be1}.  This 
is a contradiction to the identity $\mu_\B=1$ which follows
from Theorem \ref{finite-index}, since the index of the
Longo-Rehren $Q$-system is equal to the $\mu$-index of
$\{\A(I)\}$ and the $\mu$-index is multiplicative with
respect to a tensor product.  We have thus obtained the 
non-degeneracy of the braiding on the unitary braided
tensor category of finite
dimensional representations of the local conformal
net $\{\A(I)\}$.

The following is \cite[Theorem 3.21]{BE1}, which is
proved directly from the definition of the $\a$-induction.

\begin{theorem}
\label{alpha-sigma}
Let $\{\A(I)\}$ be a completely rational local
conformal net and $\{\B(I)\}$ be its extension with
finite index.  Let $N,M,\sys,\iota$ be as above and consider
the $\a^\pm$-induction.  We regard a representation
$\sigma$ of $\{\B(I)\}$ as an endomorphism of $M$.
Then we have
$\langle \a^\pm_\la,\sigma\rangle=
\langle\la,\bar\iota\cdot\sigma\cdot\iota\rangle$.
\end{theorem}

We now have the following important consequence.

\begin{theorem}
Let $\{\A(I)\}$ be a completely rational local
conformal net and $\{\B(I)\}$ be its extension with
finite index.  Let $I, N,M,\sys,\iota$ be as above and consider
the $\a^\pm$-induction.  Then
the ambichiral $M$-$M$ sectors are identified with
the localized endomorphisms of $\{B(I)\}$ in $I$.
\end{theorem}

An outline of the proof of this theorem is as follows.
By Theorem \ref{alpha-sigma}, we know that the
irreducible decomposition of any representation of
$\{\B(I)\}$ occurs among the ambichiral $M$-$M$
sectors.  Then
Theorems \ref{finite-index} and \ref{ambichiral}
give the conclusion.

Note that the above theorem
implies that the unitary modular tensor category
of finite dimensional representations of $\{\B(I)\}$ is
``smaller'' than that of $\{\A(I)\}$, and this is opposite
to the case of classical representations.  In this sense,
$\alpha$-induction is similar to restriction of representations
rather than induction to some extent.

Theorem \ref{mod-inv} says that any irreducible
extension $\{\B(I)\}$ of a completely rational
local conformal net $\{\A(I)\}$ produces a
modular invariant.
(The index $[\B(I):\A(I)]$ is automatically
finite in this situation.  See \cite[page 39]{ILP} 
and \cite[Proposition 2.3]{KL1}.)  Since we have
only a small number of modular invariants usually,
this gives a severe restriction on an extension
$\{\B(I)\}$ for a given $\{\A(I)\}$.

We also have computations of the quantum doubles
related to $\a$-induction in \cite{BEK3}.

The same mathematical structure as $\alpha$-induction
has been studied in the context of anyon condensation \cite{BS}.
See Table 1 on page 440 in \cite{Kn}.

\subsection{Examples and construction methods}

We now discuss how to construct local conformal nets.
One way to construct a local conformal net is from a
Kac-Moody Lie algebra, but from our viewpoint, it is
easier to use a loop group for a compact Lie group.
Consider a connected and simply connected Lie group,
say, $SU(N)$.
Let $L(SU(N))$ be the set of all the $C^\infty$-maps from
$S^1$ to $SU(N)$.  We fix a positive integer $k$ called
a {\sl level}.  Then we have finitely many irreducible
projective unitary
representations of $L(SU(N))$ called {\sl positive
energy representations} at level $k$.
(See \cite{PS} for details.) 
We have one distinguished representation, called the
{\sl vacuum representation}, among them.
For each interval $I\subset S^1$, we
denote the set of $C^\infty$-maps from $S^1$ to $SU(N)$
such that the image outside the interval $I$ is always the
identity matrix by $L_I(SU(N))$.  Then setting $\A(I)$
to be the von Neumann algebra generated by the image of
$L_I(SU(N))$ by the vacuum representation,
we have a local conformal net $\{\A(I)\}$, which is 
labelled as $SU(N)_k$.
(See \cite{W}, \cite{FG} for details.) 
A similar construction for other Lie groups has been
done in \cite{TL}.  These examples correspond to
the so-called {\sl Wess-Zumino-Witten models}, and
this name is also often attached to these local conformal
nets.  The local conformal nets corresponding to the
Wess-Zumino-Witten
model $SU(N)_k$ are completely rational by \cite{X2}.
For the case of $SU(2)_k$, the irreducible 
representations  are labelled as
$\{\la_0,\la_1,\dots,\la_k\}$ and the fusion rules are
given as 
$$\la_l\cdot\la_m=\la_{|l-m|}\oplus\la_{|l-m|+2}\oplus
\la_{|l-m|+4}\oplus\cdots\oplus\la_{\min(l+m,2k-l-m)}$$
by \cite{W}, where $\la_0$ is the vacuum sector.
The dimensions are given as
$d_{\la_m}=\sin((m+1)\pi/(k+2))/\sin(\pi/(k+2))$.
The statistical phase of $\la_l$ is given as
$\exp(\pi i l(l+2)/(2k+4))$, which follows from
the above fusion rules and the spin-statistics theorem
in \cite[Theorem 3.13]{GL}.  We also have
$$S_{\la_l,\la_m}=\sqrt{\frac{2}{k+2}}
\sin\left(\frac{(l+1)(m+1)\pi}{k+2}\right).$$

Another construction of a local conformal net is from
a lattice $\Lambda$ in the Euclidean space $\R^n$, 
that is, an additive subgroup of $\R^n$ which is
isomorphic to $\Z^n$ and spans $\R^n$ linearly.
A lattice is called {\sl even} when we have
$(x,y)\in {\mathbb{Z}}$ and $(x,x)\in 2{\mathbb{Z}}$ 
for the inner products of $x,y\in \Lambda$.
One obtains a local conformal net from an even lattice
$\Lambda$.  This is like a loop group construction
for $\R^n/\Lambda$. (See \cite{KL4}, \cite{DX} for details.)
The local conformal nets arising from even lattices are
also completely rational by \cite{DX}.
Let $\Lambda^*=\{x\in \R^n\mid
(x,y)\in\Z \textrm{ for all }y\in\Lambda\}$, the
dual lattice of $\Lambda$.  Then
the irreducible representations of the local conformal
net arising from $\Lambda$ are labelled with the 
elements of $\Lambda^*/\Lambda$ and all have 
dimension $1$.  It is holomorphic if and only if
we have $\Lambda^*=\Lambda$.

The above complete rationality
results imply complete rationality of many 
examples, but still complete rationality of
many examples arising from representations of loop
groups as in \cite{TL} is open, so we have the following
problem.

\begin{problem}
Prove complete rationality of local conformal nets
arising from positive energy representations of
loop groups corresponding to Wess-Zumino-Witten models.
\end{problem}

Another construction of a local conformal net is
from a vertex operator algebra.  We see this
construction in the next Chapter.

We next show methods to obtain new local conformal
nets from known ones.

\begin{example}{\rm
For two local conformal nets $\{\A(I)\}$ and $\{\B(I)\}$,
we construct a new one
$\{\A(I)\otimes \B(I)\}$.  This is called the
{\sl tensor product} of local conformal nets.
Both the Hilbert space and the vacuum vector of the
tensor product of local conformal nets are those of the
tensor products.  Each irreducible representation of
the tensor product local conformal net is a tensor 
product of two irreducible representations of the
two local conformal nets, up to unitary equivalence.
That is, each finite dimensional representation
of $\{\A(I)\otimes \B(I)\}$ is of the form $\la\otimes\mu$,
where $\la$ and $\mu$ are finite dimensional representations
of $\{\A(I)\}$ and $\{\B(I)\}$, respectively.  We
also have $\Hom(\la_1\otimes\mu_1,\la_2\otimes\mu_2)=
\Hom(\la_1,\la_2)\otimes\Hom(\mu_1,\mu_2)$.  This
representation category of $\{\A(I)\otimes \B(I)\}$
is written as $\Rep(\A)\boxtimes\Rep(\B)$ and called
the {\sl Deligne product} of $\Rep(\A)$ and $\Rep(\B)$,
which was introduced in \cite{De}. 
(Also see \cite[Definition 1.1.15]{BK}.)
}\end{example}

\begin{example}{\rm
The next construction is called the
{\sl simple current extension}.
This is an extension of a local conformal net
$\{\A(I)\}$ with something similar to a semi-direct
product with a group.  
(See Example \ref{ex-simple} for the initial
appearance of this type of construction.)
Suppose some irreducible
representations of $\{\A(I)\}$ have dimension $1$
and they are closed under the conjugation and the
tensor product.  If they further have all statistical
phases $1$, then they make a group of DHR {\sl auto}morphisms
and a local $Q$-system by \cite[Lemma 4.4]{R2} and
\cite[Corollary 3.7]{BE2}.
An automorphism used in this construction is called
a {\sl simple current} in physics literatures and
this is the source of the name of the construction.
This method also gives a realization of local conformal
nets arising from even lattices as follows.
There is an important conformal field theory called 
{\sl Free bosons} in physics literatures,
and we have the corresponding local conformal net
$\{\A(I)\}$.  Its representation theory with tensor
product structure is given by $\R$ with the
additive structure.
The representation theory of the
$n$th tensor power of$\{\A(I)\}$ is
identified with $\R^n$.  Its simple current extension
with $\Lambda\subset \R^n$ gives the local conformal
net corresponding to the even lattice $\Lambda$.
}\end{example}

\begin{example}
\label{ex-orbif1}{\rm
The next one is called the {\sl orbifold construction}.
An automorphism of a local conformal net $\{\A(I)\}$ on $H$
is a unitary operator $U$ on $H$ satisfying 
$U\A(I)U^*=\A(I)$ for all intervals $I$ and
$U\Omega=\Omega$.  (In this case, $U$ automatically commutes
with the action of $\Diff(S^1)$.  See \cite{CW}.)
We then consider a group $G$ of automorphisms of a
local conformal net $\{\A(I)\}$ and define a subnet by
$\B(I)=\{x\in \A(I)\mid UxU^*=x, U\in G\}$.
Replacing $H$ with the closure of $\B(I)\Omega$, which
is independent of $I$, we obtain
a new local conformal net $\{\B(I)\}$.  This
construction is called the {\sl orbifold construction}.
(See Example \ref{ex-orbif} for the initial
appearance of this type of construction.)
We usually consider a finite group $G$.
See \cite{X4} for details.
}\end{example}

Another construction is called the {\sl coset construction}.
Suppose we have two local conformal nets $\{\A(I)\}$, $\{\B(I)\}$ 
where the latter is a subnet of the former.
Then the family of von Neumann algebras
$\A(I)'\cap \B(I)$ on the Hilbert space
$\overline{(\A(I)'\cap \B(I))\Omega}$ gives
a new local conformal net. (This Hilbert space is again
independent of $I$.)  This construction is called
the coset construction.  (See \cite{X3} for details.)

There was a conjecture or an expectation that all
completely rational local conformal nets would be given
by combinations of the above construction methods,
but this is unlikely to be true as we see in the later
sections.

Theorem \ref{finite-index} implies that if 
$\{\B(I)\}$ is completely rational and $\{\A(I)\}$ is its
orbifold theory by a finite group $G$, then the 
index $[\B(I):\A(I)]$  is the order of $G$ and 
this implies complete rationality of $\{\A(I)\}$.
(This was first proved in \cite[Proposition 4.2]{X4}.  Now this
is a consequence of general results in \cite[Theorem 5.3]{LX}
and \cite[Theorem 33]{KLM}.)
The corresponding statement in theory of vertex
operator algebras is still open.

Also when we perform the coset construction for
$\{\A(I)\subset \B(I)\}$ with completely rationality
of $\{\B(I)\}$ and finiteness of the index
$[\B(I):\A(I)\vee (\A(I)'\cap \B(I))]$, we have complete
rationality of $\{\A(I)'\cap \B(I)\}$.

We also have a construction based on an extension by a $Q$-system.
We show this construction in the next section.

\subsection{Classification of local extensions of $SU(2)_k$}

In the next section, we deal with classification theory
of local conformal nets, and it is important to classify
all irreducible local extensions of certain completely
rational local conformal nets.  As an easier example, we 
classify all local extensions of the local conformal nets
$SU(2)_k$ first here.  We first note that we have only 
finitely many irreducible
extensions for a given completely local
rational net.  This is because any such extension is
given by the corresponding local $Q$-system, and
for any endomorphism
$\bigoplus_{i=1}^k n_i \la_i$ appearing in the $Q$-system,
we have a bound for each $n_i$ by \cite[page 39]{ILP}
and we have only finitely many equivalence classes of
$Q$-system for each such endomorphism by \cite{IK}.
We list this as follows.

\begin{theorem}
For any local conformal net, we
have only finitely many irreducible extensions.
\end{theorem}

We now consider the case of $SU(2)_k$. 
Recall that the irreducible superselection sectors of
the local conformal net $SU(2)_k$ are labelled as
$\{\la_0,\la_1,\dots,\la_k\}$ with the fusion rules
$$\la_l\cdot\la_m=\la_{|l-m|}\oplus\la_{|l-m|+2}\oplus
\la_{|l-m|+4}\oplus\cdots\oplus\la_{\min(l+m,2k-l-m)}.$$
For the $S$- and $T$-matrices given before, the modular
invariant matrices have been completely classified in
\cite{CIZ}.  Theorems \ref{main-be1} and \ref{mod-inv}
give that we have $\theta=\bigoplus_\la Z_{0,\la}\la$
for the dual canonical endomorphism giving an extension.
We then have only the following possibilities for $\theta$.
\begin{align*}
\theta&=\id,\quad
\text{for the type $A_{k+1}$ modular invariant at level $k$},\\
\theta&=\la_0\oplus\la_{4n-4},\quad
\text{for the type $D_{2n}$ modular invariant at level $k=4n-4$},\\
\theta&=\la_0\oplus\la_6,\quad
\text{for the type $E_6$ modular invariant at level $k=12$},\\
\theta&=\la_0\oplus\la_{10}\oplus\la_{18}\oplus\la_{28},\quad
\text{for the type  $E_8$ modular invariant at level $k=28$}.
\end{align*}
As a sample, we list the modular invariant matrix labelled with
$E_6$ for $SU(2)_{10}$ as follows.
$$\left(\begin{array}{ccccccccccc}
1 & 0 & 0 & 0 & 0 & 0 & 1 & 0 & 0 & 0 & 0 \\
0 & 0 & 0 & 0 & 0 & 0 & 0 & 0 & 0 & 0 & 0 \\
0 & 0 & 0 & 0 & 0 & 0 & 0 & 0 & 0 & 0 & 0 \\
0 & 0 & 0 & 1 & 0 & 0 & 0 & 1 & 0 & 0 & 0 \\
0 & 0 & 0 & 0 & 1 & 0 & 0 & 0 & 0 & 0 & 1 \\
0 & 0 & 0 & 0 & 0 & 0 & 0 & 0 & 0 & 0 & 0 \\
1 & 0 & 0 & 0 & 0 & 0 & 1 & 0 & 0 & 0 & 0 \\
0 & 0 & 0 & 1 & 0 & 0 & 0 & 1 & 0 & 0 & 0 \\
0 & 0 & 0 & 0 & 0 & 0 & 0 & 0 & 0 & 0 & 0 \\
0 & 0 & 0 & 0 & 0 & 0 & 0 & 0 & 0 & 0 & 0 \\
0 & 0 & 0 & 0 & 1 & 0 & 0 & 0 & 0 & 0 & 1 
\end{array}\right).$$

The case $\theta=\id$ is trivial.
For the case $\theta=\la_0\oplus\la_{4n-4}$, we have 
a unique realization as a simple current extension by
\cite[Lemma 4.4]{R2} and \cite{IK}.
For the remaining two cases,
we have local $Q$-systems on the unitary modular tensor categories
arising from $SU(2)_{10}$ and $SU(2)_{28}$, and they
correspond to the so-called
``conformal embeddings'' $SU(2)_{10}\subset SO(5)_1$
and $SU(2)_{28}\subset (G_2)_1$.  (The former is an
inclusion of a local conformal net labelled with
$SU(2)_{10}$ into another labelled with $SO(5)_1$.)
The uniqueness of each local $Q$-system follows from
\cite[Theorem 5.3]{KL2}.  This completes the
classification of irreducible local extensions of the
local conformal nets $SU(2)_k$ and they are labelled
as in the above list of $\theta$.

Another way to realize the last two cases of the $Q$-systems
is given purely combinatorially as in \cite{BEK2}.
The locality follows from \cite[Proposition 3.2]{BE4}
and the uniqueness again follows from 
\cite[Theorem 5.3]{KL2}.

\subsection{Classification of local conformal nets with $c<1$}

We next discuss classification theory of local conformal
nets.  The most desirable result would be a complete
listing of all completely rational local conformal nets,
but this would be impossible, since any finite group
produces an orbifold local conformal net from a tensor power
of a holomorphic local conformal net which almost remembers
the original finite group.
On the one hand, the history of classification  theory of
von Neumann algebras with amenability suggests that
we have some kind of algebraic complete invariant
for completely rational local conformal nets.
On the other hand, the above examples arising from
holomorphic local conformal nets suggest that simple
representation theoretic invariants would not work.
Furthermore, there are many different examples
of holomorphic local conformal nets, so the unitary
modular tensor category of the finite dimensional
representations is far from a complete invariant.
(A holomorphic vertex operator algebra is known to
have a central charge equal to a multiple of $8$ as
in \cite{Z}.  The corresponding statement for holomorphic
local conformal nets has not been known.)
Some different holomorphic local conformal nets are
distinguished by the central charge or the
{\sl vacuum character},  ${\mathrm{Tr}}(q^{L_0-c/24})$.
In theory of vertex operator algebras, there are known
examples of different vertex operator algebras with 
the same central charge and the same vacuum character.
We see the difference from a part of the binary operations
$v_n w$ as in \cite{LY}.  We have the corresponding
local conformal nets in \cite{KS}, but it is not
known whether they are really different as local
conformal nets or not, though clearly they should be
different.

It may be that unitary vertex operator algebra with 
$C_2$-cofiniteness could be a complete invariant
of completely rational local conformal nets, as we
see in the next Chapter.
The planar algebra \cite{J3} of Jones gives one
formulation of a complete invariant for type II$_1$
hyperfinite subfactors with finite index and finite
depth, and the invariant consists of countably many
finite dimensional vector spaces with countably
many multi-linear operations with countably many
compatibility conditions, so this has formal
similarity to vertex operator algebras we see
later.  In this section, we explain classification
results we have today.

The {\sl Virasoro algebra} is an infinite dimensional
Lie algebra generated by a central element $c$ and
countably many generators $L_n$, $n\in\Z$ subject to
the following relations.
\begin{equation}\label{vir-alg}
[L_m,L_n]=(m-n) L_{m+n} + \frac{(m^3-m)\delta_{m,-n}}{12}c,
\end{equation}
where $\delta$ is Kronecker's $\delta$.
This is a unique central extension of the complexification
of the Lie algebra corresponding to the infinite dimensional
Lie group $\Diff(S^1)$.  The central element $c$ is called
the {\sl central charge}.

Suppose we have a local conformal net $\{\A(I)\}$.  It
has a projective unitary representation of
$\Diff(S^1)$, which implies that we have a unitary
representation of the Virasoro algebra, where
unitarity means we have $L_n^*=L_{-n}$.
It is known by \cite{FQS}, \cite{GKO} that
if we have an irreducible unitary representation
of the Virasoro algebra, then the central charge $c$ is
mapped to a positive real number and the set of
all possible such values is 
$$\left\{1-\frac{6}{m(m+1)}\mid 
m=3,4,5,\dots\right\}\cup [1,\infty).$$
The value of $c$ in the image is also called the
central charge.  (See \cite[Section 2.2]{CW},
\cite[Theorem A.1]{Ca} for
example.  Also see \cite[Theorem 3.4]{CKLW}.)
We note that since early days of subfactor theory,
this restriction of the values of the central charge
is similar to that of the index value.
The unitary representation of the Virasoro algebra
arising from a local conformal net is not irreducible
in general, still we can show that the image of $c$ is
a scalar as in \cite[Proposition 3.5]{KL1}.
We call the value the central charge of the
local conformal net and simply write $c$ for this value. 
This is a numerical invariant
of a local conformal net.

For a local conformal net $\{\A(I)\}$, consider the
projective unitary representation $U$ of
$\Diff(S^1)$ associated with it.
For an interval $I\subset S^1$, set $\B(I)$ to be
the von Neumann algebra generated by the image
$U(\Diff(I))$.  By restricting the Hilbert space
to $\overline{\B(I)\Omega}$, we see $\{\B(I)\}$ is
a local conformal net and this depends only on the
value of $c$.  We call this the {\sl Virasoro net}
with the central charge $c$ and write
$\{\Vir_c(I)\}$ for this.  (See 
\cite[Proposition 3.5]{KL1}.)

Another way to construct this Virasoro net is as follows.
We simply write $L_n$ for the image of $L_n$ under the
unitary representation of the Virasoro algebra.  We see
that they are closable operators, so we denote their
closures again by $L_n$.  The Fourier expansion
$\sum_{n\in\Z} L_n z^{-n-2}$ makes sense as an operator-valued
distribution on $S^1$ and called the
{\sl stress-energy tensor}.  (It is a matter of convention
that the exponent of $z$ is $-n-2$.)
Constructing ``observables'' with test functions
supported in $I\subset S^1$, we obtain the Virasoro
net $\{\Vir_c(I)\}$.  (See \cite{BSM}.)

For the case $c=1-6/m(m+1)$, the coset construction of \cite{GKO} 
adapted to the operator algebraic setting shows that
the Virasoro net $\{\Vir_c(I)\}$ is given from
the coset construction for the embedding
$SU(2)_{m-1}\subset SU(2)_{m-2}\times SU(2)_1$
as in \cite[Corollary 3.3]{KL1}.
Complete rationality of $SU(2)_k$, which was shown in
\cite{X2}, gives that of the Virasoro nets with $c<1$
as in \cite[Section 4.3]{X3} and \cite[Section 3.5.1]{L5}.

For the central charge $c=1-6/m(m+1)$, $(m=2,3,4,\dots)$,
we have $m(m-1)/2$  irreducible superselection sectors
$\la_{(p,q)}$ of the Virasoro net $\{\Vir_c(I)\}$ with
$(p,q)$, $1\le p\le m-1$, $1\le q\le m$ with the identification
$\la_{(p,q)}=\la_{(m-p,m+1-q)}$.
They have fusion rules as follows.
$$
\la_{(p,q)}\la_{(p',q')}=
\bigoplus_{r=|p-p'|+1, r+p+p':{\rm odd}}
^{\min(p+p'-1, 2m-p-p'-1)\hphantom{x}}
\bigoplus_{s=|q-q'|+1, s+q+q':{\rm odd}}
^{\hphantom{x}\min(q+q'-1, 2(m+1)-q-q'-1)}
\la_{(r,s)}.
$$
The $\mu$-index of the Virasoro net $\{\Vir_c(I)\}$ is
$$\frac{m(m+1)}{8\sin^2{\frac{\pi}{m}} \sin^2{\frac{\pi}{m+1}}}$$
by \cite[Lemma 3.6]{X5}.
The statistical phase of the superselection sector $\la_{(p,q)}$ is 
$$\exp 2\pi i \left(\dfrac{(m+1)p^2-mq^2-1+m(m+1)(p-q)^2}
{4m(m+1)}\right).$$

If $c<1$ for a general local conformal net
$\{\A(I)\}$, it is an irreducible extension of the Virasoro net
$\{\Vir_c(I)\}$.  Recall that such an extension
produces a modular invariant of the unitary modular
tensor category arising from  $\{\Vir_c(I)\}$.
The unitary modular tensor categories arising from the
finite dimensional representations of 
$\{\Vir_c(I)\}$ have been studied well in other
contexts and their modular invariants have been
completely classified in \cite[Tables 2 and 3]{CIZ}.
We can verify the unitary representations of
$SL(2,\Z)$ studied in \cite{CIZ} are the same
as those arising from the unitary modular tensor categories 
of the finite dimensional representations of 
$\{\Vir_c(I)\}$ by \cite[Theorem 3.13]{GL} and the fusion rules in
\cite[Section 3]{KL1}, so we have some modular invariant
listed in \cite[Tables 2 and 3]{CIZ}.  Furthermore, the
modular invariants
in \cite{CIZ} are labelled as type I and II, but it
is easy to see that we have only type I modular
invariants by Theorem \ref{alpha-sigma}.

The modular invariants in \cite[Tables 2 and 3]{CIZ} 
are labelled with pairs of the $A$-$D$-$E$ Dynkin
diagrams whose Coxeter numbers differ by $1$, and
the type I modular invariants are labelled with
those of the $A_n$-$D_{2n}$-$E_{6,8}$ diagrams.
If such a modular invariant $Z$ arises from a
local extension of $\{\Vir_c(I)\}$, then its
dual canonical endomorphism must be of the
form $\bigoplus_{\la}Z_{0,\la}\la$ by
Theorem \ref{main-be1}.  As in the previous section,
the case of simple current extensions are easy to handle,
and we also have local $Q$-systems on the
unitary modular tensor categories
arising from $SU(2)_{10}$ and $SU(2)_{28}$.
By copying these $Q$-systems,
we see that each such $\bigoplus_{\la}Z_{0,\la}\la$
arising from the type I modular invariants of
$\{\Vir_c(I)\}$ does have a corresponding $Q$-system.
For example, when $c=21/22$, the unitary fusion category
arising from the irreducible sectors
$\{\la_{1,1},\la_{1,3},\la_{1,5},\dots,
\la_{1,11}\}$ is isomorphic to that arising from
the irreducible sectors 
$\{\la_0,\la_2,\la_4,\dots,\la_{10}\}$ of $SU(2)_{10}$,
so the $Q$-system arising from $\la_0\oplus\la_6$ on
the latter can be copied to the one arising
from $\la_{1,1}\oplus\la_{1,7}$ on the former.  Then
the locality criterion in \cite[Proposition 3.2]{BE4}
gives the locality of the $Q$-system, hence the
locality of the extension of $\{\Vir_c(I)\}$.
Furthermore, the 2-cohomology vanishing result
\cite[Theorem 5.3]{KL2} implies that such a 
$Q$-system is unique for each $\bigoplus_{\la}Z_{0,\la}\la$.
(See \cite[Theorem 4.1]{KL1} and \cite[Remark 2.5]{KL1} based
on \cite{KL2} for more details.  See \cite{IK} for a
general theory of 2-cohomology of subfactors.)
We thus have the following theorem.

\begin{theorem}
\label{classif}
The complete list of
local conformal nets with $c<1$ is as follows.
\begin{enumerate}
\item The Virasoro nets $\Vir_c$ ($c<1$).
\item The simple current extensions of the Virasoro nets
with $c<1$ by $\Z/2\Z$.
\item Four exceptionals at $c=21/22$, $25/26$,
$144/145$, $154/155$.
\end{enumerate}
\end{theorem}

The first class in the list is not exciting at all, and neither is
the second class.  Two of the four exceptionals in the third
class had been conjectured to arise from the coset construction,
and it has been proved in \cite[Section 6.2]{KL1}
that this is indeed the case.
The case $c=21/22$ has been proved to arise from a more complicated
coset construction in \cite{Kt}.  The final remaining case
$c=144/145$ arises as an ``extension by a $Q$-system'' 
because we directly construct a local $Q$-system with the
dual canonical endomorphism being in the unitary modular tensor
category of the finite dimensional representations of
$\{\Vir_c(I)\}$ and 
this has not been constructed by any other method so far.
This construction of the exceptionals has been generalized
to some infinite series in \cite{X6} under the name of
the {\sl mirror extensions}.  Note that the
above classification has some 
formal similarity to the classification
of subfactors with index less than 4, Theorem \ref{ocn}.

Here we have used the classification result of modular invariants.
Originally modular invariants arise in the setting of full conformal
field theory as the coefficients of modular invariant partition
functions.  Note that the above use of the modular invariants
is quite different from this and this is why we have only type I
modular invariants.

\subsection{Miscellaneous remarks}

We discuss some related problems here.

Conformal field theory on Riemann surfaces as in \cite{S}
has been well-studied and conformal blocks play a prominent
role there.  It is not clear
at all how to formulate this in our approach, so we have the
following open problem.

\begin{problem}
Formulate conformal field theory on Riemann surfaces with
operator algebraic methods.
\end{problem}

See \cite{FFRS1}, \cite{FFRS2}, 
\cite{FuRS1}, \cite{FuRS2} for some related results in
the Euclidean field theory approach.

We have operator algebraic versions of $N=1$ and $N=2$
superconformal field theories as in 
\cite{CKL1} and \cite{CHKLX}, and there we 
have superconformal nets rather than local conformal
nets.  We also have connections to noncommutative
geometry in \cite{CHKL}, \cite{CHKLX}.
(See \cite{FGR1}, \cite{FGR2} for an early work on
connections between superconformal field theory
and noncommutative geometry.)

Also see \cite{BDH1}, \cite{BDH2}, \cite{BDH3} for
another approach to local conformal nets.

\section{Vertex operator algebras}

We have a different axiomatization of chiral conformal 
field theory from a local conformal net and it is a theory
of {\sl vertex
operator algebras}.  It is a direct axiomatization of Wightman
fields on the circle $S^1$.  In physics literatures,
certain operator-valued distributions are called
vertex operators and this is the origin of
the name ``vertex operator algebra''.  We explain this theory
in comparison to that of local conformal nets.  We emphasize 
relations to local conformal nets rather than a general
theory of vertex operator algebras.

A certain amount of the theory has been 
devoted to a single example called the {\sl Moonshine
vertex operator algebra}, so we explain the background
of the Moonshine conjecture for which it was constructed.
(See \cite{G2} for  more details.) 
We start with the following very general problem.

\begin{problem}
Suppose a finite group $G$ is given.  
Realize it as the automorphism
group of some ``interesting'' algebraic structure.
\end{problem}

This is not a precisely formulated mathematical problem since
the word ``interesting'' is ambiguous, of course. 
One precise formulation is as follows.

\begin{problem}
Suppose a finite group $G$ is given. Realize it as the Galois
group of an extension over $\Q$.
\end{problem}

This is called the {\sl inverse Galois problem} and still open today.
Note that $\Q$ is the smallest field of characteristic $0$.
The hyperfinite II$_1$ factor is the ``smallest'' infinite dimensional
factor in some appropriate sense, and we can define a Galois 
group for its extension naturally.  Then we consider
a direct analogue of the above problem for this von Neumann
algebraic version, but it turns out that a solution for
all finite groups is easily given, and this  is
not a particularly exciting problem.  We look for more interesting
analogues of the above Galois theory problem.

Our ``interesting'' algebraic structure now is that of
a vertex operator algebra, which is infinite dimensional.
(We have a certain Galois type theory for vertex operator
algebras \cite{DM}.)
It looks that a natural vector space for a finite group 
is finite dimensional, but somehow in the Moonshine conjecture
and related topics, infinite dimensional vector spaces
naturally appear in connection to finite groups.

Among finite groups, finite simple groups are clearly
fundamental.  Today we have a complete list of 
finite simple groups as follows.
(See \cite{FLM}, \cite{G2} and texts cited there for
more details.)
\begin{enumerate}
\item Cyclic groups of prime order.
\item Alternating groups of degree $5$ or higher.
\item $16$ series of groups of Lie type over finite fields.
\item $26$ sporadic finite simple groups.
\end{enumerate}

The third class consists of matrix groups such as
$PSL(n, {\mathbb F}_q)$.  The last class consists of
exceptional structures, and the first ones were found
by Mathieu in the 19th century.
The largest group among the 26 groups in terms of the
order is called the {\sl Monster group}, and its order is
$$2^{46}\cdot3^{20}\cdot5^9\cdot7^6\cdot11^2\cdot13^3\cdot
17\cdot19\cdot23\cdot29\cdot31\cdot41\cdot47\cdot59\cdot71,$$
which is approximately $8\times10^{53}$.
This group was first constructed in \cite{Gr} 
as the automorphism group of some commutative, but
nonassociative algebra of 196884 dimensions.
From the beginning, it has been known that the smallest
dimension of a non-trivial irreducible representation
of the Monster group is 196883.

Now we turn to a different topic of the classical
{\sl $j$-function}.  This is a function of a complex
number $\tau$ with ${\mathrm{Im}}\;\tau>0$ given
as follows.
\begin{align*}
j(\tau)&=\frac{(1+240\sum_{n>0}\sigma_3(n)q^n)^3}
{q\prod_{n>0}(1-q^n)^{24}}\\
&=q^{-1}+744+196884q+
21493760q^2+864299970 q^3+\cdots,
\end{align*}
where $\sigma_3(n)$ is a sum of the cubes of the divisors
of $n$ and we set $q=\exp(2\pi i \tau)$.

This function has modular invariance property
$$j(\tau)=j\left(\displaystyle\frac{a\tau+b}{c\tau+d}\right),$$
for
$$\left(\begin{array}{cc}
a & b \\ c& d\end{array}\right)\in SL(2,{\mathbb{Z}}),$$
and this property and the condition that the Laurent
series of $q$ start with $q^{-1}$ determine the $j$-function
uniquely except for the constant term.
The constant term $744$ has been chosen for a historic
reason and this has no significance, so
we set $J(\tau)=j(\tau)-744$ and use this from now on.

McKay noticed that the first non-trivial coefficient of
the Laurent expansion of the $J$-function satisfies the equality
$196884=196883+1$, where $1$ the dimension of the trivial
representation of the Monster group and
$196883$ is the smallest dimension of  its non-trivial
representation.  People suspected it is simply a coincidence,
but it has turned out that all the coefficients of the
Laurent expansion of the $J$-function with small exponents
are linear combinations of the dimensions of irreducible
representations of the Monster group with ``small'' positive
integer coefficients.  (We have $1$ as the dimension
of the trivial representation, so it is trivial that any
positive integer is a sum of the dimensions of irreducible
representations of the Monster group, but it is highly
non-trivial that we have small multiplicities.)

Based on this, Conway-Norton \cite{CN} formulated what is
called the {\sl Moonshine conjecture} today, which has been
proved by Borcherds \cite{B}.

\begin{conjecture}
\begin{enumerate}
\item
We have some graded infinite dimensional $\C$-vector space
$V=\bigoplus_{n=0}^\infty V_n$ ($\dim V_n<\infty$)
with some natural algebraic structure and its automorphism
group is the Monster group.
\item Each element $g$ of the Monster group acts
on each $V_n$ linearly.  The Laurent series
$$\sum_{n=0}^\infty ({\mathrm{Tr}}\;g|_{V_n})q^{n-1}$$
arising from the trace value of the $g$-action on $V_n$
is a classical function called a Hauptmodul corresponding
to a genus 0 subgroup of $SL(2,{\mathbb{R}})$.
(The case $g$ is the identity element is the $J$-function.)
\end{enumerate}
\end{conjecture}

The above Laurent series is called the McKay-Thompson series.
The first statement is vague since it does not specify
the ``natural algebraic structure'', but 
Frenkel-Lepowsky-Meurman \cite{FLM} introduced the axioms
for {\sl vertex operator algebras} and constructed an example
$V^\natural$, called the Moonshine vertex operator algebra,
corresponding to the first statement of the above Conjecture.
This was the starting point of the entire theory.

\subsection{Basic definitions}

There are various, slightly different
versions of the definition of vertex
operator algebras, so we fix our definition here.
We follow \cite{CKLW}.  (See also \cite{Kc}, \cite{FLM}.)

Let $V$ be a $\C$-vector space. We say that a formal series $a(z) =
\sum_{n\in\Z} a_{(n)} z^{-n-1}$ with coefficients
$a_{(n)} \in \End(V)$ is a {\sl field} on $V$, 
if for any $b \in V$,  we have $a_{(n)}b = 0$ for all
sufficiently large $n$. 

\begin{definition}{\rm
A $\C$-vector space $V$ is a {\sl vertex algebra}
if we have the following properties.
\begin{enumerate}
\item (State-field correspondence) For each $a\in V$, we have a
field $Y(a,z)=\sum_{n\in\Z} a_{(n)} z^{-n-1}$ on $V$.
\item (Translation covariance)
We have a linear map $T\in \End(V)$ such that we have
$[T,Y(a,z)]=\frac{d}{dz}Y(a,z)$ for all $a\in V$.
\item (Existence of the vacuum vector)
We have a vector $\Omega\in V$ with
$T\Omega=0$, $Y(\Omega,z)=\id_V$, $a_{(-1)}\Omega=a$.
\item (Locality) For all $a,b\in V$, we have
$(z-w)^N[Y(a,z),Y(b,w)]=0$ for a sufficiently large
integer $N$.
\end{enumerate}
We then call $Y(a,z)$ a {\sl vertex operator}.
}\end{definition}

A vertex operator is an algebraic version of the Fourier
expansion of an operator-valued distribution on the circle.
The state-field correspondence means that any vector in $V$
gives an operator-valued distribution.  The locality axiom
is one representation of the idea that
$Y(a,z)$ and $Y(b,w)$ commute for $z\neq w$.
(Recall that a distribution $T$ on $\R$ has
$\supp\; T\subset \{0\}$ if and only if there
exists a positive integer $N$ with $x^N T=0$.)

The following {\sl Borcherds identity} is a consequence of the
above axioms, where $a,b,c\in V$ and $m,n,k\in\Z$.
\begin{align*}
&\sum_{j=0}^\infty \left(\begin{array}{c} m \\ j \end{array}\right)
(a_{(n+j)}b)_{(m+k-j)} c\\
&=\sum_{j=0}^\infty (-1)^j
\left(\begin{array}{c} n \\ j \end{array}\right)
a_{(m+n-j)}b_{(k+j)} c-
\sum_{j=0}^\infty (-1)^{j+n}
\left(\begin{array}{c} n \\ j \end{array}\right)
b_{(n+k-j)}a_{(m+j)} c.
\end{align*}

\begin{definition}{\rm
We say a linear subspace $W\subset V$ is a {\sl vertex subalgebra} if
we have $\Omega\in W$ and $a_{(n)}b\in W$ for all $a,b\in W$ and $n\in\Z$.
(In this case, $W$ is automatically $T$-invariant.)
We say a linear subspace $J\subset V$ is an {\sl ideal} if it is $T$-invariant
and we have $a_{(n)}b\in J$ for all $a\in V$, $b\in J$ and $n\in\Z$.
A vertex algebra is said to be {\sl simple} if any ideal in $V$ is either
$0$ or $V$. A {\sl (antilinear) homomorphism}
from a vertex algebra $V$ to a vertex
algebra $W$ is an (anti)linear map $\phi$ satisfying 
$\phi(a_{(n)}b)=\phi(a)_{(n)}\phi(b)$ for all $a,b\in V$ and $n\in\Z$.
We similarly define an {\sl automorphism}.
}\end{definition}

If $J$ is an ideal of $V$, we also have 
$a_{(n)}b\in J$ for all $a\in J$, $b\in V$ and $n\in\Z$.

We next introduce conformal symmetry in this context.

\begin{definition}{\rm
Let $V$ be a $\C$-vector space and $L(z)=\sum_{n\in\Z}L_n z^{-n-2}$ be
a field on $V$.  If the endomorphisms $L_n$ satisfy the Virasoro algebra
relations
$$[L_m,L_n]=(m-n)L_{m+n}+\frac{(m^3-m)\delta_{m+n,0}}{12}c,$$
with central charge $c\in\C$, then we say $L(z)$ is a {\sl Virasoro
field}.  If $V$ is a vertex algebra and
$Y(\omega,z)=\sum_{n\in\Z} L_n z^{-n-2}$ is a Virasoro
field, then we say $\omega\in V$ is a {\sl Virasoro vector}.
A Virasoro vector $\omega$ is called a {\sl conformal vector} if
$L_{-1}=T$ and $L_0$ is diagonalizable on $V$.  (The latter means that
$V$ is an algebraic direct sum of the eigenspaces of $L_0$.)
Then the corresponding vertex operator $Y(\omega,z)$ is called the
{\sl energy-momentum field} and $L_0$ the {\sl conformal Hamiltonian}.
A vertex algebra with a conformal vector is called a {\sl
conformal vertex algebra}.  We then say $V$ has central charge $c\in\C$.
}\end{definition}

\begin{definition}{\rm
A nonzero element $a$ of a conformal vertex algebra in $\Ker(L_0-\a)$ is
said to be a {\sl homogeneous} element of {\sl conformal weight}
$d_a=\a$.  We then set $a_n=a_{(n+d_a-1)}$ for $n\in\Z-d_a$.
For a sum $a$ of homogeneous elements, we extend $a_n$ by linearity.
}\end{definition}

\begin{definition}{\rm
A homogeneous element $a$ in a conformal vertex algebra $V$ and the
corresponding field $Y(a,z)$ are called {\sl quasi-primary} if $L_1 a=0$
and {\sl primary} if $L_na=0$ for all $n>0$.
}\end{definition}

\begin{definition}{\rm
We say that a conformal vertex algebra $V$ is of {\sl CFT type}
if we have $\Ker(L_0-\a)\neq0$ only for $\a\in\{0,1,2,3,\dots\}$
and $V_0=\C\Omega$.
}\end{definition}

\begin{definition}{\rm
We say that a conformal vertex algebra $V$ is a
{\sl vertex operator algebra} if we have the following.
\begin{enumerate}
\item We have $V=\bigoplus_{n\in\Z} V_n$, where $V_n=\Ker(L_0-n)$.
\item We have $V_n=0$ for all sufficiently small $n$.
\item We have $\dim(V_n)<\infty$ for $n\in\Z$.
\end{enumerate}
}\end{definition}

\begin{definition}{\rm
An {\sl invariant bilinear form} on a vertex operator algebra $V$ is a
bilinear form $(\cdot,\cdot)$ on $V$ satisfying
$$(Y(a,z)b,c)=(b,Y(e^{zL_1}(-z^{-2})^{L_0}a,z^{-1})c)$$
for all $a,b,c\in V$.
}\end{definition}

\begin{definition}{\rm
For a vertex operator algebra $V$ with a conformal vector $\omega$,
an automorphism $g$ as a vertex algebra is called a {\sl VOA 
automorphism} if we have $g(\omega)=\omega$.
}\end{definition}

\begin{definition}{\rm
Let $V$ be a vertex operator algebra and suppose we have a
positive definite inner product $(\cdot\mid\cdot)$, where this
is supposed to be antilinear in the first variable.  We
say the inner product is {\sl normalized} if we have $(\Omega\mid\Omega)=1$.
We say that the inner product is {\sl invariant} if there
exists a VOA antilinear automorphism $\theta$ of $V$ such that
$(\theta\cdot\mid\cdot)$ is an invariant bilinear form on $V$.
We say that $\theta$ is a {\sl PCT operator} associated with
the inner product.
}\end{definition}

If we have an invariant inner product, we automatically have
$(L_na\mid b)=(a\mid L_{-n}b)$ for $a,b\in V$ and also
$V_n=0$ for $n<0$.  The PCT operator $\theta$ is unique and
we have $\theta^2=1$ and $(\theta a\mid \theta b)=(b\mid a)$
for all $a,b\in V$. (See \cite[Section 5.1]{CKLW} for details.)

\begin{definition}{\rm
A {\sl unitary} vertex operator algebra is a pair of
a vertex operator algebra and a normalized invariant
inner product.
}\end{definition}

See \cite{DL} for details of unitarity.
A unitary vertex operator algebra is simple
if and only if we have $V_0=\C\Omega$.
(See \cite[Proposition 5.3]{CKLW}.)

For a unitary vertex operator algebra $V$,
we write $\Aut_{(\cdot\mid\cdot)}(V)$ for the automorphism
group fixing the inner product.

\begin{definition}{\rm
A {\sl unitary subalgebra} $W$ of a unitary vertex operator algebra 
$(V,(\cdot\mid\cdot))$ is a vertex subalgebra with $\theta W\subset W$
and $L_1 W\subset W$.
}\end{definition}

\subsection{Modules and modular tensor categories}

We introduce a notion of a module of a vertex operator algebra, which
corresponds to a representation of a local conformal net, as follows.
(Also see \cite[Definition 1.2.3]{Z}.)

\begin{definition}{\rm
Let $M$ be a $\C$-vector space and suppose we have a field 
$Y^M(a,z)=\sum_{n\in\Z} a^M_{(n)} z^{-n-1}$, $a^M_{(n)}\in\End(M)$,
on $M$ for any $a\in V$, where the map $a\mapsto Y^M(a,z)$ is linear and
$V$ is a vertex algebra.
We say $M$ is a module over $V$ if we have $Y^M(\Omega,z)=\id_M$ and
the following Borcherds identity for $a,b\in V$, $c\in M$, $m,n,k\in\Z$.
\begin{align*}
&\sum_{j=0}^\infty \left(\begin{array}{c} m \\ j \end{array}\right)
(a_{(n+j)}b)^M_{(m+k-j)} c\\
&=\sum_{j=0}^\infty (-1)^j
\left(\begin{array}{c} n \\ j \end{array}\right)
a_{(m+n-j)}^M b_{(k+j)}^M c-
\sum_{j=0}^\infty (-1)^{j+n}
\left(\begin{array}{c} n \\ j \end{array}\right)
b_{(n+k-j)}^M a_{(m+j)}^M c.
\end{align*}
}\end{definition}

In this section, we consider only simple
vertex operator algebras of CFT type.

The fusion rules on modules have been introduced in \cite{FHL} and they
give the tensor product operations on modules.

We have a natural notion of irreducible module and it is of the form
$M=\bigoplus_{n=0}^\infty M_n$, where every $M_n$ is finite
dimensional and $L_0$ acts on $M_n$ as a scalar $n+h$ for
some constant $h$.   The vertex operator algebra $V$ itself is
a module of $V$ with $h=0$,
and if this is the only irreducible module, then
we say $V$ is {\sl holomorphic}.

We define the formal power series, the
{\sl character} of $M$, by
$\ch(M)=\sum_{n=0}^\infty \dim(M_n) q^{n+h-c/24}$, where $c$ is 
the central charge.  We introduce the following important notion.

\begin{definition}{\rm 
If the quotient space $V/\{v_{(-2)} w\mid v, w\in V\}$ 
is finite dimensional, we say that the vertex
operator algebra $V$ is {\sl $C_2$-cofinite}.  
}\end{definition}

It has been proved in  \cite{Z}
that under the $C_2$-cofiniteness condition and small
other conditions which automatically hold in the unitary case,
we have only finitely
many irreducible modules $M_1,M_2,\dots,M_k$ up to isomorphism,
their characters are absolutely convergent for $|q|<1$, and 
the linear span of $\ch(M_1), \ch(M_2),\dots,\ch(M_k)$ is
closed under the action of $SL(2,\Z)$ on the upper half
plane through the fractional linear transformation on
$\tau$ with $q=\exp(2\pi i\tau)$.

Under the $C_2$-cofiniteness assumption and small
other assumptions which again
automatically hold in the unitary case,
Huang \cite{H1,H2} further showed that the $S$-matrix
defined by the transformation $\tau\mapsto -1/\tau$ on the
characters satisfy the Verlinde formula (\ref{verl}) with respect
to the fusion rules and the tensor 
category of the modules is modular.

Note the characters and their modular transformations make sense
in the setting of operator algebras since $L_0$ acts on
each representation space, so we have the following
conjecture.

\begin{conjecture}
For a completely rational local conformal net, we have
convergent characters for all irreducible representations
and they are closed under modular transformations of $SL(2,\Z)$.
Furthermore, the $S$-matrix defined with braiding gives
transformation rules of the characters under the transformation
$\tau\mapsto -1/\tau$.
\end{conjecture}

This conjecture was made in \cite[page 625]{FG}, and a completely
rational local conformal net satisfying this is called {\sl modular}
in \cite{KL3}.

For an inclusion $\{\A(I)\subset\B(I)\}$ of  local conformal nets,
the $\alpha$-induction produces $\B(I)$-$\B(I)$ morphisms which
do not arise from a representation of $\{\B(I)\}$.  (That is, the
non-ambichiral $\B(I)$-$\B(I)$ morphisms.)  To some extent, they
are similar to {\sl twisted modules} of a vertex operator algebra,
but they look more general, so we have the following problem.
(See \cite[Section 5]{O}, \cite{Pa} for an abstract treatment
in the setting of tensor categories.)

\begin{problem}
Develop the theory of $\a$-induction for vertex operator algebras.
\end{problem}

\subsection{Examples and construction methods}

The construction methods of local conformal nets
we have explained had been known in the context of
vertex operator algebras earlier except for
an extension by a $Q$-system.  We explain other
constructions for vertex operator algebras.
The last construction based on a $Q$-system
has now been also introduced in the context
of vertex operator algebras in \cite{HKL}.

\begin{example}
We have a simple unitary vertex operator algebra $L(c,0)$ with
central charge $c$ arising from  a unitary representation
of the Virasoro algebra with central charge $c$ and
the lowest conformal energy $0$. See \cite[Section 4.1]{DL}
for details.
\end{example}

\begin{example}
Let $\g$ be a simple complex Lie algebra and $V_{\g_k}$ be the conformal
vertex algebra generated by the unitary representation of the affine
Lie algebra associated with $\g$ having level $k$ and lowest conformal
energy $0$.  Then $V_{\g_k}$ is a simple unitary vertex operator
algebra.  (See \cite{FZ}.  Also see \cite[Section 4.2]{DL} for unitarity.)
\end{example}

\begin{example}
Let $\Lambda\subset \R^n$ be an even lattice.  That is, it is
isomorphic to $\Z^n$ as an abelian group and linearly spans the
entire $\R^n$, and we have
$(x,y)\in {\mathbb{Z}}$ and $(x,x)\in 2{\mathbb{Z}}$ 
for all $x,y\in\Lambda$,
where $(\cdot,\cdot)$ is the standard Euclidean inner product.
There is a general construction
of a unitary vertex operator algebra from such a lattice and
we obtain $V_\Lambda$ from $\Lambda$.  The central
charge is the rank of $\Lambda$.  (See \cite{FLM}.
Also see \cite[Section 4.4]{DL} for unitarity.)
\end{example}

\begin{example}
If we have two unitary vertex operator algebras $(V,(\cdot\mid\cdot))$
and $(W,(\cdot\mid\cdot))$, then the tensor product $V\otimes W$ has
a natural inner product with which we have a unitary vertex 
operator algebra.
\end{example}

\begin{example}
\label{ex-simple}
For a unitary vertex operator algebra with certain modules called
{\sl simple currents} satisfying some nice compatibility condition,
we can extend the unitary vertex operator algebra.  This is 
called a {\sl simple current extension}.  This was first studied
in \cite{SY}.  Also see \cite[Section 4]{Ho},
\cite{FuRS3}, \cite{HKL}.
\end{example}

\begin{example}
\label{ex-orbif}
For a unitary vertex operator algebra $(V,(\cdot\mid\cdot))$ and
$G\subset \Aut_{(\cdot\mid\cdot)}(V)$, the fixed point
subalgebra $V^G$ is a unitary vertex operator algebra.
This is called an {\sl orbifold subalgebra}.  This was first
studied in \cite{DVVV}.
\end{example}

\begin{example}
Let $(V,(\cdot\mid\cdot))$ be a unitary vertex operator algebra and 
$W$ its subalgebra.  Then 
$$W^c=\{b\in V\mid [Y(a,z),Y(b,w)]=0
\textrm{ for all }a\in W\}$$
is a vertex subalgebra of $V$.  This is called a
{\sl coset subalgebra}.  This is also called the
{\sl commutant} of $W$ in $V$ and was introduced in \cite{FZ}.
If $W$ is unitary, then $W^c$ is also unitary.  
\end{example}

\subsection{Moonshine vertex operator algebras}

A rough outline of the construction of the Moonshine vertex
operator algebra is as follows.
We have an exceptional even lattice in dimension $24$ called
the {\sl Leech lattice}.  It is the unique $24$-dimensional
even lattice
$\Lambda$ with $\Lambda=\Lambda^*$ and having no vectors
$x\in \Lambda$ with $(x,x)=2$.
(See \cite{CS}, \cite[page 304]{FLM} for details.)
We have a corresponding unitary
vertex operator algebra $V_\Lambda$.
The involution  $x\to -x$ on $\Lambda$ induces an 
automorphism of $V_\Lambda$  of order 2.  Its fixed
point vertex operator subalgebra has a non-trivial
simple current extension of order 2.  Taking this
extension is called the {\sl twisted orbifold construction}
and we obtain $V^\natural$ with this.  This is the
Moonshine vertex operator algebra.

Miyamoto \cite{M} has given a new construction of
$V^\natural$ as follows. 
The most fundamental vertex operator algebra among
the Virasoro vertex operator algebras
has central charge $1/2$ and
is denoted by $L(1/2,0)$.  It has been known that 
the Moonshine vertex operator algebra contains the
48th tensor power of $L(1/2,0)$ in \cite{DMZ}.
Conversely, we start with the 48th tensor power of
$L(1/2,0)$ and construct the Moonshine vertex operator
algebra as its extension.
Based on analogy to lattice theory, a finite tensor
power of $L(1/2,0)$ contained in another vertex
operator algebra is called a {\sl Virasoro frame}.
An extension of a Virasoro frame is called
a {\sl framed vertex operator algebra} \cite{DGH}.
The moonshine vertex operator algebra is constructed
as a framed vertex operator algebra.  It has been
conjectured that this is the unique vertex operator
algebra $V$ having only trivial irreducible modules
with $c=24$ and $V_1=0$.

The local conformal net corresponding
to the Moonshine vertex operator algebra has been
constructed in \cite{KL4} and its automorphism group
in the operator algebraic sense is the Monster group.
This local conformal net is an extension of the
48th tensor power of the Virasoro net with $c=1/2$ and
this extension is a simple current extension of a
simple current extension of the tensor power of
the Virasoro net with $c=1/2$.

The following is \cite[Conjecture 3.4]{X7}.

\begin{conjecture}
A holomorphic local conformal net 
with $c=24$ and the eigenspace of
$L_0$ with eigenvalue $1$ being $0$ is unique up to 
isomorphism.
\end{conjecture}

The following is a problem corresponding to
\cite[Conjecture 3.5]{X7} which arose from \cite{Sch},
where Schellekens gave 71 possible Lie algebra
structures as an invariant for classifying holomorphic
vertex operator algebras with $c=24$.

\begin{problem}
Classify all holomorphic local conformal nets with $c=24$.
\end{problem}

\subsection{Local conformal nets and vertex operator algebras}

We now consider relations between vertex operator algebras
and local conformal nets.  Both are supposed to be
mathematical axiomatizations of the same physical theory,
so we might expect the two sets of axioms are equivalent in
the sense that we have a canonical bijective correspondence
between the mathematical objects satisfying one set of
axioms and those for the other.  However,
both axiomatizations are broad and may contain some
weird examples, so it is expected that we have to impose
some more conditions in order to obtain such a nice bijective
correspondence.

In principle, when we have some idea, example or construction
on local conformal nets or vertex operator algebras,
one can often ``translate'' it to the other side.
For example, the local conformal net corresponding
to the Moonshine vertex operator algebra has been
constructed in \cite{KL4} and its automorphism group
in the operator algebraic sense is the Monster group.
Also, a construction of holomorphic framed vertex
operator algebras in \cite{LY} has been translated to
the setting of local conformal nets in \cite{KS}.
For the converse direction, 
with the results in \cite{HKL},
Theorem \ref{classif} on local conformal nets
implies the corresponding
classification of vertex operator algebras with
$c<1$ as explained in \cite[page 351]{K1}.
Such a translation has been done on a case-by-case
basis.  It is sometimes easy, sometimes difficult,
and sometimes still unknown.

Here we deal with a construction of a local conformal
net from a unitary vertex operator algebra with some
extra nice properties.

\begin{definition}{\rm
Let $(V, (\cdot\mid\cdot))$ be a unitary vertex operator
algebra.  We say that $a\in V$ (or $Y(a,z)$) satisfies 
{\sl energy-bounds} if we have positive integers $s,k$
and a constant $M>0$ such that we have
$$\|a_nb\|\le M(|n|+1)^s\|(L_0+1)^kb\|,$$
for all $b\in V$ and $n\in\Z$.
If every $a\in V$ satisfies energy-bounds, we say
$V$ is {\sl energy-bounded}.
}\end{definition}

We have the following, which is \cite[Proposition 6.1]{CKLW}.

\begin{proposition}
If $V$ is generated by a family of homogeneous elements
satisfying energy-bounds, then $V$ is energy-bounded.
\end{proposition}

Roughly speaking, we need norm estimates for $(a_n b)_m c$ from
those for $a_n(b_m c)$ and $b_m(a_n c)$.  This is essentially
done with the Borcherds identity.

We also have the following, which is \cite[Proposition 6.3]{CKLW}.

\begin{proposition}
If $V$ is a simple unitary 
vertex operator algebra generated by $V_1$ and $F\subset V_2$
where $F$ is a family of quasi-primary $\theta$-invariant 
Virasoro vectors, then $V$ is energy-bounded.
\end{proposition}

We have certain commutation relations for elements in $V_1$ and
$F$, and this implies energy-bounds for them.  Then the above
Proposition follows from the previous one.

For a unitary vertex operator algebra $(V,(\cdot\mid\cdot))$,
define a Hilbert space $H$ by the completion of $V$ with respect
to the inner product $(\cdot\mid\cdot)$.  For any $a\in V$ and
$n\in\Z$, we regard $a_{(n)}$ as a densely defined operator on
$H$.  By the invariance of the scalar product, the operator
$a_{(n)}$ has a densely defined adjoint, so it is closable.
Suppose $V$ is energy-bounded and let $f(z)$ be a smooth 
function on $S^1=\{z\in\C\mid |z|=1\}$ with Fourier coefficients
$$\hat f_n=\int_{-\pi}^\pi f(e^{i\theta})e^{-in\theta}
\frac{d\theta}{2\pi}$$
for $n\in\Z$.  For every $a\in V$, we define the operator
$Y_0(a,f)$ with domain $V$ by
$$Y_0(a,f)b=\sum_{n\in\Z} \hat f_n a_n b$$
for $b\in V$.  The convergence follows from the energy-bounds
and $Y_0(a,f)$ is a densely defined operator.  This is again
closable.  We denote by $Y(a,f)$ the closure of $Y_0(a, f)$
and call it a {\sl smeared vertex operator}.

We define  $\A_{(V,(\cdot\mid\cdot))}(I)$ 
to be the von Neumann algebra generated by the
(possibly unbounded) operators $Y(a,f)$ with
$a\in V$, $f\in C^\infty(S^1)$ and $\supp\;f\subset I$.
(For a family of closed operators $\{T_i\}$, we apply the
polar decomposition to each $T_i$ and consider the
von Neumann algebra generated by the partial isometry
part of $T_i$ and the spectral projections of the
self-adjoint part of $T_i$.)
The family $\{\A_{(V,(\cdot\mid\cdot))}(I)\}$ clearly
satisfies isotony.  We can verify that 
$(\bigvee_I \A_{(V,(\cdot\mid\cdot))}(I))\Omega$
is dense in $H$.  A proof 
of conformal covariance is nontrivial, but can be done
as in \cite{GW} and \cite{TL} by studying the representations
of the Virasoro algebra and $\Diff(S^1)$.  We also have the vacuum
vector $\Omega$ and the positive energy condition.
However, locality is not clear at all from our construction, 
so we make the following definition.

\begin{definition}{\rm
We say that a unitary vertex operator algebra
$(V,(\cdot\mid\cdot))$ is {\sl strongly local} if
it is energy-bounded and we have
$\A_{(V,(\cdot\mid\cdot))}(I)\subset 
\A_{(V,(\cdot\mid\cdot))}(I')'$ for all intervals 
$I\subset S^1$.
}\end{definition}

Difficulty in having strong locality is seen as follows.
It is well-known that if $A$ and $B$ are unbounded self-adjoint
operators, having $AB=BA$ on a common core does not imply
commutativity of the spectral projections of $A$ and $B$.  
That is, having commutativity of spectral projections from
certain algebraic commutativity relations is a nontrivial
task.

A strongly local unitary vertex operator algebra
produces a local conformal net through the above
procedure by definition.
The following is \cite[Theorem 6.9]{CKLW}.

\begin{theorem}
Let $V$ be a strongly local unitary vertex operator algebra
and $\{\A_{(V,(\cdot\mid\cdot))}(I)\}$ the corresponding
local conformal net.  Then we have
$\Aut(\A_{(V,(\cdot\mid\cdot))})=\Aut_{(\cdot\mid\cdot)}(V)$.
If $\Aut(V)$ is finite, then we have 
$\Aut(\A_{(V,(\cdot\mid\cdot))})=\Aut(V)$.
\end{theorem}

We now have the following theorem for a criterion of strong
locality \cite[Theorem 8.1]{CKLW}.

\begin{theorem}
Let $V$ be a simple unitary energy-bounded vertex operator algebra
and $F\subset V$.  Suppose $F$ contains only quasi-primary
elements, $F$ generates $V$ and $\A_{(F,(\cdot\mid\cdot))}(I)\subset 
\A_{(F,(\cdot\mid\cdot))}(I')'$ for some interval $I$, where
$\A_{(F,(\cdot\mid\cdot))}(I)$ is defined similarly to
$\A_{(V,(\cdot\mid\cdot))}(I)$.  We then have 
$\A_{(F,(\cdot\mid\cdot))}(I)=\A_{(V,(\cdot\mid\cdot))}(I)$
for all intervals $I$,
which implies strongly locality of $\{\A_{(V,(\cdot\mid\cdot))}(I)\}$.
\end{theorem}

From this, we can prove
the following result, \cite[Theorem 8.3]{CKLW}.

\begin{theorem}
Let $V$ be a simple unitary vertex operator algebra
generated by $V_1 \cup F$ where $F\subset V_2$ is a family of
quasi-primary $\theta$-invariant Virasoro vectors, then
$V$ is strongly local.
\end{theorem}

We also have the following result, \cite[Theorem 7.1]{CKLW}

\begin{theorem}
Let $V$ be a simple unitary strongly local vertex operator algebra
and $W$ its subalgebra.  Then $W$ is also strongly local.
\end{theorem}

The following is \cite[Corollary 8.2]{CKLW}.

\begin{theorem}
Let $V_1, V_2$ be simple unitary strongly local vertex operator algebras.
Then $V_1\otimes V_2$ is also strongly local.
\end{theorem}

We list some examples of strongly local vertex operator
algebras following \cite{CKLW}.

\begin{example}
The unitary vertex algebra $L(c, 0)$ is strongly local.
\end{example}

\begin{example}
Let $\g$ be a complex simple Lie algebra and let $V_{\g_k}$
be the corresponding level
$k$ unitary vertex operator algebra.
Then $V_{\g_k}$ is generated by $(V_{\g_k} )_1$
and hence it is strongly local.
\end{example}

The following is \cite[Theorem 8.15]{CKLW}.
This construction was first made in \cite{KL4}.

\begin{example}
The moonshine vertex operator algebra $V^\natural$
is a simple unitary strongly
local vertex operator algebra.  Hence the automorphism
group of the corresponding local conformal net is
the Monster group.
\end{example}

The following is \cite[Conjecture 8.17]{CKLW}.

\begin{conjecture}
Let $\Lambda$ be an even lattice and $V_\Lambda$ be the
corresponding unitary vertex operator algebra.  Then
$V_\Lambda$ is strongly local.
\end{conjecture}

The following is (a part of ) \cite[Theorem 9.2]{CKLW} which is
given by extending the methods in \cite{FJ}.

\begin{theorem}Let $V$ be a simple unitary strongly local 
vertex operator algebra and $\{\A_{(V,(\cdot\mid\cdot))}(I)\}$
be the corresponding local conformal net.
Then one can recover the vertex operator algebra
structure on $V$, which is an algebraic direct sum of
the eigenspaces of the conformal Hamiltonian,
from the local conformal net
$\{\A_{(V,(\cdot\mid\cdot))}(I)\}$.
\end{theorem}

This is proven by constructing the smeared vertex operators 
from abstract considerations using only the local conformal
net $\{\A_{(V,(\cdot\mid\cdot))}(I)\}$.

We note the $C_2$-cofiniteness has some formal similarity to complete
rationality as follows.
By considering $V/\{v_{(-n)} w\mid v, w\in V\}$, we can
define $C_n$-cofiniteness, but $C_1$-cofiniteness is
trivial with codimension $0$ and $C_n$-cofiniteness for
$n>2$ follows from $C_2$-cofiniteness.  In the definition
of the $\mu$-index, we can split $S^1$ into $2n$ intervals,
and consider a subfactor generated by the factors corresponding
to alternating intervals which is contained into the
commutant of the factor corresponding to the other alternating
intervals.  Let $\mu_n$ be the index of this subfactor.
The Haag duality implies we always have $\mu_1=1$.  The
finiteness of $\mu_2$ implies finiteness of all other
$\mu_n$.  (See \cite{KLM} for more details.)
Based on this analogy, we have the following conjecture.

\begin{conjecture}
We have a bijective correspondence between 
completely rational local conformal nets and
unitary $C_2$-cofinite vertex operator algebras.
We also have equivalence of unitary fusion categories for
finite dimensional representations of a 
completely rational local conformal net and
modules of the corresponding vertex operator
algebras.  We further have coincidence of
the corresponding characters of the
finite dimensional  representations
of a completely rational
local conformal net and modules of the
corresponding vertex operator algebra.
\end{conjecture}

We remark that a relation between local conformal
nets and unitary vertex operator algebras is somehow
similar to that between  Lie groups and Lie algebras.
The relation between loop groups and Kac-Moody Lie
algebras is somewhere between the two relations.

Based on the above conjectured correspondence,
we also list the following problem.

\begin{problem}
For a given finite group $G$, construct a local
conformal net whose automorphism group is $G$.
The construction should be ``as natural as possible''.
\end{problem}

We believe this problem has a positive solution and
it would produce the Moonshine local conformal net
if the group $G$ is the Monster group.  This is based
on the same reason as the one given after Problem 
\ref{prob-tensor}.

\section{Other types of conformal field theory}

Besides the $2$-dimensional chiral conformal field theory, 
we also have other types of $2$-dimensional conformal
field theory.  Here we briefly mention other settings.

\subsection{Full conformal field theory}

A full conformal field theory is a quantum field theory
on $2$-dimensional Minkowski space 
$\M=\{(x,t)\mid x,t\in\R\}$.  We present the 
formulation following \cite{KL2}.

We have light ray coordinates $\xi_\pm=t\pm x$, and
set $\LL_\pm=\{\xi\mid\xi_\pm=0\}$, the two light 
ray lines.  A double cone is an open subset of $\M$ of
the form $\OO=I_+\times I_-$ where $I_\pm\subset\LL_\pm$
are bounded intervals.  We set $\K$ to be the set of
double cones.  The group $PSL(2,\R)$  acts
on $\R\cup\{\infty\}$ by fractional linear transformations,
so the actions restricts to a local action on $\R$ as in 
\cite{BGL}.  In particular, if $F\subset \R$
has a compact closure, then there exists a connected
neighborhood $\U$ of the identity in $PSL(2,\R)$ such
that we have $gF\subset \R$ for all $g\in\U$.  We regard 
this as a local action of the universal covering
group $\widetilde{PSL(2,\R)}$ on $\R$.  We then
have a local product action of 
$\widetilde{PSL(2,\R)}\times \widetilde{PSL(2,\R)}$
on $\M=\LL_+\times\LL_-$.

\begin{definition}{\rm
A local M\"obius covariant net $\{\A(\OO)\}$ is
an assignment of a von Neumann algebra $\A(\OO)$
on a fixed Hilbert space $H$ to $\OO\in\K$ satisfying
the following properties.
\begin{enumerate}
\item (Isotony) For $\OO_1\subset\OO_2$, we have
$\A(\OO_1)\subset\A(\OO_2)$.
\item (M\"obius covariance) There exists a unitary
representation $U$ of 
$\widetilde{PSL(2,\R)}\times \widetilde{PSL(2,\R)}$
on $H$ such that for every double cone $\OO$, we have
$$U(g)\A(\OO)U(g)^*=\A(g\OO),\quad g\in\U,$$
where $\U\subset 
\widetilde{PSL(2,\R)}\times \widetilde{PSL(2,\R)}$
is any connected neighborhood of the
identity with $g\OO\subset\M$ for all $g\in\U$.
\item (Locality) If $\OO_1$ and $\OO_2$ are spacelike
separated, then we have $[\A(\OO_1),\A(\OO_2)]=0$.
\item (Positive energy condition) The one-parameter
unitary subgroup of $U$ corresponding to time
translations has a positive generator.
\item (Existence of the vacuum vector)
There exists a unit $U$-invariant vector $\Omega$
with $\overline{\bigcup_{\OO\in\K}\A(\OO)\Omega}=H$.
\item (Irreducibility) The von Neumann algebra
generated by all $\A(\OO)$ is $B(H)$.
\end{enumerate}
}\end{definition}

Let $\G$ be the quotient of 
$\widetilde{PSL(2,\R)}\times \widetilde{PSL(2,\R)}$
modulo the relation $(r_{2\pi},r_{-2\pi})=(\id,\id)$,
where $r_\theta$ is a rotation by $\theta$.  We
then see that the representation $U$ of
$\widetilde{PSL(2,\R)}\times \widetilde{PSL(2,\R)}$
factors through a representation of $\G$.
(See \cite[Proposition 2.1]{KL2}.)  Then a
local M\"obius covariant net $\{\A(\OO)\}$ 
extends to a local $\G$-covariant net on
the {\sl Einstein cylinder} $\E=\R\times S^1$, the
cover of $S^1\times S^1$ obtained by lifting
the time coordinate from $S^1$ to $\R$.
We use the same symbol $\{\A(\OO)\}$ for the
net on $\M$ and its extension on $\E$.
We have various properties for 
local M\"obius covariant nets as in
\cite[Proposition 2.2]{KL2}.
We can then define a {\sl local conformal net}
on $\M$ by requiring appropriate conformal 
covariance as in \cite[pages 68--69]{KL2}.

For a local M\"obius covariant net  $\{\A(\OO)\}$,
and a bounded interval $I\subset \LL_+$, we
define $\A_+(I)=\bigcap_{\OO=I\times J} \A(\OO)$,
where the intersection is taken over all
intervals $J\subset \LL_-$.  We also have
$\A_-(I)$.  We can restrict $\A_\pm(I)$ to
the Hilbert spaces $H_\pm=\overline{\A_\pm(I)\Omega}$,
which is independent of $I$.  Then we can
regard $\{\A_\pm(I)\}$ as local conformal nets
and $\{\A(\OO)\}$ is an extension of $\{\A_+(I)\otimes
\A_-(J)\}$ on $H_+\otimes H_-\isom H$  as in
\cite[Proposition 2.3, Corollary 2.4]{KL2}.
We also write $\A_L$ and $\A_R$ for $\A+$ and $\A_-$,
respectively.
We also define  a notion of a representation of
$\{\A(\OO)\}$, and the $\mu$-index and
complete rationality of $\{\A(\OO)\}$ 
as in \cite[Section 2.1]{KL2}.

Suppose we have completely rational local conformal
nets $\{\A_L(I)\}$ and $\{\A_R(I)\}$ and
an extension $\{\B(I\times J)\}$ with
$\A_L(I)\otimes \A_R(J)\subset \B(I\times J)$.
Let $\theta=\sum_{i,j} Z_{i,j}\la_i^L\otimes\la_j^R$
be the dual canonical endomorphism of the subfactor
$\A_L(I)\otimes \A_R(J)\subset \B(I\times J)$
where $\{\la_i^L\}$ and $\{\la_j^R\}$ are the
irreducible representations of $\{\A_L(I)\}$
and $\{\A_R(I)\}$, respectively.
Let $S_L, S_R$ be the $S$-matrices of 
$\{\A_L(I)\}$ and $\{\A_R(I)\}$, respectively,
and let $T_L, T_R$ be the $T$-matrices of 
$\{\A_L(I)\}$ and $\{\A_R(I)\}$, respectively.
Then M\"uger has the following.  (See also
\cite[Proposition 6.6]{BKL}.)

\begin{theorem}
\label{full-modular}
The following are equivalent.
\begin{enumerate}
\item The local conformal net $\{\B(\OO)\}$ has only
the trivial irreducible representation.
\item The $\mu$-index of $\{\B(\OO)\}$ is $1$.
\item We have
$T_L Z=ZT_R$ and $S_L Z=ZS_R$.
\end{enumerate}
\end{theorem}

In particular, if we can naturally identify $\{A_L(I)\}$
and $\{A_R(I)\}$ and have $\mu$-index equal to $1$, then
the matrix $Z$ is a modular invariant.
If both  $\{A_L(I)\}$ and $\{A_R(I)\}$ have the same
central charge, we call it the central charge of
$\{\B(\OO)\}$.  
Study of a local conformal net $\{\B(\OO)\}$ is reduced
to that of the following.
\begin{enumerate}
\item The two local conformal nets
$\{A_L(I)\}$ and $\{A_R(I)\}$.
\item The $Q$-system corresponding to the dual
canonical endomorphism for the subfactor
$\A_L(I)\otimes\A_R(J)\subset\B(I\times J)$.
\end{enumerate}
The $Q$-system above is one for the representation
category $\Rep(\A_L)\boxtimes\Rep(\A_R)^\opp$, where
$\opp$ means the unitary modular tensor category where the
braiding is reversed.

For the case the central charge is less
than $1$, we have the following classification theorem.
(See \cite[Theorem 5.5]{KL2}.)

\begin{theorem}
The local conformal nets $\{\B(\OO)\}$ with central
charge less than $1$ which are maximal with respect
to inclusions are in a bijective correspondence
the modular invariants listed in \cite{CIZ}.
\end{theorem}

The modular invariants in \cite{CIZ} are labelled with 
pairs of the $A$-$D$-$E$ Dynkin diagrams with Coxeter
numbers differing by $1$.  That is we have the pairs
$(A_{n-1},A_n)$, $(D_{2n+1},A_{4n})$, $(A_{4n}, D_{2n+2})$,
$(D_{2n+2},A_{4n+2})$, $(A_{4n+2},D_{2n+3})$, $(A_{10}, E_6)$,
$(E_6,A_{12})$, $(A_{16}, E_7)$, $(E_7, A_{18})$,
$(A_{28},E_8)$ and $(E_8,A_{30})$.  The uniqueness
for each pair follows from \cite[Theorem 5.3]{KL2}.

For realization of these modular invariants, we appeal to the
following result \cite[Corollary 1.6]{R3}.

\begin{theorem}
Let $\{\A(I)\}$ be a completely rational local conformal net.
Let $(\theta,v,w)$ be a $Q$-system (without locality assumed)
where $\theta$ is an object in the representation category
$\Rep(\A)$.  Let $(Z_{\la,\mu})$ be the modular invariant 
arising from the $\a$-induction associated with $(\theta,v,w)$
as in  Theorem \ref{mod-inv}.  Then we have a $Q$-system
for the representation $\bigoplus Z_{\la,\mu}\la\oplus \mu$,
hence a $2$-dimensional local conformal net extending
$\{\A(I)\otimes\A(J)\}$.
\end{theorem}

It is expected that the $N=2$ full superconformal field
theory is related to Calabi-Yau manifolds \cite{Ge}, so
we also have the following problem.

\begin{problem}
Construct an operator algebraic object corresponding
to a Calabi-Yau manifold in the setting of 
$N=2$ full superconformal field theory
and study the mirror symmetry in this context.
\end{problem}

\subsection{Boundary conformal field theory}

We present our setting for boundary conformal field theory
based on \cite{LR2}.  (Also see \cite{CKL2},
\cite[Section 6.4]{BKL}.)

Let $\M_+=\{(t,x)\in\M \mid x>0\}$ be the half 
2-dimensional Minkowski space.  Let $\K_+$ be the set
of double cones $\OO$ whose closures are contained in
$\M_+$.  A double cone $\OO\in\K_+$ is represented
as $I\times J$ where $I,J$ are bounded intervals
in $\R$ with $I < J$.  We fix  a completely rational
local conformal net $\{\A(I)\}$ and restrict it to
a net on $\R$ by removing the point $\infty$.

The universal cover $\widetilde{PSL(2,\R)}$
acts globally in the universal cover of $S^1$.
The product action on the chiral lines of $\M$
gives a local action of 
$\widetilde{PSL(2,\R)}\times\widetilde{PSL(2,\R)}$
on $\M$.  We deal with the local action of
$PSL(2,\R)$ obtained by restricting the local
action of $PSL(2,\R)\times PSL(2,\R)$ to
the diagonal.  This action restricts to
local actions of $PSL(2,\R)$ on $\M_+$ and
its boundary, the time-axis.

\begin{definition}{\rm
An assignment of a von Neumann algebra
$\B_+(\OO)$ on a fixed Hilbert space $H_\B$
to each double cone $\OO\in \K_+$ is called
a {\sl boundary net} if it satisfies the
following.
\begin{enumerate}
\item (Isotony) For $\OO_1\subset \OO_2$, we have
$\B_+(\OO_1)\subset \B_+(\OO_2)$.
\item (Locality) If $\OO_1$ and $\OO_2$ are spacelike separated
in $M_+$, then we have $[\B_+(\OO_1),\B_+(\OO_2)]=0$.
\item (M\"obius covariance) There exists a unitary
representation $U$ of $\widetilde{PSL(2,\R)}$ on
the Hilbert space $H_\B$ such that
we have $U(g)\B_+(\OO)U(g)^*=\B_+(g\OO)$ for
every $\OO\in\K_+$ with $g\in \widetilde{PSL(2,\R)}$
having a path of elements 
$g_s\in \widetilde{PSL(2,\R)}$
connecting the identity
of $\widetilde{PSL(2,\R)}$ and $g$ 
satisfying $g_s\OO\in\K_+$ for all $s$.
\item (Positive energy condition) The generator of
the translation one-parameter subgroup of $U$ is
positive.
\item (Existence of the vacuum vector)
We have a unit vector $\Omega\in H_\B$ 
such that $\C\Omega$ are the $U$-invariant 
vectors and we have $\overline{\B_+(\OO)\Omega}=H_\B$
for each $\OO\in \K_+$.
\end{enumerate}
}\end{definition}

Furthermore, we have the following.

\begin{definition}{\rm
Let $\{\A_+(\OO)\}$ be the boundary 
net on $\M_+$ given by
$\A_+(\OO)=\A(I)\vee\A(J)$, where $\OO=I\times J$.
Then a {\sl boundary net $\{\B_+(\OO)\}$ 
associated with $\{\A(I)\}$} is a boundary net
$\{\B_+(\OO)\}$ satisfying the following conditions.
\begin{enumerate}
\item (Joint irreducibility)
There is a representation $\pi$ of $\{\A(I)\}$
on $H_\B$ such that we have $\pi(\A_+(\OO))\subset\B_+(\OO)$
and $U(g)\pi(\A_+(\OO))U(g)^*=\pi(\A_+(g\OO))$
for doubles cones $\OO,g\OO\in K_+$.
\item For each double cone $\OO$, the von Neumann algebra
generated by $B_+(\OO)$ and all algebras $\pi(A(I))$ is
$B(H_\B)$.
\end{enumerate}
}\end{definition}

The above axioms imply that the inclusion
$\pi(\A_+(\OO))\subset \B_+(\OO)$ is irreducible.

Starting with a boundary net
$\{\B_+(\OO)\supset \pi(\A_+(\OO))\}$, we define
the generated net $\{\B^{\gen}(I)\supset \pi(\A(I))\}$
with $I\subset \R$ by
$$\B^\gen(I)=\bigvee_{\OO\in\K_+,\OO\subset W_I} \B_+(\OO)
\supset \pi(\A(I)).$$
where $W_I=\{(t,x)\mid t\pm x\in I\}$ is the left wedge
such that its intersection with the $t$-axis is $I$.
The following is \cite[Proposition 2.5]{LR2}.

\begin{theorem}
The net $\{\B^\gen(I)\}$ is isotonous and covariant
in the sense that we have $U(g) \B^\gen(I) U(g)^*=
\B^\gen(gI)$.
We also have $\pi(\A(I))\subset \B^\gen(I)
\subset (\pi(\A(I'))'$.
\end{theorem}

The net $\{\B^\gen(I)\}$ may not satisfy locality,
though we have relative locality in the sense
$[\pi(\A(I_1)),\B^\gen(I_2)]=0$ for $I_1,I_2$ with
$I_1\cap I_2=\varnothing$.
We say $\{\B^\gen(I)\}$ is a {\sl non-local
extension} of $\{\A(I)\}$ in this case.
(The name ``non-local'' means ``possibly non-local''.)

For a given non-local extension
$\{\B(I)\supset\A(I)\}$ on $\R$, we define
$\B_+^\ind(\OO)=\B(L)\cap \B(K)'$,
where $\OO=I\times J$ and $L\subset K$ with
$L\cap K'=I \cup J$, or equivalently
$\OO=W_L\cap W'_K$.  

The dual net is defined by $\B^\dual_+(\OO)=\B_+(\OO')'$ 
and we have $\B_+^\dual(\OO)=\B_+(\OO)$ if and only if 
$\{\B_+(\OO)\}$ satisfies the Haag duality, where
$\OO'$ is the causal complement of $\OO$.

We have $(\B_+^\ind)^\gen(I)=\B(I)$ and
$(\B^\gen)^\ind_+(\OO)=\B_+^\dual(\OO)=\B_+(\OO)$ if
$\{\B_+(\OO)\}$ already has the Haag duality.
We then have a bijective correspondence between
boundary nets $\{\B_+(\OO)\}$ associated with
$\{\A(I)\}$ with Haag duality and non-local
extensions $\{\B(I)\}$ of $\{\A(I)\}$.

When $\{\A(I)\}$ is the Virasoro net with $c<1$, 
we can classify all irreducible non-local
extensions, hence all boundary nets
associated with $\{\Vir_c(I)\}$ with Haag
duality as follows.

Let $G_1$ be one of the $A$-$D$-$E$ Dynkin diagrams
with Coxeter number $m$.  Let $G_2$ be 
one of the $A$-$D$-$E$ Dynkin diagrams
with Coxeter number $m+1$.  Let $v_1, v_2$ be
vertices of $G_1,G_2$, respectively.  For a
vertex $v$ of a graph, we denote the orbit
of $v$ under the graph automorphisms by $[v]$.
Then we have the following theorem
(\cite[Theorem 3.1]{KLPR}.)

\begin{theorem}
Irreducible non-local extensions of the
Virasoro nets $\{\Vir_c(I)\}$ with $c<1$ are
in a bijective correspondences to
the quadruples $(G_1,[v_1],G_2,[v_2])$
as above.
\end{theorem}

We can pass from a boundary conformal field theory to a 
full conformal field theory and also back by removing
and adding the boundary.
See \cite{LR3}, \cite{CKL2}, \cite{BKL} for these
relations between full and boundary
conformal field theories.

We also have results on the
phase boundaries and topological defects
in the operator algebraic
setting.  See \cite{BKLR1}, \cite{BKLR2} for details.
See \cite{FuRS1}, \cite{PZ} for earlier works on topological defects.


\begin{thebibliography}{999}  

\bibitem{AH}
M. Asaeda and U. Haagerup,
Exotic subfactors of finite depth with Jones indices
${(5+\sqrt{13})}/{2}$ and ${(5+\sqrt{17})}/{2}$,
\textit{Comm. Math. Phys.} {\bf 202} (1999), 1--63.

\bibitem{BS} 
F. A. Bais, J. K. Slingerland,
Condensate induced transitions between topologically ordered phases,
\textit{Phys. Rev. B} {\bf 79} (2009), 045316.

\bibitem{BK} 
B. Bakalov and A. Kirillov, Jr.,
``Lectures on tensor categories and modular functors'',
American Mathematical Society, Providence (2001).

\bibitem{BDH1} 
A. Bartels, C. L. Douglas and A. Henriques,
Conformal nets I: coordinate-free nets,
\textit{Int. Math. Res. Not.} {\bf 2015} (2015), 4975--5052. 

\bibitem{BDH2}
A. Bartels, C. L. Douglas and A. Henriques,
Conformal nets II: conformal blocks,
arXiv:1409.8672.

\bibitem{BDH3}
A. Bartels, C. L. Douglas and A. Henriques,
Conformal nets III: fusion of defects,
arXiv:1310.8263.

\bibitem{Ba} 
H. Baumg\"artel, ``Operator algebraic methods in 
quantum field theory. A series of lectures'', Akademie Verlag, 
Berlin (1995).

\bibitem{BPZ}
A. A. Belavin, A. M. Polyakov and A. B. Zamolodchikov, 
Infinite conformal symmetry in two-dimensional quantum field theory, 
\textit{Nucl. Phys.} {\bf 241} (1984), 333--380.

\bibitem{BMPS}
S. Bigelow, S. Morrison, E. Peters and N. Snyder,
Constructing the extended Haagerup planar algebra,
\textit{Acta Math.} {\bf 209} (2012) 29--82.

\bibitem{BKL}
M. Bischoff, Y. Kawahigashi and R. Longo,
Characterization of 2D rational local conformal nets and its
boundary conditions: the maximal case, arXiv:1410.8848.

\bibitem{BKLR1}
M. Bischoff, Y. Kawahigashi, R. Longo and K.-H. Rehren,
Phase boundaries and algebraic conformal QFT,
arXiv:1405.7863.

\bibitem{BKLR2}
M. Bischoff, Y. Kawahigashi, R. Longo and K.-H. Rehren,
Tensor categories of endomorphisms and inclusions of von Neumann algebras,
\textit{SpringerBriefs in Mathematical Physics}, {\bf 3} 
Springer Verlag, Berlin (2015).

\bibitem{BE1}
J. B\"ockenhauer and D. E. Evans, 
Modular invariants, graphs and $\alpha$-induction
for nets of subfactors I.
\textit{Comm. Math. Phys.} {\bf 197} (1998), 361--386.

\bibitem{BE2}
J. B\"ockenhauer and D. E. Evans, 
Modular invariants, graphs and $\alpha$-induction
for nets of subfactors II.
\textit{Comm. Math. Phys.} {\bf 200} (1999), 57--103.

\bibitem{BE3}
J. B\"ockenhauer and D. E. Evans, 
Modular invariants, graphs and $\alpha$-induction
for nets of subfactors III.
\textit{Comm. Math. Phys.} {\bf 205} (1999), 183--228.

\bibitem{BE4}
J. B\"ockenhauer and D. E. Evans, 
Modular invariants from subfactors:
Type I coupling matrices and intermediate subfactors,
\textit{Comm. Math. Phys.} {\bf 213} (2000), 267--289.

\bibitem{BEK1}
J. B\"ockenhauer, D. E. Evans and Y. Kawahigashi,
On $\alpha$-induction, chiral projectors and modular
invariants for subfactors,
\textit{Comm. Math. Phys.} {\bf 208} (1999), 429--487.

\bibitem{BEK2}
J. B\"ockenhauer, D. E. Evans and Y. Kawahigashi,
Chiral structure of modular invariants for subfactors,
\textit{Comm. Math. Phys.} {\bf 210} (2000),  733--784. 

\bibitem{BEK3}
J. B\"ockenhauer, D. E. Evans and Y. Kawahigashi,
Longo-Rehren subfactors arising from $\alpha$-induction,
\textit{Publ. Res. Inst. Math. Sci.} {\bf 37} (2001), 1--35. (

\bibitem{B}
R. E. Borcherds,
{\it Monstrous moonshine and monstrous Lie superalgebras},
\textit{Invent. Math.} {\bf 109} (1992), 405--444.

\bibitem{BGL}
R. Brunetti, D. Guido and R. Longo,
Modular structure and duality in conformal quantum field theory, 
\textit{Comm. Math. Phys.} {\bf 156} (1993), 201--219.

\bibitem{BSM}
D. Buchholz and H. Schulz-Mirbach,
Haag duality in conformal quantum field theory,
\textit{Rev. Math. Phys.} {\bf 2} (1990), 105--125.

\bibitem{BZ} G. Burde and H. Zieschang,
``Knots'', Walter de Gruyter \& Co. (1985).

\bibitem{CE}
D. Calaque and P. Etingof, 
Lectures on tensor categories,
``Quantum groups'', 1--38, 
\textit{IRMA Lect. Math. Theor. Phys.} {\bf 12},
Eur. Math. Soc., Z\"urich (2008). 

\bibitem{CIZ}
A. Cappelli, C. Itzykson and J.-B. Zuber,
The $A$-$D$-$E$ classification of minimal and
$A^{(1)}_1$ conformal invariant theories,
\textit{Comm. Math. Phys.} {\bf 113} (1987), 1--26.

\bibitem{Ca} 
S. Carpi, 
On the representation theory of Virasoro nets,
\textit{Comm. Math. Phys.} {\bf 244} (2004), 261--284. 

\bibitem{CHKL}
S. Carpi, R. Hillier, Y. Kawahigashi and R. Longo,
Spectral triples and the super-Virasoro algebra,
\textit{Comm. Math. Phys.} {\bf  295} (2010), 71--97. 

\bibitem{CHKLX}
S. Carpi, R. Hillier, Y. Kawahigashi, R. Longo and F. Xu,
$N=2$ superconformal nets,
\textit{Comm. Math. Phys.} {\bf 336} (2015), 1285--1328.

\bibitem{CKL1}
S. Carpi, Y. Kawahigashi and R. Longo,
Structure and classification of superconformal nets,
\textit{Ann. Henri Poincar\'e} {\bf 9} (2008), 1069--1121. 

\bibitem{CKL2}
S. Carpi, Y. Kawahigashi and R. Longo,
How to add a boundary condition, 
\textit{Comm. Math. Phys.} {\bf 322} (2013), 149--166. 

\bibitem{CKLW}
S. Carpi, Y. Kawahigashi, R. Longo and M. Weiner,
From vertex operator algebras to conformal nets and back,
\textit{Mem. Amer. Math. Soc.} {\bf 254} (2018), no. 1213, vi+85 pp.

\bibitem{CW} S. Carpi and M. Weiner,
On the uniqueness of diffeomorphism symmetry
in conformal field theory, 
\textit{Comm. Math. Phys.} {\bf 258} (2005), 203--221.

\bibitem{C}
A. Connes,
Classification of injective factors cases II$_1$, II$_\infty$,
III$_\lambda$, $\lambda\neq1$,
\textit{Ann. of Math.} {\bf 104} (1976), 73--115.

\bibitem{CN}
J. H. Conway and S. P. Norton,
Monstrous moonshine, \textit{Bull. London Math. Soc.}
{\bf 11} (1979), 308--339.

\bibitem{CS}
J. H. Conway and N. J. A. Sloane,
``Sphere packings, lattices and groups'' (third edition),
Springer Verlag, Berlin (1998).

\bibitem{DFK}
C. D'Antoni, K. Fredenhagen and S. K\"oster, 
Implementation of conformal covariance by diffeomorphism symmetry,
\textit{Lett. Math. Phys.} {\bf 67} (2004), 239--247.

\bibitem{DLR}
C. D'Antoni, R. Longo and F. R\v adulescu, 
Conformal nets, maximal temperature and models from free probability. 
\textit{J. Operator Theory} {\bf 45} (2001), 195--208. 

\bibitem{De}
P. Deligne, 
Cat\'egories tannakiennes, 
in: P. Cartier et al. (Eds.), Grothendieck Festschrift, vol. II,
Birkha\"user, Basel (1991), pp. 111--195.

\bibitem{DMS}
P. Di Francesco, P. Mathieu and D. S\'en\'echal,
``Conformal Field Theory'', Springer Verlag, Berlin (1996).

\bibitem{DVVV}
R. Dijkgraaf, C. Vafa, E. Verlinde and H. Verlinde, 
The operator algebra of orbifold models,
\textit{Comm. Math. Phys.} {\bf 123} (1989), 485--526.

\bibitem{DGH}
C. Dong, R. L. Griess, Jr. and G. H\"ohn,
Framed vertex operator algebras, codes and the Moonshine module,
\textit{Comm. Math. Phys.} {\bf 193} (1998), 407--448.

\bibitem{DL}
C. Dong and X. Lin,
Unitary vertex operator algebras. 
\textit{J. Algebra} {\bf 397} (2014), 252--277. 

\bibitem{DM}
C. Dong and G. Mason,
On quantum Galois theory,
\textit{Duke Math. J.} {\bf 86} (1997), 305--321. 

\bibitem{DMZ}
C. Dong, G. Mason and Y. Zhu,
Discrete series of the Virasoro algebra and the moonshine module,
Proc. Symp. Pure. Math., Amer. Math. Soc.
{\bf 56} II (1994), 295--316.

\bibitem{DX}
C. Dong and F. Xu,
Conformal nets associated with lattices and their orbifolds,
\textit{Adv. Math.} {\bf 206} (2006), 279--306.

\bibitem{DHR1} S. Doplicher, R. Haag and J. E. Roberts,
Local observables and particle statistics, I. 
\textit{Comm. Math. Phys.} {\bf 23} (1971), 199--230.

\bibitem{DHR2} S. Doplicher, R. Haag and J. E. Roberts,
Local observables and particle statistics, II.
\textit{Comm. Math. Phys.} {\bf 35} (1974), 49--85.

\bibitem{Dr}
V. Drinfel$'$d,
Quantum groups,
Proceedings of the International Congress of Mathematicians, 798--820,
American  Mathematical Society (1987). 

\bibitem{ENO} P. Etingof, D. Nikshych and V. Ostrik,
On fusion categories, 
\textit{Ann. of Math.} {\bf 162} (2005) 581--642.

\bibitem{EG}
D. E. Evans and T. Gannon, 
The exoticness and realisability of
twisted Haagerup-Izumi modular data, 
\textit{Comm. Math. Phys.} {\bf 307} (2011), 463--512. 

\bibitem{EK} D. E. Evans and Y. Kawahigashi,
``Quantum Symmetries on Operator Algebras'',
Oxford University Press, Oxford (1998).

\bibitem{F}
K. Fredenhagen,
Generalizations of the theory of superselection sectors,
in: Kastler, D.(Ed.), \textit{The Algebraic Theory of
Superselection Sectors},
World Scientific, (1990), 379--387.

\bibitem{FJ} K. Fredenhagen and M. J\"or\ss,
Conformal Haag-Kastler nets, pointlike localized fields and the
existence of operator product expansion,
\textit{Comm. Math. Phys.}  {\bf 176} (1996), 541--554.

\bibitem{FRS1}
K. Fredenhagen, K.-H. Rehren and B. Schroer,
Superselection sectors with braid group statistics
and exchange algebras,
I. \textit{Comm. Math. Phys.}  {\bf 125} (1989), 201--226;

\bibitem{FRS2}
K. Fredenhagen, K.-H. Rehren and B. Schroer,
Superselection sectors with braid group statistics
and exchange algebras,
II. \textit{Rev. Math. Phys.} {\bf Special issue} (1992), 113--157.

\bibitem{FHL}  
I. Frenkel, Y.-Z. Huang and J. Lepowsky, 
On axiomatic approaches to vertex
operator algebras and modules, 
\textit{Mem. Amer. Math. Soc.} {\bf 104} (1993),  no. 494.

\bibitem{FLM} I. Frenkel, J. Lepowsky and A. Meurman,
``Vertex operator algebras and the Monster'',
Academic Press (1988).

\bibitem{FZ} I. Frenkel and Y. Zhu,
Vertex operator algebras associated to representations 
of affine and Virasoro algebras, \textit{Duke Math. J.}
{\bf 66} (1992), 123--168. 

\bibitem{FQS}
D. Friedan, Z. Qiu and S. Shenker,
Details of the non-unitarity proof for highest weight
representations of the Virasoro algebra,
\textit{Comm. Math. Phys.}  {\bf 107} (1986), 535--542.

\bibitem{FFRS1}
J. Fr\"ohlich, J. Fuchs, I. Runkel and C. Schweigert,
Correspondences of ribbon categories,
\textit{Adv. Math.} {\bf 199} (2006), 192--329.

\bibitem{FFRS2}
J. Fr\"ohlich, J. Fuchs, I. Runkel and C. Schweigert,
Duality and defects in rational conformal
field theory, \textit{Nucl. Phys. B} {\bf 763} (2007), 354--430.

\bibitem{FGR1}
J. Fr\"ohlich, O. Grandjean and A. Recknagel, 
Supersymmetric quantum theory and differential geometry,
\textit{Comm. Math. Phys.} {\bf 193} (1998),  527--594. 

\bibitem{FGR2}
J. Fr\"ohlich, O. Grandjean and A. Recknagel, 
Supersymmetric quantum theory and non-commutative geometry,
\textit{Comm. Math. Phys.} {\bf 203} (1999), 119--184. 

\bibitem{FuRS1}
J. Fuchs, I. Runkel, C. Schweigert, 
TFT construction of RCFT correlators I: partition
functions, \textit{Nucl. Phys. B} {\bf 646} (2002),
353--497.

\bibitem{FuRS2}
J. Fuchs, I. Runkel, C. Schweigert, 
TFT construction of RCFT correlators II: unoriented worldsheets,
\textit{Nucl. Phys. B} {\bf 678} (2004),
511--637.

\bibitem{FuRS3}
J. Fuchs, I. Runkel, C. Schweigert, 
TFT construction of RCFT correlators. III. 
Simple currents, 
\textit{Nuclear Phys. B} {\bf 694} (2004), 277--353. 

\bibitem{FG}
F. Gabbiani and J. Fr\"ohlich, 
Operator algebras and conformal field theory. 
\textit{Comm. Math. Phys.}  {\bf 155} (1993),  569--640. 

\bibitem{G1} T. Gannon,
The classification of affine $SU(3)$ modular invariant 
partition functions,
\textit{Comm. Math. Phys.}  {\bf 161} (1994), 233--263. 

\bibitem{G2} T. Gannon,
``Moonshine Beyond The Monster: The Bridge Connecting
Algebra, Modular Forms And Physics'',
Cambridge University Press (2006).

\bibitem{Ge} D. Gepner, 
Exactly solvable string compactifications on manifolds of $SU(N)$ holonomy, 
\textit{Phys. Lett. B} {\bf 199} (1987), 380--388.

\bibitem{GKO}
P. Goddard, A. Kent and D. Olive,
Unitary representations of the Virasoro and
super-Virasoro algebras,
\textit{Comm. Math. Phys.}  {\bf 103} (1986), 105--119.

\bibitem{GW}
R. Goodman and N. R. Wallach,
Projective unitary positive-energy representations of
$\Diff(S^1)$,
\textit{J. Funct. Anal.} {\bf 63}, (1985) 299--321.

\bibitem{Gr}
R. L. Griess, Jr.,
The friendly giant,
\textit{Invent. Math.} {\bf 69} (1982), 1--102.

\bibitem{GL}
D. Guido and R. Longo, The conformal spin and statistics theorem,
\textit{Comm. Math. Phys.}  {\bf 181} (1996), 11--35.

\bibitem{Ha} R. Haag,
``Local Quantum Physics'', Springer (1996).

\bibitem{Hi} F. Hiai,
Minimizing indices of conditional expectations onto a subfactor,
\textit{Publ. Res. Inst. Math. Sci.} {\bf 24} (1988), 673--678. 

\bibitem{Ho}
G. H\"ohn, 
Genera of vertex operator algebras and three-dimensional 
topological quantum field theories, 
in ``Vertex operator algebras in mathematics and physics'',
89--107, Fields Inst. Commun. {\bf 39},
Amer. Math. Soc. (2003). 

\bibitem{H1} Y.-Z. Huang,
Vertex operator algebras and the Verlinde conjecture,
\textit{Commun. Contemp. Math.} {\bf 10} (2008), 103--154. 

\bibitem{H2} Y.-Z. Huang,
Rigidity and modularity of vertex tensor categories,
\textit{Commun. Contemp. Math.} {\bf 10} (2008), 871--911. 

\bibitem{HKL} Y.-Z. Huang, A. Kirillov, Jr. and J. Lepowsky,
Braided tensor categories and extensions of vertex operator algebras,
\textit{Comm. Math. Phys.}, {\bf 337} (2015), 1143--1159.

\bibitem{I1} M. Izumi,
Subalgebras of infinite $C^*$-algebras with finite Watatani indices.
II. Cuntz-Krieger algebras, \textit{Duke Math. J.} {\bf 91}
(1998), 409--461. 

\bibitem{I2} M. Izumi,
The structure of sectors associated with the Longo-Rehren
inclusions. I. General Theory, 
\textit{Comm. Math. Phys.} {\bf 213} (2000), 127--179.

\bibitem{I3} M. Izumi,
The structure of sectors associated with Longo-Rehren inclusions.
II. Examples,
\textit{Rev. Math. Phys.} {\bf 13} (2001), 603--674.

\bibitem{ILP} M. Izumi, R. Longo and S. Popa
A Galois correspondence for compact groups of automorphisms
of von Neumann algebras with a generalization to Kac algebras,
\textit{J. Funct. Anal.} {\bf 10} (1998), 25--63.

\bibitem{IK} M. Izumi and H. Kosaki,
On a subfactor analogue of the second cohomology,
\textit{Rev. Math. Phys.} {\bf 14} (2002), 733--757.

\bibitem{J1}
V. F. R. Jones, Index for subfactors, 
\textit{Invent. Math.} {\bf 72} (1983), 1--25.

\bibitem{J2}
V. F. R. Jones,
A polynomial invariant for knots via von Neumann algebras,
\textit{Bull. Amer. Math. Soc.} {\bf 12} (1985), 103--111. 

\bibitem{J3}
V. F. R. Jones, 
Planar algebras, I,
arXiv:math/9909027.

\bibitem{JMS} 
V. F. R. Jones, S. Morrison and N. Snyder, 
The classification of subfactors of index at most 5,
\textit{Bull. Amer. Math. Soc. (N.S.)} {\bf 51} (2014), 277--327.

\bibitem{Kc}
V. Kac, ``Vertex algebras for beginners'' (Second edition),
University Lecture Series, {\bf 10}, 
American Mathematical Society (1998).

\bibitem{K1}
Y. Kawahigashi, 
Conformal field theory and operator algebras, 
in ``New Trends in Mathematical Physics'', Springer (2009), 345--356.

\bibitem{K2}
Y. Kawahigashi,
From operator algebras to superconformal field theory,
\textit{J. Math. Phys.} {\bf 51} (2010), 015209, 20 pp.

\bibitem{KL1}
Y. Kawahigashi and R. Longo, 
Classification of local conformal nets. Case $c < 1$,
\textit{Ann. of Math.} {\bf 160} (2004), 493--522.

\bibitem{KL2}
Y. Kawahigashi and R. Longo,
Classification of two-dimensional local conformal nets with $c<1$
and 2-cohomology vanishing for tensor categories,
\textit{Comm. Math. Phys.} {\bf 244} (2004), 63--97.

\bibitem{KL3}
Y. Kawahigashi and R. Longo,
Noncommutative spectral invariants and black hole entropy,
\textit{Comm. Math. Phys.} {\bf 257} (2005), 193--225.

\bibitem{KL4}
Y. Kawahigashi and R. Longo,
Local conformal nets arising from framed vertex operator algebras,
\textit{Adv. Math.} {\bf 206} (2006), 729--751. 

\bibitem{KLM}
Y. Kawahigashi, R. Longo  and M. M\"uger, 
Multi-interval subfactors and modularity of representations
in conformal field theory,
\textit{Comm. Math. Phys.} {\bf 219} (2001), 631--669.

\bibitem{KLPR}
Y. Kawahigashi, R. Longo, U. Pennig and  K.-H. Rehren,
The classification of non-local chiral CFT with $c<1$,
\textit{Comm. Math. Phys.}  {\bf 271} (2007), 375--385.

\bibitem{KSW}
Y. Kawahigashi, N. Sato and M. Wakui, 
(2+1)-dimensional topological quantum field theory
from subfactors and Dehn surgery formula for 3-manifold invariants,
\textit{Adv. Math.} {\bf 15} (2005), 165--204. 

\bibitem{KS}
Y. Kawahigashi and N. Suthichitranont,
Construction of holomorphic local conformal framed nets, 
\textit{Internat. Math. Res. Notices} 
{\bf 2014} (2014), 2924--2943. 

\bibitem{KO}
A. Kirillov, Jr. and V. Ostrik, 
On a $q$-analogue of the McKay correspondence and the 
ADE classification of $sl_2$ conformal field theories,
\textit{Adv. Math.} {\bf 171} (2002), 183--227. 

\bibitem{Kn} 
L. Kong, Anyon condensation and tensor categories.
\textit{Nucl. Phys. B} {\bf 886} (2014), 436--482. 

\bibitem{K} 
H. Kosaki,
Extension of Jones' theory on index to arbitrary factors. 
\textit{J. Funct. Anal.} {\bf 66} (1986), 123--140.

\bibitem{Kt}
S. K\"oster, 
Local nature of coset models, 
\textit{Rev. Math. Phys.} {\bf 16} (2004), 353--382. 

\bibitem{LY}
C. H. Lam and H. Yamauchi, 
On the structure of framed vertex operator algebras and
their pointwise frame stabilizers,
\textit{Comm. Math. Phys.} {\bf 277} (2008), 237--285.

\bibitem{L1} R. Longo,
Index of subfactors and statistics of quantum fields, I.
\textit{Comm. Math. Phys.}  {\bf 126} (1989), 217--247.

\bibitem{L2} R. Longo,
Index of subfactors and statistics of quantum fields, II.
Correspondences, braid group statistics and Jones polynomial,
\textit{Comm. Math. Phys.}  {\bf 130} (1990), 285--309.

\bibitem{L3} R. Longo,
Minimal index and braided subfactors, 
\textit{J. Funct. Anal.} {\bf 109} (1992), 98--112. 

\bibitem{L4}
R. Longo, A duality for Hopf algebras and for subfactors,
\textit{Comm. Math. Phys.}  {\bf 159} (1994), 133--150.

\bibitem{L5}
R. Longo, Conformal subnets and intermediate subfactors,
\textit{Comm. Math. Phys.}  {\bf 237} (2003), 7--30.

\bibitem{LR1}
R. Longo  and K.-H. Rehren, 
Nets of Subfactors,
\textit{Rev. Math. Phys.} {\bf 7} (1995), 567--597.

\bibitem{LR2}
R. Longo  and K.-H. Rehren, 
Local fields in boundary conformal QFT,
\textit{Rev. Math. Phys.} {\bf 16} (2004), 909--960.

\bibitem{LR3}
R. Longo  and K.-H. Rehren,
How to remove the boundary in CFT -- an operator algebraic procedure,
\textit{Comm. Math. Phys.} {\bf 285} (2009), 1165--1182.

\bibitem{LX}
R. Longo  and F. Xu,
Topological sectors and a dichotomy in conformal field theory,
\textit{Comm. Math. Phys.} {\bf 251} (2004), 321--364.

\bibitem{Ma}
T. Masuda,
An analogue of Longo's canonical endomorphism for bimodule theory
and its application to asymptotic inclusions,
\textit{Internat. J. Math.} {\bf 8} (1997), 249--265.

\bibitem{M}
M. Miyamoto, A new construction of the moonshine vertex
operator algebra over the real number field,
\textit{Ann. of Math.} {\bf 159} (2004), 535--596.

\bibitem{O1}
A. Ocneanu,
Quantized group, string algebras and Galois theory for algebras,
in {\em Operator algebras and applications, Vol. 2 (Warwick, 1987)},
(ed. D. E.  Evans and M. Takesaki), London Mathematical Society
Lecture Note Series {\bf 36},
Cambridge University Press, Cambridge, (1988), 119--172.

\bibitem{O2}
A. Ocneanu,
``Quantum symmetry, differential geometry of finite graphs and
classification of subfactors'', University of Tokyo Seminary Notes
{\bf 45}, (Notes recorded by Y. Kawahigashi), (1991).

\bibitem{O3}
A. Ocneanu,
Paths on Coxeter diagrams:
from Platonic solids and singularities to minimal models and subfactors,
(Notes recorded by S. Goto), in {\em Lectures on operator theory},
(ed. B. V. Rajarama Bhat et al.),
The Fields Institute Monographs, AMS Publications (2000), 243--323.

\bibitem{O}
V. Ostrik,
Module categories, weak Hopf algebras and modular invariants,
\textit{Transformation Groups} {\bf 8} (2003), 177--206.

\bibitem{Pa}
B. Pareigis, 
On braiding and dyslexia, \textit{J. Alg.} {\bf 171} (1995), 413--425.

\bibitem{PZ}
V. B. Petkova and J.-B. Zuber, 
Generalised twisted partition functions,
\textit{Phys. Lett. B} {\bf 504} (2001), 157--164. 

\bibitem{PP}
M. Pimsner and S. Popa, 
Entropy and index for subfactors. 
\textit{Ann. Sci. \'Ecole Norm. Sup.} (4) {\bf 19} (1986),  57--106. 

\bibitem{P1}
S. Popa, 
Classification of amenable subfactors of type II,
\textit{Acta Math.} {\bf 172} (1994), 163--255. 

\bibitem{P2}
S. Popa, 
Symmetric enveloping algebras, amenability and AFD properties
for subfactors, \textit{Math. Res. Lett.} {\bf 1} (1994), 409--425.

\bibitem{PS} A. Pressley and G. Segal, 
``Loop groups'',
Oxford University Press (1986). 

\bibitem{R1}
K.-H. Rehren,
Braid Group Statistics and their superselection rules,
in: Kastler, D.(Ed.), \textit{The Algebraic Theory of
Superselection Sectors},
World Scientific, (1990), 333--355.

\bibitem{R2}
K.-H. Rehren,
Space-time fields and exchange fields,
\textit{Comm. Math. Phys.} {\bf  132} (1990), 461--483. 

\bibitem{R3}
K.-H. Rehren,
Canonical tensor product subfactors,
\textit{Comm. Math. Phys.} {\bf  211} (2000), 395--406. 

\bibitem{R4}
K.-H. Rehren,
Algebraic conformal quantum field theory in perspective,
arXiv:1501.03313.

\bibitem{RT}
N. Reshetikhin and V. G. Turaev, 
Invariants of 3-manifolds via link polynomials and quantum groups,
\textit{Invent. Math.} {\bf 103} (1991),  547--597. 

\bibitem{Sch} 
A. N. Schellekens, 
Meromorphic $c=24$ conformal field theories,
\textit{Comm. Math. Phys.} {\bf 153} (1993), 159--185.

\bibitem{SY}
A. N. Schellekens and S. Yankielowicz,
Extended chiral algebras and modular invariant partition functions,
\textit{Nuclear Phys. B} {\bf 327} (1989), 673--703. 

\bibitem{S} 
G. Segal, 
The definition of conformal field theory,
in: \textit{Topology, geometry and quantum field theory}, 421--577, 
London Math. Soc. Lecture Note Ser. {\bf 308},
Cambridge Univ. Press (2004). 

\bibitem{SW} R. F. Streater and A. S. Wightman,
``PCT, spin and statistics, and all that'',
Princeton Landmarks in Physics,
Princeton University Press (2000).

\bibitem{T1} M. Takesaki,
``Theory of operator algebras. I'',
Springer-Verlag, Berlin (2002).

\bibitem{T2} M. Takesaki,
``Theory of operator algebras. II'',
Springer-Verlag, Berlin (2003).

\bibitem{T3} M. Takesaki,
``Theory of operator algebras. III'',
Springer-Verlag, Berlin (2003).

\bibitem{TL} V. Toledano Laredo, 
Integrating unitary
representations of infinite-dimensional Lie groups,
\textit{J. Funct. Anal.} {\bf 161} (1999), 478--508.

\bibitem{Tu} V. Turaev, 
``Quantum invariants of knots and 3-manifolds'' 
(Second revised edition), Walter de Gruyter \& Co.
(2010).

\bibitem{W} A. Wassermann,
Operator algebras and conformal field theory III: Fusion
of positive energy representations of $SU(N)$ using bounded operators,
\textit{Invent. Math.} {\bf 133} (1998), 467--538.

\bibitem{X1}
F. Xu,
New braided endomorphisms from conformal inclusions,
\textit{Comm. Math. Phys.} {\bf 192} (1998), 347--403.

\bibitem{X2}
F. Xu,
Jones-Wassermann subfactors for disconnected intervals,
\textit{Comm. Math. Phys.}{\bf 2} (2000), 307--347.

\bibitem{X3}
F. Xu,
Algebraic coset conformal field theories I,
\textit{Comm. Math. Phys.} {\bf 211} (2000), 1--44.

\bibitem{X4}
F. Xu,
Algebraic orbifold conformal field theories,
\textit{Proc. Nat. Acad. Sci. U.S.A.}
{\bf 97} (2000), 14069--14073.

\bibitem{X5}
F. Xu, On a conjecture of Kac-Wakimoto,
\textit{Publ. RIMS, Kyoto Univ.} {\bf 37} (2001), 165--190.

\bibitem{X6}
F. Xu,
Mirror extensions of local nets,
\textit{Comm. Math. Phys.} {\bf 270} (2007), 835--847.

\bibitem{X7}
F. Xu,
An application of mirror extensions. (English summary) 
\textit{Comm. Math. Phys.} {\bf 290} (2009), 83--103. 

\bibitem{Z}
Y. Zhu, Modular invariance of characters of
vertex operator algebras,
\textit{J. Amer. Math. Soc.} {\bf 9} (1996), 237--302.

\end{thebibliography}
\end{document}